\documentclass[a4paper]{iopart}

\usepackage[english]{babel}
\usepackage{caption}
\usepackage[letterpaper,top=2cm,bottom=2cm,left=3cm,right=3cm,marginparwidth=1.75cm]{geometry}

\usepackage{amsmath}
\usepackage{physics}
\usepackage{graphicx}
\usepackage[table]{xcolor}
\usepackage[utf8]{inputenc}
\usepackage{color}
\usepackage{bbm} 
\usepackage[english]{babel}
\usepackage{graphicx}
\usepackage{placeins}
\usepackage{multirow}
\usepackage{physics}
\usepackage{csquotes}
\usepackage{amsfonts,amsmath,amssymb,stmaryrd}
\usepackage{graphicx}
\usepackage{subfigure}  
\usepackage{epsfig}
\usepackage{mathrsfs}
\usepackage{verbatim}
\usepackage{centernot}
\usepackage{ulem}
\usepackage{longtable}
\usepackage{wasysym}
\usepackage{nicefrac}
\usepackage{cite}  
\usepackage[colorlinks=true, allcolors=blue]{hyperref}
\usepackage{bm}	
\renewcommand{\vec}[1]{\bm{#1}}

\begin{document}
\title[]{From Architectures to Applications: A  Review of Neural Quantum States }

\author{Hannah Lange$^{1,2,3}$, Anka Van de Walle$^{2,3}$, Atiye Abedinnia$^{4}$ and Annabelle Bohrdt$^{2,4}$\footnote{Corresponding author email: annabelle.bohrdt@ur.de}}
\address{$^{1}$Max-Planck-Institute for Quantum Optics, Hans-Kopfermann-Str.1, Garching D-85748, Germany}
\address{$^{2}$Munich Center for Quantum Science and Technology, Schellingstr. 4, Munich D-80799, Germany}
\address{$^{3}$Ludwig-Maximilians-University Munich, Theresienstr. 37, Munich D-80333, Germany}
\address{$^{4}$University of Regensburg, Universitätsstr. 31, Regensburg D-93053, Germany}

\begin{abstract}
  \begin{itemize}
  Due to the exponential growth of the Hilbert space dimension with system size, the simulation of quantum many-body systems has remained a persistent challenge until today. Here, we review a relatively new class of variational states for the simulation of such systems, namely neural quantum states (NQS), which overcome the exponential scaling by compressing the state in terms of the network parameters rather than storing all exponentially many coefficients needed for an exact parameterization of the state. We introduce the commonly used NQS architectures and their various applications for the simulation of ground and excited states, finite temperature and open system states as well as NQS approaches to simulate the dynamics of quantum states. Furthermore, we discuss NQS in the context of quantum state tomography.
  \end{itemize}
\end{abstract}

Quantum many-body systems are of great interest for many research areas, including physics, biology and chemistry. However, their simulation has remained challenging until today, due to the exponential growth of the Hilbert space with the system size, making it exceedingly difficult to parameterize the wave functions of large systems using exact methods. 
One common approach to overcome this problem are variational methods, where a certain functional form of the quantum state is assumed, with free parameters to be optimized to obtain the best possible representation of the state under investigation. A well established method based on variational wave functions are tensor networks (TN) \cite{ORUS2014117,Zwolak2004, Evenbly2014,Vidal2007,Shi2006,Cincio2008,verstraete2004renormalization,Klumper_1993}, among them variants that can be contracted efficiently, like matrix product states (MPS) \cite{SCHOLLWOCK201196,White1992, Schollwock2005}, and which hence allow an efficient evaluation of observables. MPS are restricted to states that obey the area law of entanglement \cite{Hastings_2007,Eisert2010}, and are hence particularly well suited for one-dimensional gapped quantum systems, although extensions to higher dimensional systems are possible \cite{ORUS2014117,verstraete2004renormalization,Orus2019}. Another class of methods for the numerical simulation of quantum systems, quantum Monte Carlo (MC) algorithms \cite{becca_sorella_2017,Ceperley1986,Foulkes2001}, suffers from the sign problem \cite{PAN2024879, troyer2005computational} and slow convergence for large system sizes close to critical points or other challenging statistical physics problems \cite{carrasquilla2020machine}.\\

The ability of sufficiently large neural networks to represent any continuous function \cite{Cybenco1989,HORNIK1991251,Kim2003,Roux2008} motivated their use for the simulation of quantum states, and was pioneered by Carleo and Troyer \cite{Carleo2017} in 2017. To date, these so-called neural quantum states (NQS) have been shown to overcome many problems that are inherent to some conventional methods such as MPS: $(i)$ Some works have demonstrated that NQS are capable of representing volume-law entangled states \cite{Sharir2022,Deng2017,Gao2017,denis2023comment,Levine2019} and can hence in principle be used for a broad range of quantum systems \cite{Sharir2022,Gao2017,Lu2019,Levine2019,luo2023gauge,Deng2017,Huang2021}. In particular, it has been shown that in some cases mappings between NQS and efficiently contractable TNs can be established, e.g. in Ref. \cite{Sharir2020} the authors find that TNs are a subset of the considered NQS \cite{Sharir2022}. $(ii)$ They can be designed to be particularly well suited for two-dimensional problems. Most prominently, some architectures like convolutional neural networks were specifically designed for two-dimensional data; $(iii)$ In many cases, they allow for an efficient evaluation of operators, in some cases even global operators like the momentum \cite{lange2023neural}. 

NQS are typically used for two distinct tasks: First, they have appeared in the field of quantum state tomography, where they are used for quantum state reconstruction of states prepared in experiments, allowing the estimation of observables that can not be accessed in experiments \cite{Torlai2020_, Iouchtchenko_2023}. Second, they can be used as simulation tools for quantum systems, with a Hamiltonian driven optimization similar to TNs. In this setting no training data is needed \cite{Carleo2017}. Furthermore, NQS simulations can not only be applied to represent ground states, but also excited states, finite temperature states, the time evolution of quantum states or open systems.\\

The goal of this article is to give an overview of the current state of NQS, i.e. existing NQS architectures, their training and their performance in quantum state simulation and tomography in comparison to conventional methods.  Previously published overviews on neural quantum states can be found in Refs. \cite{carrasquilla2020machine, Dawid2022,Carleo2019,Carrasquilla2021, Carrasquilla2021,Melko2024} in the more general context of neural network applications in quantum physics, or more specifically on NQS and their optimization in Refs. \cite{Jia2019,Yang2019,vivas2022neuralnetwork,Reh2023}. Furthermore, another review on NQS \cite{medvidović2024neuralnetwork} appeared after the publication of the first version of this work. \\

The outline of this review is as follows: We start with an overview of existing NQS architectures and their application to physical systems as well as commonly used design choices. The second part considers applications of NQS, namely the simulation of ground and excited states, finite temperature states, time evolution and open quantum systems. Furthermore, we review quantum state tomography and hybrid simulation schemes with NQS.

\vspace{1cm}
\tableofcontents
\newpage

\section{A Short Introduction to Neural Quantum States (NQS)}

In most cases, neural quantum states (NQS) are used to represent a quantum state 
\begin{align}
    \ket{\psi} = \sum_{\vec{\sigma}}\psi(\vec{\sigma}) \ket{\vec{\sigma}},
\end{align}
for complex $\psi(\vec{\sigma})$ in a certain basis choice $\ket{\vec{\sigma}}$ that is given by the $d_l$ different local configurations, e.g. the spin configurations or Fock space configurations $\vec{\sigma}$. For a system with $N$ sites, $\vec{\sigma}=(\sigma_1,\sigma_2,\dots,\sigma_N)$ consists of e.g. $\sigma_i = 0,1$ for spin systems ($d_l=2$) or $\sigma_i = 0,1,2,\dots n_{\mathrm{max}}$ for bosonic systems ($d_l=n_{\mathrm{max}}+1$). 

The underlying idea of neural quantum states is to use neural networks in order to represent the wave function coefficients $\psi(\vec{\sigma})$ of the state under investigation. Hereby, the neural network is used as a variational wave function, mapping  configurations $\vec{\sigma}$ to the respective wave function coefficient $\psi_{\vec{\theta}}(\vec{\sigma})$, parameterized by the neural network parameters $\vec{\theta}$. More precisely, the input of the neural network used for the NQS representation are configurations $\vec{\sigma}$, and the output is 
\begin{align}
\psi_{\vec{\theta}}(\vec{\sigma})=\sqrt{p_{\vec{\theta}}(\vec{\sigma})} e^{i\phi_{\vec{\theta}}(\vec{\sigma})}, 
\end{align}
which is often split into its amplitude $p_{\vec{\theta}}(\vec{\sigma}) = \vert \psi_{\vec{\theta}}(\vec{\sigma}) \vert^2$ and its phase $\phi_{\vec{\theta}}(\vec{\sigma}) = \mathrm{Im}\left( \mathrm{log}\,\psi_{\vec{\theta}}(\vec{\sigma}) \right)$. To feed an input $\vec{\sigma}$ into the network, $\vec{\sigma}$ can be one-hot encoded, i.e. the $d_l$ different local configurations are encoded binary, resulting in a matrix $\vec{\sigma}\in \mathbb{R}^{N\times d_l}$ for every configuration $\vec{\sigma}$ of length $N$. Note that this is not necessary for spin-$1/2$ systems where the values of the spins are normally mapped to a sequence of $\pm 1$ or $0,1$. Furthermore, the input is often embedded into a space of dimension $d_h$, i.e. using a trainable or physically inspired projection the input is projected onto the $d_h$ dimensional space that the network is operating on.

The main difficulty of variational approaches is to come up with a good representation $\psi_{\vec{\theta}}(\vec{\sigma})$ of the true wave function coefficients. Here, the great strength of neural networks, namely their expressive power, comes into play: Neural networks with at least one hidden layer, a sufficient number of parameters and an arbitrary non-linear activation function have the ability to represent continuous functions of any -- potentially very complicated -- form \cite{Cybenco1989,HORNIK1991251,Kim2003,Roux2008}. This makes them promising candidates for a successful representation $\psi_{\vec{\theta}}$ that is close to the exact wave function $\psi$. In order to obtain this representation  $\psi_{\vec{\theta}}$, the network parameters $\vec{\theta}$ are adjusted during the training of the NQS, i.e. starting from some initialization of the neural network parameters $\vec{\theta}_0$, the network parameters are adjusted such that $\psi_{\vec{\theta}}$ approximates the true wave function $\psi$ at the end of the training.\\

\begin{figure}[t]
\centering
\includegraphics[width=0.98\textwidth]{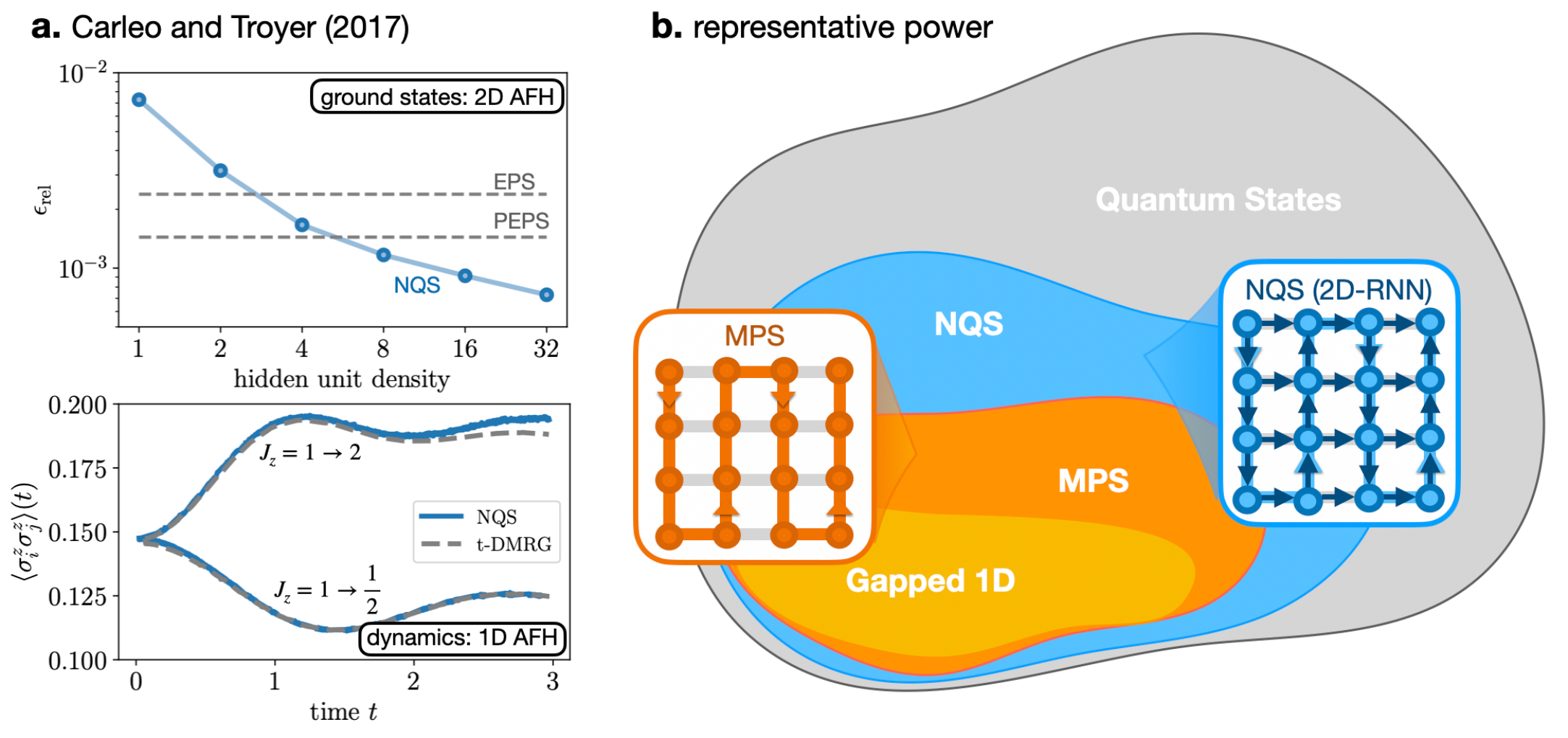}
\caption{Neural quantum states (NQS): a)  In  the seminal work bei Carleo and Troyer \cite{Carleo2017}, NQS are used for ground state representations (top) and dynamics simulations of spin systems (bottom), here for antiferromagnetic Heisenberg (AFH) models on a $10\times 10$ lattice (top) and on an $80$ site chain. Figure adapted from Ref. \cite{Carleo2017}. b) Representative power of Matrix product states (MPS) and NQS. NQS have been proven to have strictly the same or higher expressive power than practically usable variational tensor networks, see e.g. Ref. \cite{Sharir2022}. This results from the high connectivity between sites of NQS, here exemplary shown for a two-dimensional recurrent neural network (2D-RNN, see blue inset) as considered in Ref. \cite{Wu2023} on the top right, compared to MPS, for which information is passed through the system in a 1D manner (see orange inset). Figures adapted from Refs. \cite{Dawid2022,Sharir2022}. } 
\label{fig:Representability}    
\end{figure}

NQS have been first proposed by Carleo and Troyer in 2017 \cite{Carleo2017}. In this first work, NQS have been applied for the ground state and dynamics simulations of spin systems. Fig. \ref{fig:Representability}a shows two exemplary results from the original work for the antiferromagnetic Heisenberg (AFH) model: In the top figure, the relative ground state errors $\epsilon_\mathrm{rel}$ of the NQS energies $E_\mathrm{NQS}$ compared to exact energies $E_\mathrm{exact}$, i.e. $\epsilon_\mathrm{rel}=(E_\mathrm{NQS}-E_\mathrm{exact})/\vert E_\mathrm{exact}\vert$, are shown for a $10\times 10$ square lattice AFH. Already in this early work, NQS achieve competitive results to state-of-the art 2D methods like entangled plaquette states (EPS) and PEPS when using sufficiently many parameters (hidden units). Correspondingly, the NQS dynamics simulations of a spin chain with $80$ sites for two quenches of $J_z$ shows very good agreement with $t$-DMRG.

Starting from this paper, many works on NQS have appeared in recent years, exploring different NQS architectures, training approaches and their application to various physical systems. Furthermore, many works concern the theoretical representative power of NQS. NQS have been shown to be capable of representing a broad range of quantum states \cite{Sharir2022,Gao2017,Lu2019,Levine2019,luo2023gauge,Deng2017,Huang2021}. To compare their expressivity to more conventional variational approaches like TNs and PEPS, the relationship between them has been studied \cite{Chen2018,Glasser2018,Sharir2022,Wu2023}. Some NQS architectures have been proven to have strictly the same or higher expressive power than practically usable variational tensor networks \cite{Sharir2022}, see Fig. \ref{fig:Representability}. In particular, a range of works have shown that some NQS can encode some volume law states without exponential cost \cite{Sharir2022,Deng2017,Gao2017,denis2023comment,Levine2019}, although there are volume-law entangled states like the ground state of the Sachdev-Ye-Kitaev model that can not be represented efficiently \cite{Passetti_2023}.  Furthermore, Ref. \cite{chen2023antn} develops a combination of TNs and autoregressive NQS, which improves the capabilities compared to both the original TN and NQS. However, in contrast to TNs which are guaranteed to converge after a sufficiently long optimization, the training of NQS involves a non-convex landscape. Hence, it can be challenging to find the actual ground state, even when the NQS ansatz itself is expressive enough to capture it. Advanced training strategies to overcome this issue are discussed in Sec. \ref{sec:GS}.\\

In contrast to most machine learning applications, the training can be done in a self-contained way without the use of external data.
In general, the specific design choices for the NQS can have a significant impact on its performance, which will be the focus of Sec. \ref{sec:architectures}: Besides the choice of architecture, e.g. the way how the real and imaginary parts of the wave function coefficients $\psi_{\vec{\theta}}$ are modeled. This can be done by splitting $\psi_{\vec{\theta}}$ into amplitude $p_{\vec{\theta}}$ and phase $\phi_{\vec{\theta}}$ parts and using separate networks or separate output nodes / final layers for each part. Another possibility is to use complex network parameters to model the full $\psi_{\vec{\theta}}$ with a single network. The performance of the wave function does moreover depend on the optimization and the specific task under consideration, which is discussed in Sec. \ref{sec:applications}. 

Similar to Monte Carlo methods, observables of NQS are evaluated by generating samples $\{\vec{\sigma}\}$ from the NQS amplitudes, which are used for the estimation of the respective expectation values.
Specifically, for an operator $\hat{O}$, the expectation value can be written as 
\begin{align}
    \langle \hat{O} \rangle &= \frac{\bra{\psi_{\vec{\theta}}} \hat{O}\ket{\psi_{\vec{\theta}}} }{\langle\psi_{\vec{\theta}}\vert \psi_{\vec{\theta}}\rangle }= \sum_{\vec{\sigma},\vec{\sigma}^\prime}\frac{ \langle \psi_{\vec{\theta}} \vert \vec{\sigma} \rangle \bra{\vec{\sigma}}\hat{O}\ket{\vec{\sigma}^\prime}\langle \vec{\sigma}^\prime \vert \psi_{\vec{\theta}} \rangle}{\sum_{\vec{\sigma}^{\prime\prime}}\langle\psi_{\vec{\theta}}\vert \vec{\sigma}^{\prime\prime} \rangle \langle \vec{\sigma}^{\prime\prime}\vert \psi_{\vec{\theta}}\rangle} = \sum_{\vec{\sigma}}  P_{\vec{\theta}}(\vec{\sigma})\sum_{\vec{\sigma}^\prime} \frac{\psi_{\vec{\theta}}(\vec{\sigma}^\prime)}{ \psi_{\vec{\theta}}(\vec{\sigma})}\bra{\vec{\sigma}}\hat{O}\ket{\vec{\sigma}^\prime}
    \approx \langle O^\mathrm{loc}_{\vec{\theta}}(\vec{\sigma}) \rangle_{\vec{\sigma}}
    \label{eq:expectations}
\end{align}
with the probability for each configuration  $P_{\vec{\theta}}(\vec{\sigma})$ and the local estimator $O^\mathrm{loc}_{\vec{\theta}}(\vec{\sigma})$, defined as 
\begin{align}
P_{\vec{\theta}}(\vec{\sigma})=\frac{\vert \psi_{\vec{\theta}}(\vec{\sigma})\vert^2}{\sum_{\vec{\sigma}^{\prime\prime}}  \vert \psi_{\vec{\theta}}(\vec{\sigma}^{\prime\prime})\vert^2}\quad \mathrm{and} \quad
    O^\mathrm{loc}_{\vec{\theta}}(\vec{\sigma}) = \sum_{\vec{\sigma}^\prime} \frac{\psi_{\vec{\theta}}(\vec{\sigma}^\prime)}{ \psi_{\vec{\theta}}(\vec{\sigma})}\bra{\vec{\sigma}}\hat{O}\ket{\vec{\sigma}^\prime}
\end{align}
respectively, as well as the Monte Carlo average $\langle \cdot\rangle_{\vec{\sigma}} $. For operators involving only a limited number of matrix elements, namely local operators or global operators that do not require the calculation of higher order correlations, $ O^\mathrm{loc}_{\vec{\theta}}$ can be evaluated very efficiently \cite{Nest}. The computational cost of Eq. \eqref{eq:expectations} results from the generation of samples $\vec{\sigma}$ from $\vert \psi_{\vec{\theta}}\vert^2$, as well as from the evaluation of the wave function amplitudes for $\vec{\sigma}$ and its connected samples $\vec{\sigma}^\prime$. The computational cost of the former strongly depends on $\psi_{\vec{\theta}}$ being normalized or not, since in the latter case samples can not directly be generated from the wave function and more elaborate approaches like Metropolis sampling are needed. Normalized NQS, using so-called autoregressive architectures, are the topic of Sec. \ref{sec:autoregressive}. Autoregressive NQS can be designed to obey certain symmetries like $U(1)$ symmetries as discussed in Sec. \ref{sec:RNN}. For non-autoregressive NQS symmetries can be taken into account in the Monte Carlo sampling.  \\

\begin{table}[t]
\hspace{-0.7cm}
\begin{tabular}{|c|c|c|c|}
\hline
\rowcolor[HTML]{E7E6E6} 
 &   & \multicolumn{2}{c|}{\cellcolor[HTML]{E7E6E6}} \\ 
\rowcolor[HTML]{E7E6E6} 
 &   & \multicolumn{2}{c|}{\cellcolor[HTML]{E7E6E6}\textbf{Exemplary Works and Considered Physical Systems}} \\ 
  \rowcolor[HTML]{E7E6E6}   &    & & \\
\rowcolor[HTML]{E7E6E6} 
                 \textbf{Architecture}      &         \textbf{Features}                 & Ground and Excited States,  &    \\
                      \rowcolor[HTML]{E7E6E6} 
 &   & Finite $T$ States &Open Systems and Dynamics\\
 \rowcolor[HTML]{E7E6E6}   &    & & \\ \hline
&                &  &  \\
\textbf{Feed Forward} &                  & (frustrated) spin systems \cite{Choo2018, Cai2018}, & spin systems \cite{zhang2024paths} \\
\textbf{Neural Networks}  &    simplicity              & bosons \cite{Choo2018, Cai2018, ceven2022, Zhu2023, Saito2018} &  -- \\
(Sec. \ref{sec:FFNN}) &                  &  &  \\
 &                   &  &  \\\hline
 &                  &  &  \\
 &                 & (frustrated) spin systems \cite{Westerhout2020,Ferrari2019,Nomura2021_Dirac, Nomura_2021, Vieijra2020,Vieijra2021},  &  \\
 \textbf{Restricted }&       well studied,          & spin liquids \cite{Glasser2018,Li2021}, topologically   & spin systems in 1D \\
  \textbf{Boltzmann}  &     e.g. in terms of            &  ordered states \cite{Deng2017topological,Clark_2018,Kaubruegger2018,Lu2019,Valenti2022},   &  \cite{Carleo2017,czischek2018quenches, zhang2024paths},  ladders \cite{Hofmann2022}   \\
\textbf{Machines} &           expressivity;     & bosons \cite{Saito2017,Choo2018,McBrian_2019,Vargas2020},    &  and 2D \cite{schmitt2020quantum,Fabiani2019,Fabiani2021}\\
(Sec. \ref{sec:RBM}) &   interpretability            & fermions \cite{Normura2017}, molecules \cite{Xia2018}   &  \\
 &                 &  &  \\\hline

 &                  &  &  \\
 \textbf{Convolutional} &    incorporate             & frustrated spin systems   & 2D spin systems  \\
\textbf{Neural Networks} &    lattice              &\cite{Carleo2017,liang2018solving,choo2019two,Szabo2020,Liang2021,Li2022,Liang_2023,wang2023variational,Reh2023,chen2023efficient} on various & \cite{schmitt2020quantum,MendesSantos2023,Schmitt2022_quantumphase} \\
(Sec. \ref{sec:CNN}) &      symmetries           &  lattice geometries \cite{Fu2022,roth2021group,roth2023high,duric2024spin12,beck2024phase}  &  \\
 &                  &  &  \\\hline

 &                 &  &  \\
 \textbf{Graph} &       applicable to          &  (frustrated) spin systems  \cite{kochkov2021learning} &  \\
\textbf{Neural Networks} &     any lattice            & and bosons \cite{yang2020scalable} on various & -- \\
(Sec. \ref{sec:GraphNN}) &     geometry             &  lattice geometries &  \\
 &                 &  &  \\\hline


 &                  &  &  \\
\textbf{Transformer} &         self-attention       & (frustrated) spin systems \cite{viteritti_transformer_2023, zhang_transformer_2023, rende_optimal_2023, rende_simple_2023}, &  spin systems \\
  \textbf{Neural Networks} &        mechanism             & Rydberg states \cite{sprague2023variational, fitzek2024rydberggpt,lange2024transformer}, &  \cite{cha2021attention, Luo2022, Carrasquilla2021} \\
(Sec. \ref{sec:Transformers}) &                 & quantum chemistry \cite{Cha_2022,vonglehn2023selfattention,shang2023solving,wu2023nnqstransformer} &  \\
 &                  &  &  \\\hline
 &                  &  &  \\
 &                  & (frustrated) spin systems  &  \\
 \textbf{Autoregressive} &  perfect                & \cite{Sharir2020,Schmale2022,zhang_transformer_2023,hibat-allah_recurrent_2020,roth_iterative_2020,morawetz_u1-symmetric_2021}, spin glass \cite{hibat-allah_variational_2021},  & spin systems  \\
  \textbf{Neural Networks} &      sampling             & topologically ordered (bosonic)   &\cite{Reh2021,Luo2022},   \\
(Sec. \ref{sec:autoregressive}) &                   & states  \cite{HibatAllah2023,Doeschl2023}, Rydberg states   & Rydberg states \\
 &                  & \cite{moss_enhancing_2023,sprague2023variational,lange2024transformer}, fermions \cite{malyshev2023autoregressive,lange2023neural} & \cite{mendessantos2023wave}  \\
  &                  &  &  \\\hline
\end{tabular}
\caption{Overview of NQS architectures and their applications to ground, excited states, dynamics, finite temperature states, and open systems.}
\label{tab:architectures}
\end{table}

\begin{figure}[t]
\centering
\includegraphics[width=0.7\textwidth]{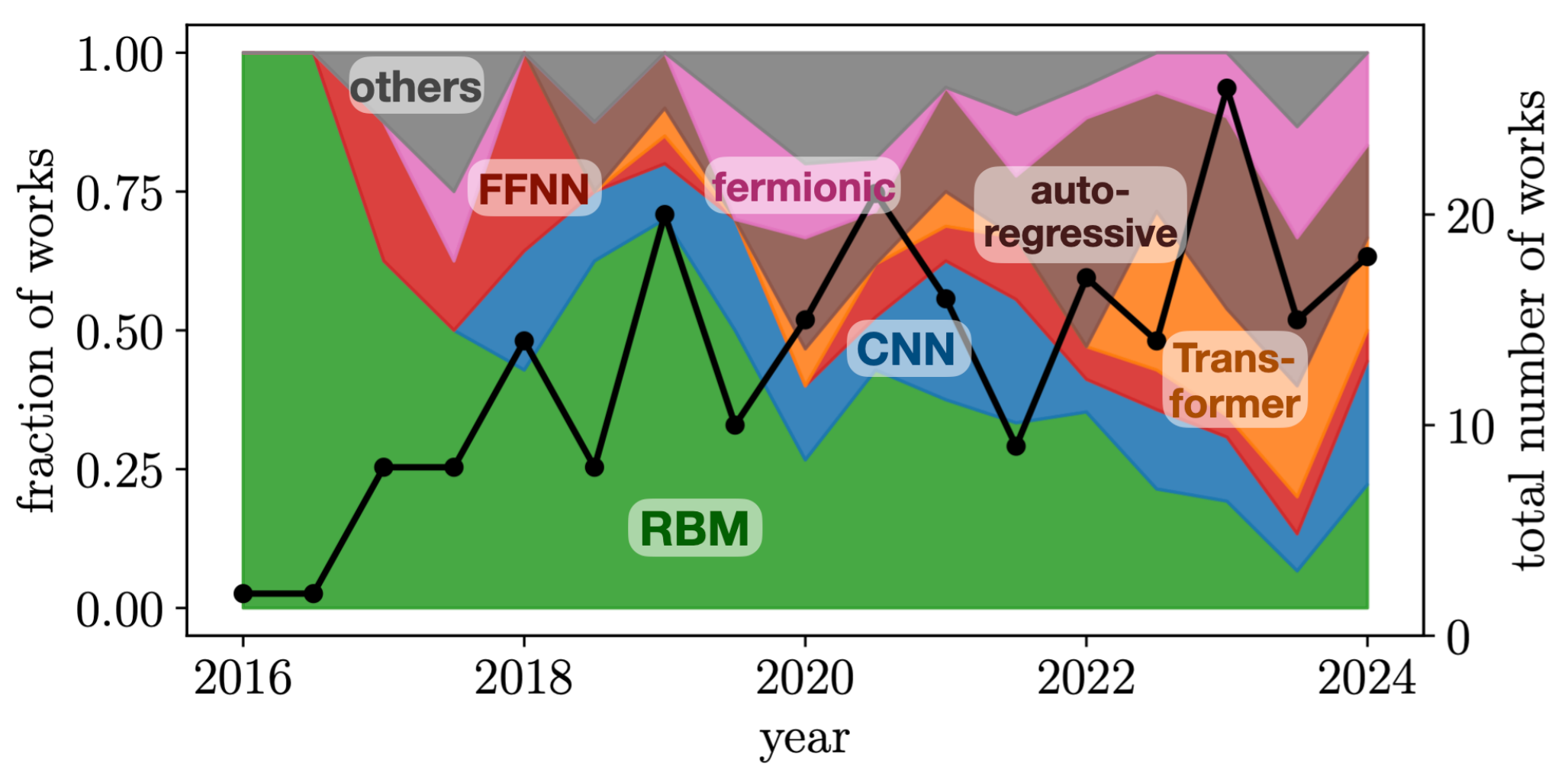}
\caption{Architectures used in the works cited in this review in a time interval of 6 months from 2016 to June 2024: The fraction of works with implementations of variants of feed forward neural networks (FFNNs, Sec. \ref{sec:FFNN}), restricted Boltzmann machines (RBMs, Sec. \ref{sec:RBM}), convolutional neural networks (CNNs, Sec. \ref{sec:CNN}), transformers (Sec. \ref{sec:Transformers}), autoregressive networks (Sec. \ref{sec:autoregressive}), fermionic networks (Sec. \ref{sec:fermions}) or other architectures. The total number of works is shown in black; note that for 2024 we consider only works until June 2024. In some works, several architectures are used, and are taken into account as separate works for each architecture here. Autoregressive transformers are counted twice.} 
\label{fig:timeline}    
\end{figure}

\section{NQS Architectures \label{sec:architectures}}

Neural network quantum states can be implemented using several techniques, including various neural network architectures and different representations of phase and amplitude parts of the wave function. Each architecture comes with its advantages and specialized training strategies, see also Ref. \cite{MEHTA20191}. Additionally, the choice of architecture can also depend on the physical model under investigation. 

In this section, we discuss commonly used architectures, their application to physical systems in the literature as well as their advantages and downsides compared to other ansätze.
In Fig. \ref{fig:timeline} the type of architectures used in the works cited in this review are shown. It can be seen that the field started with representations in terms of restricted Boltzmann machines (RBMs, Sec. \ref{sec:RBM}), as in the seminal work by Carleo and Troyer in 2017 \cite{Carleo2017}. In the first two years after that, mainly works with NQS based on convolutional neural networks (CNNs, Sec. \ref{sec:CNN}) and feed forward neural networks (FFNNs, Sec. \ref{sec:FFNN}) appeared, while in recent years autoregressive networks (Sec. \ref{sec:autoregressive}) and transformer neural networks (Sec. \ref{sec:Transformers}) have gained attention. Furthermore, after a focus on spin systems in the early stage of development of NQS architectures, the field turns towards the simulation of fermionic systems (Sec. \ref{sec:fermions}). At the time of publication of this (revised) work (June 2024), works on many different architectures appear with a similar fraction. To the best of our knowledge, this is mostly due to the fact that it is not clear at first sight and which NQS architecture is most suitable for a given physical problem is a major open question in the field. Nevertheless, most architectures have certain strengths, which we attempt to summarize in Tab. \ref{tab:architectures} and in the remaining part of this section.

\subsection{Feed Forward Neural Networks (FFNNs) \label{sec:FFNN}}

\begin{figure}[t]
\centering
\includegraphics[width=0.4\textwidth]{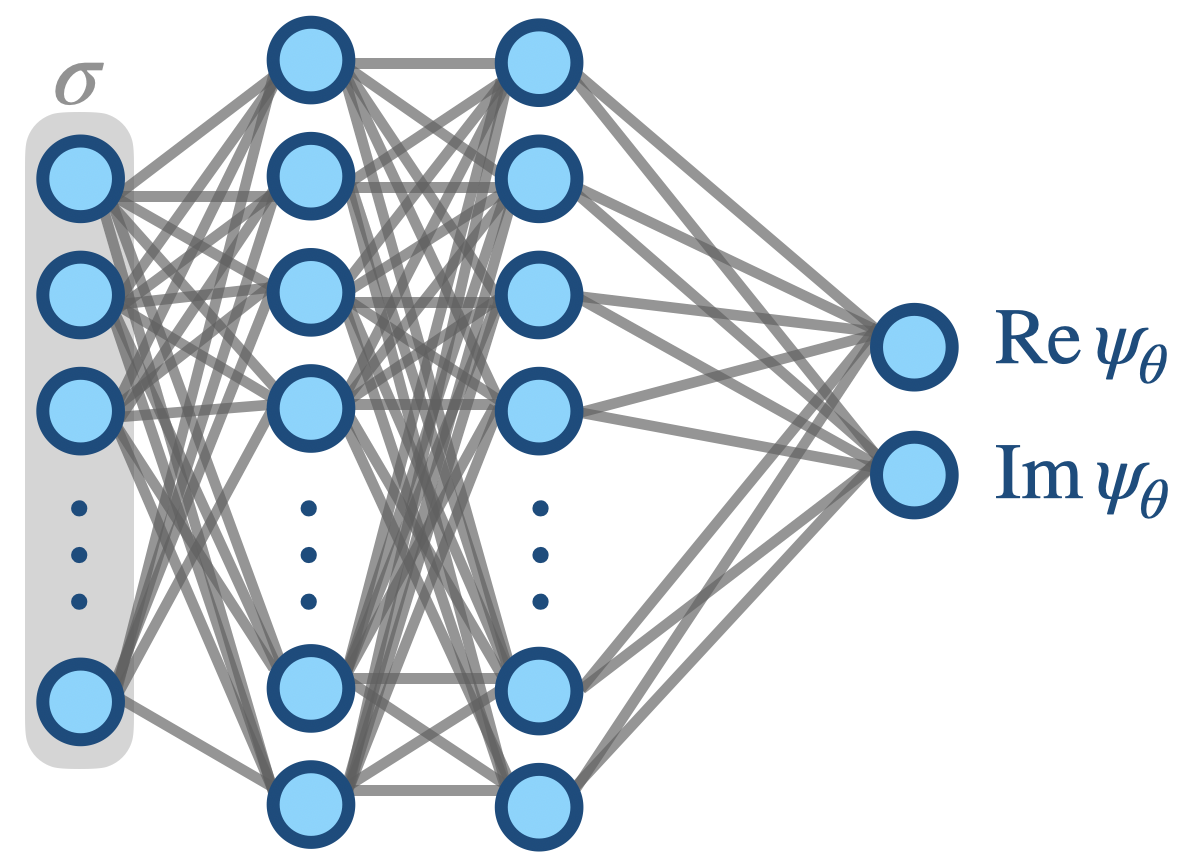}
\caption{Feed forward neural network (FFNN): The physical state $\vec{\sigma}$ (blue circles enclosed by the grey rectangle) is taken as the input. Blue circles denote the nodes of the network, grey lines the connecting weight matrices. The output of the neural network is the wave function coefficient $\psi(\boldsymbol{\sigma})$, here split into its amplitude and phase.  } 
\label{fig:FFNN}    
\end{figure}

A feed forward neural network (FFNN), often represented by the structure of a multi-layer perceptron (MLP), is the fundamental building block of artificial neural networks. It is composed of distinct layers of neurons, including an input layer that receives the data, one or more hidden layers where computations are performed, and an output layer that delivers the final result, see Fig. \ref{fig:FFNN}. Within each layer $l$, each neuron is assigned a bias $b_j^{(l)}$, and is linked to neurons in the adjacent layers through connections $w_{i,j}^{(l)}$. These weights and biases are crucial as they are iteratively adjusted during the network's training, primarily using backpropagation and optimization techniques like gradient descent. The activation functions applied to each neuron's output introduce non-linearity, enabling the network to model complex relationships. In a fully connected FFNN, every neuron in a layer is connected to all neurons in the next layer. The value of each neuron, \( a_j^{(l)} \), in layer \( l \), can be described mathematically as:
\begin{equation}
a_j^{(l)} = f\left(\sum_{i=1}^n w_{ij}^{(l)} \cdot a_i^{(l-1)} + b_j^{(l)}\right),
\end{equation}
where \( w_{ij}^{(l)} \) represents the weight from the \( i \)-th neuron in layer \( l-1 \) to the \( j \)-th neuron in layer \( l \), \( a_i^{(l-1)} \) is the activation of the \( i \)-th neuron in layer \( l-1 \), and \( b_j^{(l)} \) is the bias of the \( j \)-th neuron in layer \( l \). The function \( f \) denotes the activation function. This straightforward, yet powerful structure makes FFNNs a vital component in the field of neural networks and deep learning, and has lead to a range of applications in the context of NQS: 

Ref.~\cite{Cai2018} uses FFNNs to describe ground states of different one-dimensional systems, as well as spinless fermions and the frustrated $J_1-J_2$ spin-$1/2$ model in 2D. 
Ref.~\cite{Choo2018} explores the possibility to directly target excited states, see Sec.~\ref{sec:excited}, and compares the capabilities of FFNNs and restricted Boltzmann machines (Sec.~\ref{sec:RBM}) to represent excited states of the one-dimensional Heisenberg and Bose-Hubbard models. 
A Bose-Hubbard model on a ladder with strong magnetic flux is studied using a FFNN in Ref.~\cite{ceven2022}.
In Ref.~\cite{Zhu2023}, a FFNN is trained to represent the ground state of the one- and two-dimensional Bose-Hubbard model. By using the particle number as well as the interaction strength $U$ as additional input parameters to the network, the ground state can be directly obtained without or with little re-training for different Hamiltonian parameters.

Furthermore, FFNNs were applied to simulate quantum systems with continuous degrees of freedom. 
In Ref.~\cite{Saito2018}, a FFNN is used to simulate the ground state of the Calogero-Sutherland model in one dimension and Efimov bound states in three dimensions, where the particle positions in real space are used as input to the network. Another approach to use FFNNs to simulate continuous quantum systems was taken in Ref.~\cite{Teng2018} using Radial Basis Function (RBF) networks. RBF networks consist, as FFNNs, of an input layer, one or more hidden layers with RBFs as activation functions, and an output layer. RBFs are of the general form 
\begin{equation}
    f(x) = \phi(\lVert \vec{x} - \vec{c} \rVert),
\end{equation}
where $\vec{x}$ is a point in the input space, $\vec{c}$ is the center of the radial basis function, $\lVert \cdot \rVert$ denotes a distance measure, such as the Euclidean distance, and $\phi$ is a radial function, such as the Gaussian function. The neurons' activation is thus determined by the distance of the input $\vec{x}$ from certain points in the input space $\vec{c}$, known as centers, which are trainable network parameters. The RBF activation functions strongly depend on the distance to a center point, which allows them to capture variations in data that are radially symmetric. In Ref. \cite{Teng2018}, the input to the RBF corresponds to the quantum numbers of e.g. an undisturbed quantum harmonic oscillator, and the network parameters are optimized to represent the ground state of a quantum harmonic oscillator with an additional applied field.

\subsection{Restricted Boltzmann Machines (RBMs) \label{sec:RBM}}

\begin{figure}[t]
\centering
\includegraphics[width=0.7\textwidth]{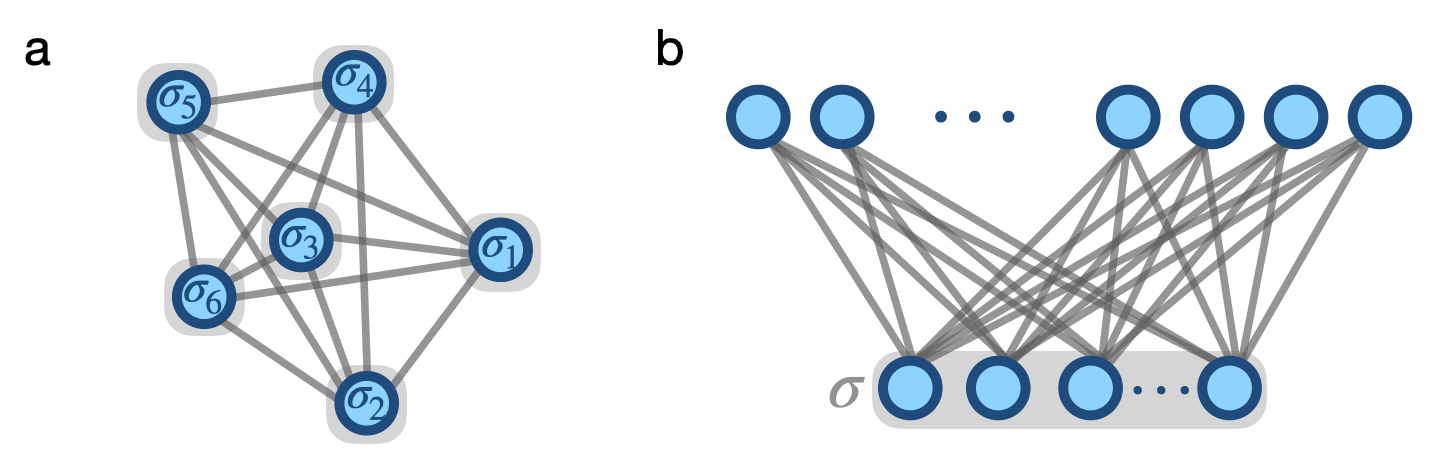}
\caption{a) Hopfield network: All-to-all connections between the physical sites are learned by the network. b) Restricted Boltzmann machine (RBM): An additional set of hidden, unphysical nodes is introduced. Furthermore, the Boltzmann machine is restricted to only inter-layer connections which connect the physical (visible) layer and the hidden layer.  } 
\label{fig:RBM}    
\end{figure}

Restricted Boltzmann machines are energy based models, i.e. they are governed by an energy function $E_{\vec{\theta}}(\vec{\sigma})$ for configurations $\vec{\sigma}$. Using statistical physics, the respective probability distribution of these models is directly related,
\begin{align}
p_{\vec{\theta}}(\vec{\sigma}) = \frac{1}{Z_{\vec{\theta}}} \mathrm{exp}(-E_{\vec{\theta}}(\vec{\sigma})),
\label{eq:RBMprob}
\end{align}
with the normalization constant $Z_{\vec{\theta}}$.
 A first example for energy based models were  Hopfield networks \cite{Hopfield} shown in Fig. \ref{fig:RBM}a, which consist of all-to-all connected nodes with connections $W_{ij}$ and the biases $b_i$, similar to an Ising model with long-range interactions and local magnetic field.  

When being used to model physical systems, the number of nodes in a Hopfield model corresponds to the number of physical sites in the system under consideration (\textit{visible nodes} $\vec{\sigma}$). In contrast, Boltzmann machines (BMs) increase the expressiveness by introducing additional, unphysical nodes (\textit{hidden nodes} $\vec{h}$) and the respective connections that increase the expressiveness of the network. With their all-to-all connections between all visible and hidden nodes, BMs are very expressive but can be hard to train. Hence, they are mostly used in their restricted version, see Fig. \ref{fig:RBM}b, were only visible-to-hidden node connections $W_{ij}$ and no hidden-to-hidden or visible-to-visible node connections are considered. 
Analogously to statistical physics, the energy of a restricted Boltzmann machine (RBM) is given by
\begin{align}
    E_{\vec{\theta}}(\vec{\sigma},\vec{h}) = -\sum_{ij}W_{ij}h_i\sigma_j-\sum_{j}b_j\sigma_j-\sum_{i}c_ih_i,
\end{align}
with the biases in the visible and hidden layers, $b_j$ and $c_i$, respectively. Using Eq. \eqref{eq:RBMprob} and summing over the hidden nodes $\mathbf{h}$, the corresponding probability for a given input configuration of physical sites $\sigma$ is given by \cite{Melko2019}
\begin{align}
    p_{\vec{\theta}}(\vec{\sigma}) = \sum_{\vec{h}} p_{\vec{\theta}}(\vec{\sigma}, \vec{h}) = \sum_{\vec{h}} \frac{\mathrm{exp}(-E_{\vec{\theta}}(\vec{\sigma},\vec{h}))}{Z_{\vec{\theta}}}.
\end{align}
An overview on the application of RBMs in physics can be found in Ref. \cite{Melko2019}.

To use RBMs for representing quantum states, apart from the amplitude $p_{\vec{\theta}}(\vec{\sigma})=\vert \psi_{\vec{\theta}}(\vec{\sigma})\vert^2$ a phase of the RBM has to be defined. This can be done in several ways, e.g. by making the network parameters $\vec{\theta}$ complex \cite{Carleo2017} or by modeling the phase with a separate RBM \cite{Torlai2018}. \\

The expressivity of the RBM ansatz is often studied in the framework of tensor networks \cite{Glasser2018,Chen2018,Clark_2018,aoki2016restricted}, using the entanglement that can be captured with an ansatz as an indicator for the representability. For RBMs, the connectivity between visible and hidden layers, that indirectly couples all sites of the physical system, allows for the entanglement entropy to scale with a subregion’s volume, in contrast to its area \cite{Deng2017}. In particular, this can make RBMs more efficient in capturing volume-law entangled states compared to e.g. MPS or PEPS. However, the efficiency of shallow RBMs to represent general quantum states has limitations, but they can be overcome with deep RBMs \cite{Roux2008,Gao2017,Liu_2019, gan2017holography}. This was further confirmed empirically e.g. in Refs. \cite{Borin2020,Pastori2019}, where random matrix product states were learned with shallow and deep RBMs using supervised approaches. Furthermore, RBMs can, due to their connectivity, straightforwardly be applied to higher dimensions, e.g. 3D systems \cite{Sfondrini2010}.\\ 

RBMs have been used for modeling a large number of physical systems, among them frustrated spin systems \cite{Westerhout2020,Ferrari2019}, spin liquids \cite{Li2021}, topologically ordered states \cite{Deng2017topological,Kaubruegger2018,Lu2019}, the Toric code \cite{Deng2017topological,Valenti2022}, Bose-Hubbard models \cite{Saito2017,Choo2018,McBrian_2019,Vargas2020}, strongly interacting fermionic systems \cite{Normura2017}, boson - fermion coupled systems such as electron - phonon coupled systems \cite{Nomura2020}, and molecules \cite{Xia2018}. In these works, often variants of RBMs are used, e.g. the correlator RBM where correlations are introduced into the RBM energy functional  based on physical insights \cite{Valenti2022}. Another modification, the convolutional RBM (CRBM) (see Sec. \ref{sec:CNN}), makes use of the fact that physical models are typically translationally invariant and feature local interactions. This is taken into account by introducing an additional convolutional layer between the visible and the hidden layers and is employed e.g. in Refs. \cite{Puente2020,Karthikconvolutional} for the simulation of Ising and Kiteav models as well as the Hubbard model. Furthermore, the implementation of symmetries was shown to improve the results \cite{Nomura_2021}. In Ref. \cite{Vieijra2020,Vieijra2021} further symmetries such as non-abelian or anyonic symmetries are considered. RBMs have also been used for the simulation of real-time dynamics \cite{Carleo2017,czischek2018quenches,schmitt2020quantum,Hofmann2022,Fabiani2019,Fabiani2021}, see also Sec. \ref{sec:dynamics}.

\subsection{Convolutional Neural Networks (CNNs) and Group CNNs \label{sec:CNN}}
\begin{figure}[t]
\centering
\includegraphics[width=0.95\textwidth]{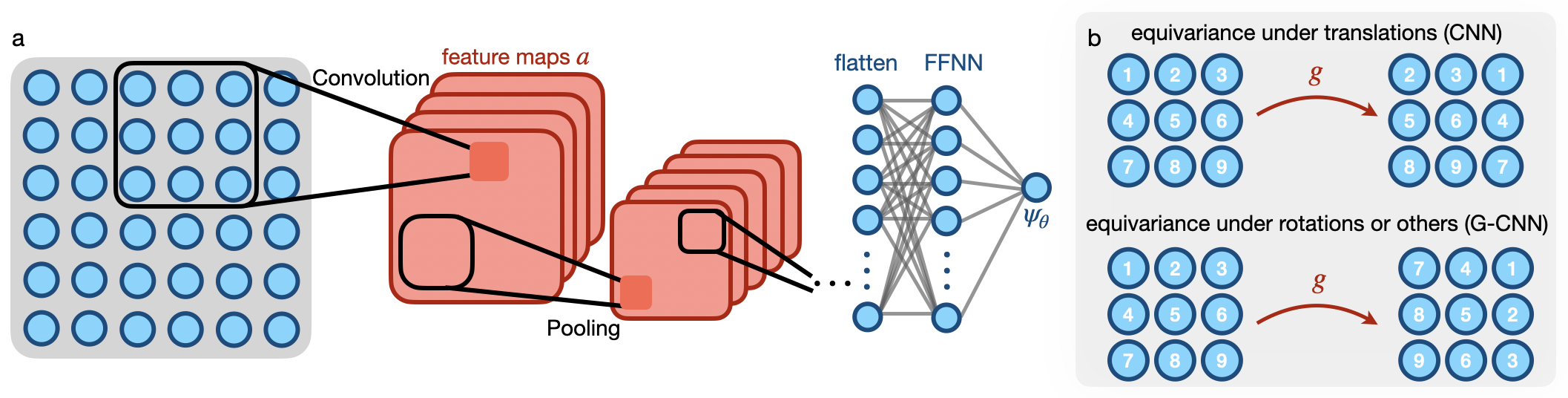}
\caption{a) Convolutional neural network (CNN) with convolutions as defined in Eq. \eqref{eq:featuremap}, pooling layers that downsample the spatial dimension and fully connected layers as final layers. b) Comparison of equivariance w.r.t. the translational group present in conventional CNNs (top) and w.r.t. other groups, e.g. $C_4$ rotations (bottom) that can be implemented in G-CNNs.} 
\label{fig:CNN}    
\end{figure}

Convolutional neural networks (CNNs) are used in processing data with a grid-like topology, most commonly two-dimensional data like images. The building blocks of CNNs are shown in Fig. \ref{fig:CNN}: First, convolutional layers employ filters or kernels to scan the input, which can detect local patterns and capture spatial relationships. Basically, each filter in the network uses the same weights for different parts of the input, making CNNs translationally invariant. This approach significantly reduces the number of parameters compared to fully connected networks. Second, pooling layers downsample the spatial dimensions, reducing computational complexity while preserving important features. The final layer of a CNN typically consists of one or more fully connected layers. 

Each convolution layer consists of feature maps $a$ and kernels $k$. The kernel $k$ is slid over the input image (or feature maps at later layers), and for each translation $(x-\tilde{x},y-\tilde{y})$, the the kernel values $k_i(x-\tilde{x},y-\tilde{y})$ are multiplied with the translated input feature values $a_i(\tilde{x},\tilde{y})$. In total, the convolution of the two-dimensional input data/feature map $a$ is

\begin{equation}
\tilde{a}(x,y) = [a * k](x,y)=\sum_{\tilde{x},\tilde{y}} \sum_{i} a_i(\tilde{x},\tilde{y}) k_i(x-\tilde{x},y-\tilde{y}).
\label{eq:featuremap}
\end{equation}
The result of this operation is a new feature map $\tilde{a}$, which is the input of the subsequent layers. The index $i$ refers to the value of the $i-th$ channel of the input feature. For an RGB image, for example, there are three channels (red, green, blue), and $a_i(\tilde{x},\tilde{y})$ is the intensity of one of these colors at pixel position $(\tilde{x},\tilde{y})$. $k_i(x,y)$ is the the value of the $i-th$ channel of the kernel at position $(x,y) $ within the kernel. \\

In the context of NQS, CNNs are regarded as a viable approach to deal with the properties of  two-dimensional systems. The application of CNNs to solve the highly frustrated $J_1-J_2$ antiferromagnetic Heisenberg model was first introduced in Ref. \cite{liang2018solving}. In these systems the sign problem remains a significant challenge for quantum Monte Carlo approaches. 
Ref. \cite{liang2018solving} demonstrates how CNNs effectively tackle the challenging problem of finding the ground state of such models. Furthermore, Ref. \cite{Levine2019} shows that deep CNNs can encode volume-law entangled states efficiently, requiring only $O(\sqrt{N})$ parameters to represent a 2D system with $N$ particles, instead of $O(N)$ for RBMs or $O(N^2)$ for fully connected networks. 
In a typical neural network, adding more layers, i.e. making the network deeper, theoretically allows the network to learn more complex features and improve its performance on tasks. However, in practice, when the network gets too deep, one faces problems such as vanishing gradients, where the gradients (which are used to update the weights in the network during training) become very small and make learning very slow or even stop it entirely. This problem is addressed e.g. in Ref. \cite{chen2023efficient} using skip connections that allow to bypass some layers and layer normalization, where inputs for all neurons within the same layer are normalized for each sample. Furthermore, a variant of stochastic reconfiguration tailored for large parameter numbers is used in this work. Besides these ground state calculations, CNNs have also been applied for dynamics simulations \cite{schmitt2020quantum,MendesSantos2023,Schmitt2022_quantumphase}, see also Sec. \ref{sec:dynamics}.

In Ref. \cite{Fu2022}, a novel approach is introduced for adapting CNNs to other common lattice structures such as triangular lattices, which are somewhat analogous to sheared square lattices, allowing the application of regular CNN filters. In the same work, the authors consider honeycomb and Kagome lattices, where the key techniques involve augmenting these lattices with strategically placed virtual vertices, effectively transforming them into grid-like structures akin to triangular lattices. This allows for the application of standard CNN convolutional kernels while preserving the unique properties of the original lattices. The method enhances information processing and exchange, expanding the receptive field and enabling the analysis of varied local structures and staggered arrangements unique to these lattices. \\

Although CNNs exhibit translational invariance, they lack the ability to learn additional types of symmetries, such as rotation or mirror symmetries. Typically, data augmentation is employed to train the model for these specific symmetries \cite{wang2023variational}. In Ref. \cite{choo2019two} the wave function was symmetrized in order to incorporate the rotational symmetries, see also Sec.~\ref{sec:designChoices}. A more intrinsic solution is the development of group convolutional neural networks (G-CNNs). These networks extend the capabilities of standard CNNs by using group theory, allowing them to automatically incorporate various symmetries, see Fig.~\ref{fig:CNN}b. The key component of G-CNNs is the group convolution operation. An equivariant convolution ensures that if the input is transformed (e.g., rotated), the output feature maps will be transformed in the same way. The group-equivariant convolution of the input/feature map $a$ with a kernel $k$ under the group $G$ evaluated at a group element $g$ corresponds to

\begin{equation}
[a *_{G} k](g) = \sum_{h \in G} a(h) k(g^{-1} h).
\end{equation}

This convolution operation is designed to be equivariant to the transformations in the group $G$. $G$ can be a group of (discrete) rotations, translations, or other transformations. Ref.~\cite{roth2021group} considers the full wallpaper group, consisting of translation, rotation and mirror symmetry.
The first convolution takes the input $\boldsymbol{\sigma} = (\sigma_1,\sigma_2,...,\sigma_N)$, where $i$ are the positions in the lattice, and transforms it as 
\begin{align}
    a^1(g) = f\left( \sum_{\vec{i}} k(g^{-1}\vec{i}) \sigma_{\vec{i}} \right),
\end{align}
where $f$ is a point-wise non-linear activation function, to obtain the value of the feature map $a(g)$ for the group element $g$. This first (embedding) layer thus generates equivariant feature maps, which are indexed with group elements, from the input. 
After repeating the application of group convolution and non-linearity for $l$ layers, the wave function coefficient $\psi(\boldsymbol{\sigma})$ is determined as
\begin{align}
    \psi(\boldsymbol{\sigma}) = \sum_g \chi_{g^{-1}} \exp(a_g^l),
\end{align}
where $\chi_g$ is the character of the symmetry operation $g$.

Ref. \cite{roth2021group} uses G-CNNs to determine the ground state energy of the $J_1-J_2$ Heisenberg model on a square and triangular lattice. Ref. \cite{roth2023high} underscores the capability of very deep G-CNNs in achieving high-accuracy results for the same lattices, and furthermore directly calculates low-lying excited states by changing the characters of the symmetry operations $\chi_g$ accordingly. For a Heisenberg model on a Kagome lattice, one of the most studied models in frustrated magnetism since it is a promising candidate to host exotic spin liquid states, Ref. \cite{duric2024spin12} presents a new ground state: the spinon pair density wave (PDW), which does not break the time-reversal and lattice symmetries. 
G-CNNs are used to study the ground state of the $J-J_d$ Heisenberg model on the Maple-Leaf lattice, which results in dimer state paramagnetic and canted magnetic order phases for different values of $J_d/J$, \cite{beck2024phase}.

\subsection{Graph Neural Networks \label{sec:GraphNN}}

\begin{figure}[t]
\centering
\includegraphics[width=0.75\textwidth]{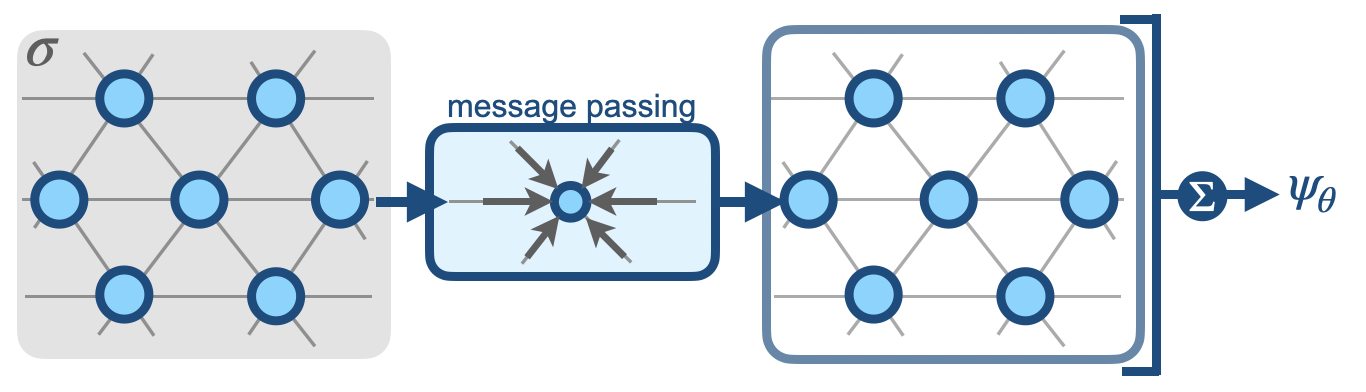}
\caption{ Graph neural network: In each permutation equivariant layer, the weights are updated through typically pairwise message passing, i.e. the value of a given node is updated according to the values of its immediate neighbors. Graph neural networks can easily be applied to any lattice geometry, here we show a hexagonal lattice as an example.} 
\label{fig:GraphNN}    
\end{figure}

Graph neural networks directly take the geometry of the underlying problem into account \cite{bronstein2021geometric}. In the case of NQS, this means that the lattice geometry of the Hamiltonian under consideration is used as the graph structure. Throughout the graph neural network, this graph structure is kept, see Fig. \ref{fig:GraphNN}. In Ref. \cite{kochkov2021learning}, a sublattice encoding, denoting the position of the site within the unit cell, is used as additional input for each site on the lattice. Subsequently, in each permutation equivariant layer of the graph neural network, the values of the nodes are updated through typically pairwise message passing. This means that the value of a given node is updated according to the values of its immediate neighbors, thus directly taking the graph structure into account. The specific details of this updating procedure are design choices, leading e.g. to graph convolutional neural networks \cite{kipf2017semisupervised} or gated graph sequence neural networks \cite{li2017gated}, where a gated recurrent unit is used. 

One advantage of the graph structure and message passing layers is that a transfer to different system sizes is straightforwardly possible, as shown in Ref. \cite{kochkov2021learning}. Ref. \cite{yang2020scalable} considers the ground state of the hard-core bosonic $t-V$ model on different lattice geometries, such as the Kagome and triangular lattices. Since this constitutes a stoquastic Hamiltonian, no sign structure has to be learned. In Ref. \cite{kochkov2021learning}, the ground state of the $J_1-J_2$ Heisenberg model on square, triangular, honeycomb and Kagome lattices is studied, in which case a non-trivial sign structure exists. The performance for using complex network weights as well as separate networks for amplitude and phase are compared. 

Permutation equivariant message passing has also been used in the context of neural network backflow transformations to simulate interacting fermions in continuous space in Refs. \cite{pescia2023,kim2023neuralnetwork}. Ref.~\cite{luo2023pairingbased} uses a graph neural network to represent a generalized pair amplitude in the context of a BCS type wave function.

\subsection{Autoregressive Networks \label{sec:autoregressive}}

Autoregressive architectures are characterized by their normalized amplitudes $p_{\vec{\theta}}=\vert \psi_{\vec{\theta}}\vert^2$. The use of autoregressive networks for NQS was first proposed by Sharir et al. \cite{Sharir2020}: At a local configuration $\sigma_i$, the authors propose to mask out the local configurations $j\geq i$, and only consider sites $\vec{\sigma}_{<i}=(\vec{\sigma}_1, \dots \vec{\sigma}_{i-1})$, such that the network represents $p_{\vec{\theta}}(\vec{\sigma}_i\vert \vec{\sigma}_{<i})$. The total probability is given by  
\begin{align}
    p_{\vec{\theta}}(\vec{\sigma}) = \prod_i^N p_{\vec{\theta}}(\vec{\sigma}_i\vert \vec{\sigma}_{<i}).
    \label{eq:autoregressive}
\end{align} 
This allows to normalize $p_{\vec{\theta}}$ by normalizing each conditional  $p_{\vec{\theta}}(\vec{\sigma}_i\vert \vec{\sigma}_{<i})$, and hence to sample directly from the amplitudes instead of more elaborate sampling procedures like Markov chain sampling needed for non-autoregressive architectures. Since the generation of many, uncorrelated samples is crucial for the training, this can yield a speed up and an improvement of the optimization. 
However, Ref. \cite{bortone2023impact} suggests that the autoregressive sampling can in some cases reduce the expressivity of the neural network wave function. Furthermore, the application of stochastic reconfiguration for optimizing autoregressive NQS can cause problems, as discussed in Sec. \ref{sec:GS}. For further reading on the potential of autoregressive networks in the context of quantum physics and NQS we refer the reader to Ref. \cite{Melko2024}.

Some architectures like CNNs and transformers (see Sec. \ref{sec:Transformers}) can be made autoregressive by masking out future inputs $\sigma_i,\dots \sigma_N$ for the $i$-th input vector \cite{Sharir2020, Schmale2022,zhang_transformer_2023,sprague2023variational,lange2024transformer,Schmale2022}.  In the following, we discuss a network architecture to which the autoregressive property is inherent due to their recurrent structure: recurrent neural networks.

\subsubsection{Recurrent Neural Networks (RNNs) \label{sec:RNN}}

\begin{figure}[t]
\centering
\includegraphics[width=0.9\textwidth]{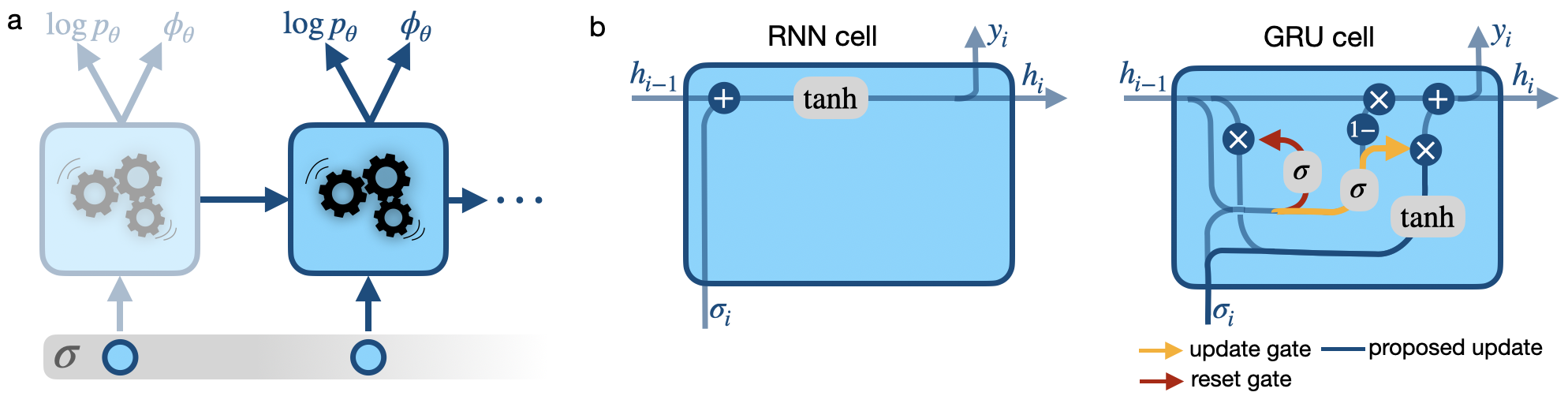}
\caption{a) Recurrent neural network (RNN). At each site denoted by the blue circles, the local configuration is passed through the RNN cell. Furthermore, information from previous sites is passed to the next cell (horizontal arrows). b) Left: A plain vanilla RNN cell, where the local configuration $\sigma_i$ and the hidden state $h_i$ are simply added and passed through a tanh-activation function. Right: Gated recurrent unit (GRU), enabling the RNN to capture longer-ranged correlations throughout the system. In a GRU, the proposed update (dark blue), reset and update gates (red and orange) decide how much of the old state is kept. } 
\label{fig:AutoregressiveNN}    
\end{figure}

Recurrent neural networks (RNNs) consist of several RNN cells, and information is passed from one cell to the next, in a recurrent manner, through the network, as schematically shown in Fig. \ref{fig:AutoregressiveNN}a. 

The first applications of RNNs to represent quantum states have considered one-dimensional spin systems \cite{hibat-allah_recurrent_2020,roth_iterative_2020}. In these cases, the RNN is constructed by $N$ cells and the information is passed from the first cell corresponding to the first spin of the 1D chain to the last cell in a recurrent fashion. At each lattice site $i$, the cell receives a local spin configuration $\vec{\sigma}_i$ and the so-called \textit{hidden} state $\vec{h}_{i-1}$ that passes information from previous lattice sites through the network. The cell then outputs the updated hidden state $\vec{h}_i$ as well as an output $\vec{y}_i$ that can be used to calculate the local conditional probability and a local phase of the state representation. Normally, each cell is represented by the same weights (weight sharing), but in some contexts the cells can also be chosen to have different weights \cite{hibat-allah_variational_2021}. In the former case, the RNN architecture is tailored to model bulk properties and hence becomes particularly effective for large systems \cite{roth_iterative_2020}. Furthermore, it is possible to iteratively retrain on larger and larger systems, which can improve the performance for large systems \cite{roth_iterative_2020}.

The local amplitude at each RNN cell is given by a conditional probability determined by the previous spin configurations $\vec{\sigma}_{<i}$, i.e. $p_{\vec{\theta}}(\vec{\sigma}_i\vert \vec{\sigma}_{<i})$. The activation function of the RNN's output layer can be chosen such that the local amplitude is normalized and hence also the total amplitude Eq. \eqref{eq:autoregressive},
making the RNN autoregressive. 

To model long-range correlations, it is crucial that the information is passed through the cells in an efficient way. This is usually done by replacing the plain vanilla RNN cells with gated recurrent units (GRU) \cite{steffen_learning_2006}, see Fig. \ref{fig:AutoregressiveNN}b, enabling a long-term memory of the RNN \cite{chung2014empirical}. In one-dimensional settings, this modification yields successful representations of spin systems like Heisenberg and transverse field Ising model \cite{roth_iterative_2020,hibat-allah_recurrent_2020}. For two-dimensional systems, the hidden states can be passed in a 1D snake through the system, similar to MPS calculations, as e.g. in Ref. \cite{roth_iterative_2020}. However, it is also possible to pass the information in a 2D fashion through the system, as proposed in Ref. \cite{graves2007multidimensional} and further improved by introducing a tensorized version of a GRU in Ref. \cite{hibat-allah_variational_2021}. Furthermore, the authors show that an imposed  $U(1)$ magnetization conservation and spatial symmetries as well as direct implementation of the Marshall sign rule for spin-$1/2$ systems improve the results, in agreement with other works \cite{morawetz_u1-symmetric_2021,hibat-allah_recurrent_2020,roth_iterative_2020}. This $U(1)$ symmetry is usually imposed by setting $p_{\vec{\theta}}(\vec{\sigma}_i\vert \vec{\sigma}_{<i})=0$ if the system with the new sampled $\vec{\sigma}_i$ violates the corresponding conservation law. For example, for the $U(1)$ particle number conservation of hardcore bosons with local states $\vec{\sigma}_i=0(1)$ corresponding to empty (occupied) sites, $p_{\vec{\theta}}(\vec{\sigma}_i=1\vert \vec{\sigma}_{<i})=0$ with if $\vec{\sigma}_{<i}$ is already in the correct particle number sector. With these modifications, RNNs have been applied in many contexts, including the Heisenberg model on square and triangular lattices \cite{hibat-allah_supplementing_2022}, prototypical states in quantum information \cite{carrasquilla_reconstructing_2019}, states with topological order \cite{HibatAllah2023,Doeschl2023} and fermionic systems using Jordan-Wigner strings \cite{lange2023neural}. In the context of real-time dynamics simulations, see also Sec. \ref{sec:dynamics}, RNNs have been used e.g. in Refs. \cite{Luo2022,Reh2021,mendessantos2023wave}. \\

In order to investigate the expressivity of the RNN ansatz, the authors of Ref. \cite{Wu2023} present a mapping from tensor networks to 1D MPS-RNNs and 2D tensorized MPS-RNNs, i.e. RNNs with linear or multilinear update rules for the hidden states and quadratic output layers. For linear update rules and one-dimensional settings, MPSs can be mapped to the 1D MPS-RNN with the same number of variational parameters, but not vice versa, making the latter potentially more expressive than MPS. The 2D version of the MPS-RNN receives hidden states from two directions, inspired from projected entangled pair states (PEPS) \cite{verstraete2004matrix}, and features multilinear updates. This architecture is shown to encode an area law of entanglement entropy, but unlike PEPS, it supports perfect sampling and hence efficient evaluations of the wave function. In particular, receiving hidden states from two directions makes the RNN more efficient in state compression compared to the class of TNs which support wave function evaluation in polynomial time \cite{Levine2019}.

\subsection{Transformers \label{sec:Transformers}}
\begin{figure}[t]
\centering
\includegraphics[width=0.95\textwidth]{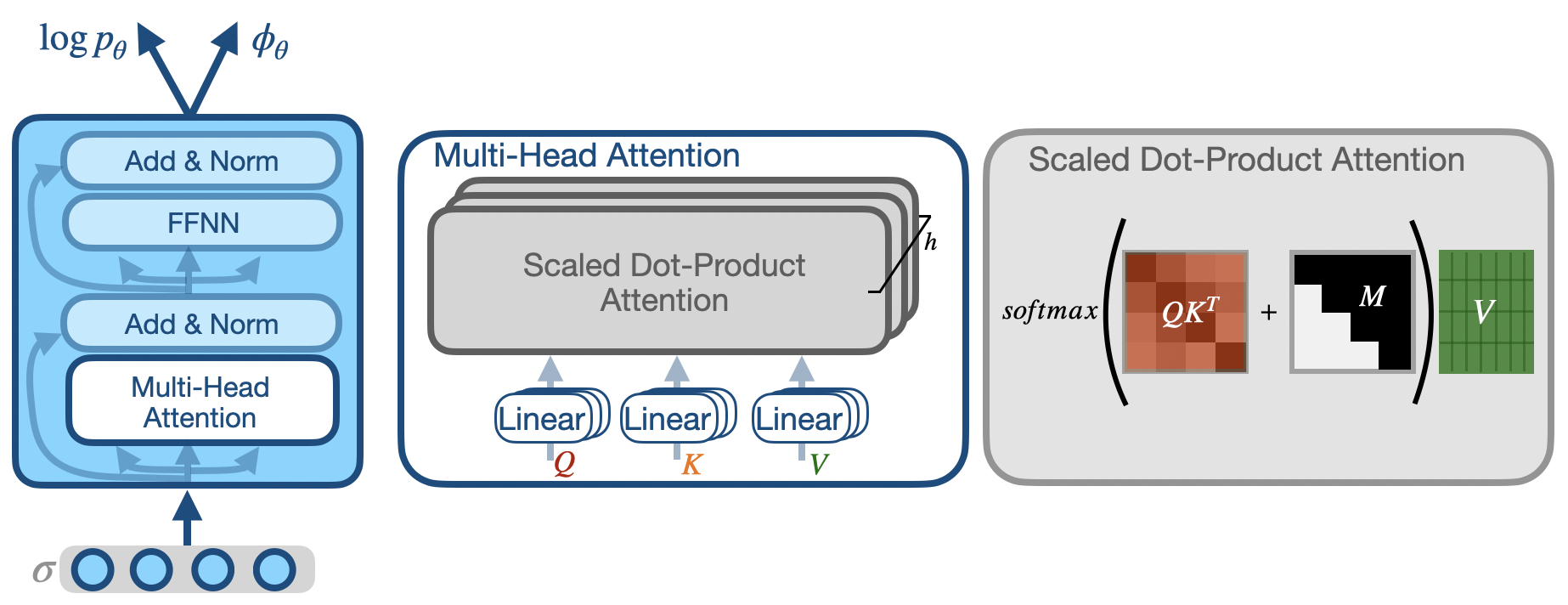}
\caption{Left: The transformer neural network wave function, adapted from Ref. \cite{vaswani2017attention}. The input is passed through the multi-head attention mechanism, addition and normalizing layers (Add and Norm) and a feed forward neural network (FFNN). Middle: The multi-head attention itself consists of $h$ scaled dot-product attention layers. Right: The scaled dot-product attention according to Eq. \eqref{eq:attention}, based on the attention map $QK^T$, which can in principle represent all-to-all correlations throughout the system. Sites of the system that have not been considered yet are masked out by adding the mask $M$, before passing it through a softmax activation layer and mutiplying with the values $V$. } 
\label{fig:Transformer}    
\end{figure}
A transformer model relies entirely on an attention mechanism which draws global dependencies between input and output \cite{vaswani2017attention}. This self-attention layer in the transformer setup generates all-to-all interactions between the sites in the system. These trainable connections can potentially represent strong connections or correlations, regardless of their position \cite{sprague2023variational, viteritti_transformer_2023}. The transformer first embeds the different given input elements into a unified feature space. This embedding corresponds to a linear projection, with trainable parameters, of the input elements with a dimension $d_i$ to elements with an embedded dimension $d_h$. The position of the inputs in the sequences are not explicitly modeled in the transformer, but efficiently transformed into abstract representations using positional encoding vectors that are added to the embedded input vectors \cite{viteritti_transformer_2023}. Each embedded input element $\sigma_i^{(e)}$ is projected on a query vector ($q_i$), key vector ($k_i$) and a value vector ($v_i$) of the same dimension $d_h$ as the embedded input, given by: 
\begin{align}
q_i = \sum_{l=1}^{d_h} W_{i,l}^q \sigma_{i,l}^{(e)} && k_i = \sum_{l=1}^{d_h} W_{i,l}^k \sigma_{i,l}^{(e)} && v_i = \sum_{l=1}^{d_h} W_{i,l}^v \sigma_{i,l}^{(e)},
\end{align}
with the matrices $W^q, W^k$ and $W^v$ to be the trainable weight matrices of dimension $d_h \cross d_h$. The query, key and value matrices are then given by $Q = (q_1, ..., q_{N})$,  $K = (k_1, ..., k_{N})$ and  $V = (v_1, ..., v_{N})$. In multi-headed attention, each query, key and value vector is mapped to $h$ vectors with trainable weight matrices, with $h$ the number of attention heads. This is indicated in Fig. \ref{fig:Transformer} (middle).  A distinction has to be made between two different classes of transformers: the encoder and the decoder. Both architectures consist of the multi-head attention mechanism and a subsequent FFNN. The encoder maps an input sequence of symbolic representations to a context vector, which is used to condition the decoder. The decoder then combines this context vector with the input snapshot to generate the final output probability. Commonly, only a decoder is used for the NQS ansatz. Unlike encoders that map the input to a context vector, the decoder model is usually made autoregressive by adding a mask $M$ to the self-attention layer, which allows connections to all previous elements in the sequence but not to subsequent elements \cite{vaswani2017attention} and enables efficient exact sampling from the model \cite{zhang_transformer_2023, sprague2023variational, Luo2022,lange2024transformer}. Then, a softmax activation function is applied to the masked dot product of the vectors $Q$ and $K$. The complete attention formalism in the decoder can be summarized by
\begin{align}
    \text{Attention}(Q, K, V) = softmax\left(\frac{QK}{\sqrt{d_h / h}} + M\right)V\, ,
    \label{eq:attention}
\end{align}
as shown in Fig. \ref{fig:Transformer} (right). In Ref. \cite{rende2024queries} different variants of the attention mechanism are compared. \\  

Transformer quantum states can learn ground state properties of various physical systems, such as the 1D transverse field Ising model and the 1D Heisenberg $J_1-J_2$ and XYZ model \cite{zhang_transformer_2023}. 
Comparable results to DMRG calculations have been obtained with a transformer decoder for a 1D frustrated spin model, with a relatively low number of parameters \cite{viteritti_transformer_2023}. In Ref. \cite{Luo2022}, transformers have shown to be able to simulate the real-time dynamics and steady state in 1D and 2D transverse field Ising and Heisenberg models \cite{Luo2022}. In Refs. \cite{Cha_2022,vonglehn2023selfattention,shang2023solving,wu2023nnqstransformer} transformers are used in the context of quantum chemistry calculations.

Small modifications of the transformer model, leading to the so-called vision transformers (ViT), inspired the use of patched transformers. This model splits the system into patches, and can be used to calculate the ground state properties of frustrated spin models \cite{viteritti_transformer_2023}. 
A large patch size enhances the efficiency of the transformer, but on the other hand the network output dimension increases exponentially with the patch size, as it encodes the probability distribution over all possible patch states. To overcome this, large patched transformers are introduced in Ref. \cite{sprague2023variational}. In this ansatz, the output of the patched transformer is passed to a patched RNN as the initial hidden state, which breaks the large inputs into smaller sub-patches, reducing the output dimension. This model has been shown to accurately capture ground state properties and phase transitions of large Rydberg systems, which can compete with quantum Monte Carlo results \cite{sprague2023variational}.
In Ref. \cite{viteritti2024transformer}, a combined architecture of a transformer network and a FFNN is used. Hereby, first the transformer maps the physical spins to a high-dimensional feature space. The authors argue that in this feature space the determination of the ground-state properties is simplified, requiring only a single FFNN layer with complex-valued parameters to parameterize the wave function in the second part of the network. The combined architecture is used for the ground state search of the Shastry-Sutherland Model, featuring a spin liquid phase besides phases with plaquette and antiferromagnetic order.

In Ref. \cite{zhang_transformer_2023}, transformer quantum states have been used to learn ground state properties of a single system, as well as to generalize to different, unseen systems. For the latter, not only the physical degrees of freedom, but also the parameters of the Hamiltonian of the system are used as input. These parameters can be formulated as new elements that have to be passed to the embedding layer. After training the transformer quantum state for a Hamiltonian with different parameters, the transformer is able to generate the ground state for unseen Hamiltonian parameters without any additional training. Although these are with slightly larger error, more accurate results can be achieved with less training then without any a priori training. In Ref. \cite{fitzek2024rydberggpt} an encoder-decoder transformer is designed to learn the distribution of measurement outcomes with the Hamiltonian parameters of the system as the input. The transformer is trained on data from different interacting Rydberg arrays and is able to generalize to systems outside of the training set. 

In Refs. \cite{rende_optimal_2023,rende_simple_2023}, a transformer with a so-called factored attention is used. In contrast to the conventional attention mechanism \eqref{eq:attention}, where the attention weights $QK^T$ and the values $V$ are calculated from both embedded inputs and the positional encoding, factored attention uses $QK^T$ that depend only on positions, and $V$ that depend only on the embeddings. Ref. \cite{rende_optimal_2023} shows that training a model with a single self-attention layer with factored attention can be mapped to solving the inverse Potts problem using the pseudo-likelihood method. This method, in combination with the patched transformer, leads to high quality results for the ground-state energy of the $J_1-J_2$ Heisenberg model \cite{rende_optimal_2023}.

\subsection{NQS for Fermionic Systems\label{sec:fermions}}

\begin{figure}[t]
\centering
\includegraphics[width=0.9\textwidth]{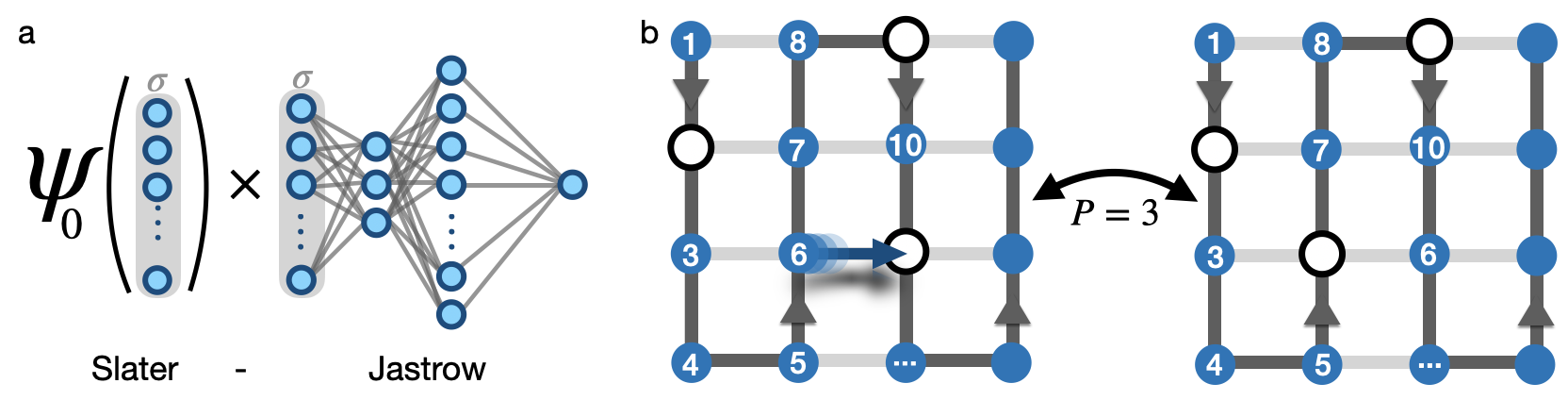}
\caption{NQS for fermionic systems: a) Fermionic architectures are typically inspired from Slater-Jastrow ansätze with an antisymmetric Slater determinant and a Jastrow factor that captures the correlations throughout the system. b) Bosonic architectures can be made fermionic by applying Jordan-Wigner strings. Effectively, this boils down to counting the number of permutations $P$ of configurations that are connected by matrix elements when calculating observables and multiplying the respective terms in Eq. \eqref{eq:JW} by $(-1)^P$. } 
\label{fig:Fermions}    
\end{figure}

NQS for the representation of fermionic quantum states can be divided into distinctly different ansätze: $(i)$ NQS ansätze that inherently incorporate the fermionic statistics and $(ii)$ bosonic NQS that are antisymmetrized by a Jordan-Wigner transformation, see Fig. \ref{fig:Fermions}a and b respectively.

\subsubsection{Fermionic Architectures }
The antisymmetry of NQS ansätze with fermionic statistics can be achieved in various ways. The most commonly used ansatz for fermionic variational wave functions is a Slater-Jastrow-inspired ansatz, where the wave function is constructed from an antisymmetric part $\psi_0$, typically a Slater determinant, and a Jastrow factor $\mathcal{J}$ capturing the correlations, i.e.
\begin{align}
    \ket{\psi_{\vec{\theta},\vec{\nu}}} = \sum_{\vec{\sigma}} \psi_{0,_{\vec{\theta}}}( \vec{\sigma}) \mathcal{J}_{\vec{\nu}}( \vec{\sigma})\ket{\vec{\sigma}},
    \label{eq:SlaterJastrow}
\end{align}
where in principle both $\psi_0$ and $\mathcal{J}$ can be parameterized by neural networks with parameters $\vec{\theta}$ and $\vec{\nu}$. This is shown schematically in Fig. \ref{fig:Fermions}a.

In the setting of first quantization, architectures like FermiNet \cite{pfau2020ferminet} and PauliNet \cite{Hermann2020,Schaetzle2021}  that use Slater determinants $\psi_0$ reach high accuracies in \textit{ab initio} molecule simulations. However, the evaluation of Slater determinants is costly in first quantization, which is overcome e.g. in Ref. \cite{Han_2019} by an antisymmetric construction of $\psi_0$ by deep neural networks. However, these approaches often come at the price of a reduced accuracy \cite{pang2022on2}. 

For quantum many-body systems, mostly second quantization is used, despite some exceptions e.g. for repulsively interacting, spin-polarized fermions \cite{Stokes2020}, where the authors chose to model $\psi_{0}( \vec{\sigma}) \mathcal{J}( \vec{\sigma})$ in Eq. \eqref{eq:SlaterJastrow} by a single neural network. In the works using second quantization, machine learning approaches are used to enhance the expressivity of the Slater-Jastrow ansatz. This can be done by employing NNs to parameterize the Jastrow factor $\mathcal{J}$.

One of the first examples is the RBM+PP architecture in Ref. \cite{Nomura2017}, with a slightly different ansatz than Eq. \eqref{eq:SlaterJastrow}, i.e.
\begin{align}
    \ket{\psi_{\vec{\theta}}} = \sum_{\vec{\sigma}} \psi_{\mathrm{ref}}( \vec{\sigma}) \mathcal{F}_{\vec{\theta}}( \vec{\sigma})\ket{\vec{\sigma}},
\end{align}
where correlations on top of a reference state $\psi_{\mathrm{ref}}$ are modeled by a generalized version of an RBM, with additional artificial neurons to mediate entanglement, that is represented by $\mathcal{F}_{\vec{\theta}}$. In this work, the authors take $\psi_{\mathrm{ref}}$ to be a pair-product state (PP) that already incorporates some of the entanglement, and test the architecture for the Fermi-Hubbard model.

Currently, usually two methods are used \cite{liu2023unifying}: $(i)$ The hidden fermion determinant state, where neural networks are used to replace the standard Slater determinant with a larger determinant which includes single particle orbitals from additional projected hidden fermions \cite{Moreno2022,gauvinndiaye2023mott}. $(ii)$ Neural backflow transformations, which add correlation by making the single particle orbitals of the Slater determinant configuration dependent, with the respective transformation learned by a neural network \cite{Feynman1956,Luo2019,Hermann2020,pfau2020ferminet,kim2023neuralnetwork,romero2024spectroscopy}. In Ref. \cite{liu2023unifying} the authors show that both $(i)$ and $(ii)$ can be written as Jastrow-like corrections to the single-particle orbitals. Furthermore, Refs. \cite{li2024emergentwignerphasesmoire,luo2024simulatingmoirequantummatter} show that using Bloch single-particle wave functions in the Slater determinant allows to simulate large fermionic systems such as moir\'{e} materials. \\

Furthermore, it is worth mentioning that in simulations of lattice models with a Slater-Jastrow variational wave function, the autocorrelation time increases drastically for large system sizes, motivating the development of a fully autoregressive Slater-Jastrow ansatz by combining a Slater determinant with an autoregressive deep neural network as a Jastrow factor \cite{Humeniuk2022}.

\subsubsection{Bosonic Architectures with Jordan Wigner Strings }
The other way to simulate fermionic systems using NQS are Jordan-Wigner (JW) transformations, which are used to map the bosonic NQS to a fermionic wave function. Hence, per se bosonic architectures can be used and no special fermionic architecture is needed. The JW transformation is given by  
\begin{align}
    \hat{c}^{(\dagger)}_j = \hat{\sigma}^{-(+)} \mathrm{exp}\left( i \pi\sum_{k< j} \hat{\sigma}_k^+\hat{\sigma}_k^- \right),
    \label{eq:JW}
\end{align}
where indices refer to a one-dimensional labeling of the fermions, the $\hat{\sigma}^{\pm}$ are the spin raising and lowering operators. The resulting annihilation (creation) operators $\hat{c}^{(\dagger)}$ fulfill the fermionic commutation relations. More precisely, for two fermions at site $i$ and $j$, exchanging these fermions yields a minus sign arising from the argument of the exponential in Eq. \eqref{eq:JW}. This rule does not have to be implemented in the NQS architecture itself, but only on the level of the calculation of expectation values. For an operator $\hat{O}$ with $\langle \hat{O}\rangle = \langle O^\mathrm{loc}_{\vec{\theta}}(\vec{\sigma}) \rangle_{\vec{\sigma}}$ (see Eq. \eqref{eq:expectations}) with
\begin{align}
O^\mathrm{loc}_{\vec{\theta}}(\vec{\sigma})=\sum_{\vec{\sigma}^\prime}\frac{\bra{\vec{\sigma}}\hat{O}\ket{\vec{\sigma}^\prime}\langle\vec{\sigma}^\prime \vert \psi_{\vec{\theta}} \rangle}{\langle\vec{\sigma} \vert \psi_{\vec{\theta}} \rangle} \eta_{\vec{\sigma}\vec{\sigma}^\prime}\, ,
\end{align}
each matrix element is multiplied by a factor $\eta_{\vec{\sigma}\vec{\sigma}^\prime}=(-1)^{P_{\vec{\sigma}\vec{\sigma}^\prime}}$ if $\vec{\sigma}^\prime$ is connected to $\vec{\sigma}$ by $P_{\vec{\sigma}\vec{\sigma}^\prime}$ two-particle permutations, see Fig. \ref{fig:Fermions}b. This method was applied to simulate molecules \cite{Barret2022}, for Fermi-Hubbard and $t-J$ models \cite{Inui2021,lange2023neural} and solid state systems \cite{Yoshioka2021}.

Despite its successful application, JW strings come with the disadvantage that the operators in Eq. \eqref{eq:JW} are highly non-local, which can cause problems for some architectures. Whether antisymmetrizing bosonic networks is as efficient as using inherently fermionic architectures is still under debate \cite{denis2023comment}.  

\subsection{Other Design Choices}\label{sec:designChoices}

Besides the architecture, other design choices can influence the performance of the NQS: \\

One choice is the way how amplitude and phase of the NQS are calculated. Hereby, amplitude and phase can be learned by two different, real-valued networks, one real-valued network with two separate output nodes or final layers for phase and amplitude or by one network with complex weights. In Ref. \cite{Viteritti2022accuarcy}, a complex-valued RBM and a RBM in which two separate real-valued networks approximate amplitude and phase, are compared for the ground state of the $J_1-J_2$ model. In a systematic study on small clusters, they show that the complex RBM outperforms the latter. 

A second design choice is how to incorporate symmetries in the NQS training in order to restrict the optimization space to states in the target symmetry sector, hence improving the performance \cite{Nomura_2021,morawetz_u1-symmetric_2021,hibat-allah_recurrent_2020,Reh2023}. Firstly, global $U(1)$ symmetries can be imposed, see e.g. Refs. \cite{morawetz_u1-symmetric_2021,hibat-allah_recurrent_2020,malyshev2023autoregressive}, by restricting to configurations $\vec{\sigma}$ that obey the respective symmetry, e.g. magnetization or particle number conservation. For autoregressive architectures, this is done by restricting the conditional probabilities to the targeted symmetry, see e.g. Sec. \ref{sec:RNN}. For non-autoregressive architectures, the Mone Carlo updates can be chosen such that all generated configurations stay in the same symmetry sector. Second, spatial symmetries can be imposed. There are different symmetrizations used in the literature\footnote{We follow Ref. \cite{Reh2023} here.}, which require to generate new samples $\vec{\sigma}^S$ that are connected by symmetry transformations $\mathcal{T}$ to the original samples $\vec{\sigma}$:
\begin{enumerate}
    \item bare-symmetry: 
    \begin{align}
        \psi^S_{\vec{\theta}}(\vec{\sigma}) = \mathrm{exp} \left(\frac{1}{\vert \mathcal{T} \vert} \sum_{\vec{\sigma}^S \in \mathcal{T}(\vec{\sigma})} \mathrm{log}\left[\psi_{\vec{\theta}}(\vec{\sigma}^S)\right]\right)
    \end{align}
    \item exp-symmetry
    \begin{align}
        \psi^S_{\vec{\theta}}(\vec{\sigma}) = \left(\frac{1}{\vert \mathcal{T} \vert} \sum_{\vec{\sigma}^S \in \mathcal{T}(\vec{\sigma})} \psi_{\vec{\theta}}(\vec{\sigma}^S)\right)
    \end{align}
    \item sep-symmetry
    \begin{align}
        \psi^S_{\vec{\theta}}(\vec{\sigma}) = \sqrt{\mathrm{exp} \left(\frac{1}{\vert \mathcal{T} \vert} \sum_{\vec{\sigma}^S \in \mathcal{T}(\vec{\sigma})} 2 \mathrm{Re}\{ \mathrm{log}\left[\psi_{\vec{\theta}}(\vec{\sigma}^S)\right]\}\right)
        \cdot \mathrm{exp} \left(i\, \mathrm{arg}\left( \sum_{\vec{\sigma}^S \in \mathcal{T}(\vec{\sigma})} \mathrm{exp}\left( i\, \mathrm{Im}\{ \mathrm{log}\left[\psi_{\vec{\theta}}(\vec{\sigma}^S)\right]\} \right)\right)\right)}.
    \end{align}
\end{enumerate}
Note that only the last keeps the autoregressive property intact since $\sum_{\vec{\sigma}^S \in \mathcal{T}(\vec{\sigma})} \vert \psi_{\vec{\theta}}^S(\vec{\sigma}^S) \vert^2 = \sum_{\vec{\sigma}^S \in \mathcal{T}(\vec{\sigma})} \vert \psi_{\vec{\theta}}(\vec{\sigma}^S) \vert^2 $. Furthermore, architectures that preserve certain symmetries explicitly can be used, such as group CNNs \cite{roth2021group} and gauge equivariant neural Networks \cite{Luo2021}.

Furthermore, the authors of Ref. \cite{Cai2018} show for the exemplary case of a FFNN that the choice of activation function can strongly influence the performance. Lastly, the number of parameters of the NQS can be varied. In Ref. \cite{Sehayek2019} it is argued for the exemplary architecture of an RBM that using overparameterized NQS and subsequently pruning small parameters of the trained model can improve the performance. The compression by pruning is also discussed in Ref. \cite{Golubeva2022}. However, increasing the number of parameters does not always yield an improvement: In Ref. \cite{dash2024efficiency} it is shown that the accuracy of an RBM increases for small widths of the hidden layer $\alpha$, but saturates at high $\alpha$. The authors observe that this behavior coincidents with a saturation of the quantum geometric tensor's rank, see Eq. \eqref{eq:S} in Sec. \ref{sec:GS}, i.e. the local dimension of the relevant manifold for the optimized NQS saturates. 
Lastly, the open-source packages allow to readily optimize code for graphics processing units (GPUs), which can highly accelerate NQS implementations, see also Sec. \ref{sec:Toolboxes}. Another direction for fast and energy efficient NQS implementations is spiking neuromorphic hardware \cite{KLASSERT2022104707,Czischek_spiking2022,Czischek2019sampling}, as implemented for a RBM in Ref. \cite{Czischek_spiking2022}.\\

For further reading on design choices beyond the discussion provided here, we refer the reader to Reh et al. \cite{Reh2023}, where the performance of RBMs, CNNs and RNNs with different symmetrization strategies are compared.

\subsection{Open-Source Toolboxes \label{sec:Toolboxes}}
There are several toolboxes that provide open-source implementations of NQS: \textit{NetKet} allows for ground state search, dynamics calculations based on TDVP and p-tVMC as well as state tomography using various architectures and comes with many implemented bosonic and fermionic Hamiltonians \cite{netket,netket_fidelity}. \textit{jVMC} \cite{jVMC}, designed for computationally efficient variational Monte Carlo, provides several architectures for ground state search and dynamics simulations as well. \textit{FermiNet} \cite{pfau2020ferminet} provides ground state simulations for atoms and molecules. All of them are based on Google's JAX library \cite{jax2018github}.
Lastly, we would like to mention \textit{QuCumber} \cite{Qucumber}, a RBM based tomography implementation.

\section{Applications of NQS \label{sec:applications}}
\subsection{Ground States}\label{sec:GS}

\subsubsection{Variational Monte Carlo}
To represent the ground state of a given system, neural quantum states are normally trained using variational Monte Carlo (VMC) \cite{McMillan1965,huang2017accelerated}. VMC is based on variational wave functions $\ket{\psi_{\vec{\theta}}}$ such as NQS, parameterized by parameters $\vec{\theta}$. To approximate ground states with NQS, the energy
\begin{align}
    E_{\vec{\theta}} = \frac{\bra{\psi_{\vec{\theta}}} \hat{H} \ket{\psi_{\vec{\theta}}}}{\langle \psi_{\vec{\theta}}\vert \psi_{\vec{\theta}} \rangle} \geq E_\mathrm{gs},
\end{align}
should be as close as possible to the ground state energy $ E_\mathrm{gs}$.
For variational wave functions $\psi_{\vec{\theta}}$, this expectation value $E_{\vec{\theta}}$ can be evaluated from samples $\vec{\sigma}$ drawn from the wave function's amplitude $\vert \psi_{\vec{\theta}}\vert^2$ according to Eq. \eqref{eq:expectations}. To approximate ground states, NQS are usually trained by minimizing the expectation value of the Hamiltonian, $E_{\vec{\theta}}=\langle \hat{H} \rangle \approx  \langle H_\mathrm{loc}(\vec{\sigma})\rangle_{\vec{\sigma}}$, i.e. parameters $\vec{\theta}$ are updated according to 
\begin{align}
    \partial_{{\vec{\theta}}_k} E_{\vec{\theta} }
    &=2\mathrm{Re}\left[\sum_{\vec{\sigma}} P_{\vec{\theta}}(\vec{\sigma})\frac{\partial_{{\vec{\theta}}_k} \psi^{*}_{\vec{\theta}} (\vec{\sigma}) }{\psi^{*}_{\vec{\theta}} (\vec{\sigma})} H^{\mathrm{loc}}_{\vec{\theta}}(\vec{\sigma})\right]-2\mathrm{Re}\left[\sum_{\vec{\sigma}^\prime} P_{\vec{\theta}}(\vec{\sigma}^\prime)\frac{\partial_{{\vec{\theta}}_k} \psi^{*}_{\vec{\theta}} (\vec{\sigma}^\prime) }{\psi^{*}_{\vec{\theta}} (\vec{\sigma}^\prime)} \sum_{\vec{\sigma}}P_{\vec{\theta}}(\vec{\sigma}) H^{\mathrm{loc}}_{\vec{\theta}}(\vec{\sigma})\right]\notag \\
     &= \frac{2}{N_s}\mathrm{Re}\left[\sum_i^{N_s} \partial_{{\vec{\theta}}_k} \mathrm{log }\psi^{*}_{\vec{\theta}} (\vec{\sigma}_i) \left[H_\mathrm{loc}(\vec{\sigma}_i)-\langle H_\mathrm{loc}(\vec{\sigma}^\prime)\rangle_{\vec{\sigma}^\prime}\right]\right]\notag \\&= 2\mathrm{Re}\langle \partial_{{\vec{\theta}}_k} \mathrm{log }\psi^{*}_{\vec{\theta}} (\vec{\sigma}) \left[H_\mathrm{loc}(\vec{\sigma})-\langle H_\mathrm{loc}(\vec{\sigma}^\prime)\rangle_{\vec{\sigma}^\prime}\right] \rangle_{\vec{\sigma}} .
    \label{eq:Egrad}
\end{align}
For autoregressive architectures, $\langle\psi_{\vec{\theta}}\vert \psi_{\vec{\theta}} \rangle = 1$ and hence the second term vanishes when calculating the derivative of Eq. \eqref{eq:expectations}. However, often $H_\mathrm{loc}(\vec{\sigma})$ is replaced by the covariate $\left(H_\mathrm{loc}(\vec{\sigma})-\langle H_\mathrm{loc}\rangle\right)$ to reduce the variance of the gradients \cite{Assaraf1999,hibat-allah_recurrent_2020}, leading to the same expression as Eq. \eqref{eq:Egrad}. Another approach is to (pre-)train the NQS with experimental or numerical data, see Sec. \ref{sec:QST}. \\

The optimization of NQS can be done with methods commonly used in machine learning, such as stochastic gradient descent, Adam \cite{kingma2017adam} and AdamW \cite{loshchilov2019adamW}. A more elaborate approach
is the stochastic reconfiguration (SR) algorithm \cite{Sorella1998,Sorella2001,becca_sorella_2017}, which incorporates the knowledge of the geometric structure of the parameter space to adjust the gradient direction \cite{Amari1998,Amari2019,Hackl2020}. The underlying idea\footnote{For the motivation of the SR algorithm from imaginary time evolution, we follow Refs. \cite{chen2023efficient,wagner2023neural}.} of SR is to perform an imaginary time evolution of the variational state $\ket{\psi_{\vec{\theta}}}$, i.e.
\begin{align}
    \ket{\psi(\tau+\delta \tau)}= \mathrm{exp}(-\delta\tau \hat{H}) \ket{\psi_{\vec{\theta}}} \underbrace{\approx}_{\mathrm{small}\, \delta\tau} \ket{\psi_{\vec{\theta}}} + \ket{\delta \psi_{\hat{H}}}.
\end{align}
For the latter equality, we have assumed small time steps $\delta \tau$, when the change of the state from the imaginary time evolution is
\begin{align}
    \ket{\delta \psi_{\hat{H}}} = - \delta\tau \hat{H}\ket{\psi_{\vec{\theta}}} = -\delta\tau \sum_{\vec{\sigma}} \psi_{\vec{\theta}}(\vec{\sigma}) H^{\mathrm{loc}}(\vec{\sigma}) \ket{\vec{\sigma}}.
\end{align}
Naturally, the evolved state $\ket{\psi(\tau+\delta\tau)}$ has less contributions from higher energy states, decreasing the energy with every imaginary time step $\delta \tau$ \cite{chen2023efficient}. In order to \textit{translate} the evolved state $\ket{\psi(\tau+\delta\tau)}$ to a parameter update $\vec{\theta} \to \vec{\theta}^\prime$, a projection onto the variational manifold of $\psi_{\vec{\theta}^\prime}$ is needed, which is done by minimizing the Fubini-Study (FS) distance $d(\psi(\tau+\delta\tau), \psi_{\vec{\theta}^\prime})$ \cite{Park2020}. Expanding also for the projected state for small $\delta \tau$, $\ket{\psi_{\vec{\theta}^\prime}}=\ket{\psi_{\vec{\theta}}} + \ket{\delta \psi_{\vec{\theta}}} $, with
\begin{align}
    \ket{\delta \psi_{\vec{\theta}}} = \sum_{k,\vec{\sigma}}\frac{\partial \psi_{\vec{\theta}}(\vec{\sigma}) }{\partial \theta_k}\delta \theta_k \ket{\vec{\sigma}} = \sum_{\vec{\sigma}} \psi_{\vec{\theta}}(\vec{\sigma}) \sum_k O_{k}(\vec{\sigma}) \delta\theta_k\ket{\vec{\sigma}} 
\end{align}
and $O_{k}(\vec{\sigma} ) = \frac{1}{\mathrm{\psi}_{\vec{\theta}(\vec{\sigma})} }\partial_{\vec{\theta}_k} \mathrm{\psi}_{\vec{\theta}} (\vec{\sigma})$ \cite{Carleo2017}, the FS distance can be written as
\begin{align}
    d(\psi(\tau+\delta\tau), \psi_{\vec{\theta}^\prime}) = \vert \vert \Bar{O} \delta \vec{\theta}-\bar{H}^{\mathrm{loc}} \vert\vert_2
\end{align}
with $\vert\vert \dots \vert\vert_2$ the $L^2$ norm, the matrix  $\Bar{O}_k(\vec{\sigma}) = \frac{1}{\sqrt{N_s}}(O_k(\vec{\sigma})-\langle O_k(\vec{\sigma})\rangle_{\vec{\sigma}})$ and the vector $\bar{H}^{\mathrm{loc}}$ defined in analogy. This results in the SR equation
\begin{align}
    \Bar{O}\delta \theta = \Bar{H}_\mathrm{loc} \quad \Leftrightarrow \quad  \delta\theta = S^{-1}\Bar{O}^\dagger \Bar{H}_\mathrm{loc} = S^{-1} F,
\end{align}
with the quantum geometric tensor 
\begin{align}
    S = \bar{O}^\dagger \Bar{O},
    \label{eq:S}
\end{align}
and the vector of forces
$$ F_k = \langle H_{\text{loc}} O_k^* \rangle - \langle  H_{\text{loc}} \rangle \langle O_k^* \rangle. $$
Hence, the SR parameter update at the $p$-th iteration is given by
\begin{align}
     \vec{\theta}(p+1)= \vec{\theta}(p) - \gamma(p) S^{-1} F,
     \label{eq:SR}
\end{align}
with a scaling parameter $\gamma(p)$\cite{Carleo2017}. \\ 

The SR update Eq. \eqref{eq:SR} hence involves an inversion of the $S$ matrix.  This inversion comes with the following problems: $(i)$ The matrix $S$ has to be estimated to a very high precision to avoid instabilities in the optimization, which requires a large number of samples. $(ii)$ $S$ is not necessarily invertible and hence, often a regularization of $S$ is needed for a stable optimization, especially if considered close to critical points or in large spin systems \cite{czischek2018quenches,Carleo2017,schmitt2020quantum}. Recent works indicate that the spectra of the $S$ matrix are distinctly different for non-autoregressive and autoregressive architectures, which can cause problems for regularizing $S$ for the latter \cite{donatella2023dynamics,lange2023neural}. $(iii)$ $S$ is a matrix of typically very large dimensions $N_{\vec{\theta}}\times N_{\vec{\theta}}$, making the inversion computationally costly since the complexity of inverting $S$ scales with $\mathcal{O}(N_sN_{\vec{\theta}}^2+N_{\vec{\theta}}^3)$ when using direct linear solvers \cite{chen2023efficient}.  To enlarge the allowed number of parameters by some orders of magnitudes, large-scale supercomputers are needed \cite{Li2022,Zhao2022}.

To overcome problem $(iii)$, several modifications of SR have been proposed. Among them are iterative solvers such as MINRES \cite{minres,Carleo2017} which avoid this scaling by iteratively computing the pseudo-inverse of $S$ \cite{Carleo2017}. However, for large $N_s$ and $N_{\vec{\theta}}$ typically also the required number of iterations grows. Another method is the sequential local optimization approach, in which SR only optimizes a portion of all parameters to reduce the time cost \cite{Zhang2023}. Other recent works have proposed modifications of the SR update rule which involve the inversion of a $N_s\times N_s$ matrix instead of $S$, with $N_s$ the number of samples to estimate the gradient, which is usually smaller than the number of parameters \cite{rende_simple_2023,chen2023efficient}. Apart from that, the performance of SR can be improved with adaptive learning rate solvers, such as the second order Runge Kutta integrator, allowing for an optimal choice of the learning rate \cite{Bukov2021}.\\

The optimization of NQS can become very difficult due to the in general very rugged and chaotic optimization landscape with many local minima \cite{Bukov2021,Park2020,Inack2022}. To overcome this problem, many techniques have been developed. Among them are \textit{variational neural annealing} that applies an artificial temperature to avoid getting stuck in local minima \cite{hibat-allah_variational_2021, hibat-allah_supplementing_2022, khandoker_supplementing_2023}, the application of \textit{symmetries} \cite{Nomura_2021,morawetz_u1-symmetric_2021,hibat-allah_recurrent_2020,Reh2023,roth2021group} to enforce a training only in the target symmetry sector, \textit{transfer learning}, i.e. the transfer learned properties of small systems to larger system sizes, \cite{Zen2020,roth_iterative_2020,Efthymiou}, and \textit{weight pruning} \cite{Sehayek2019, Golubeva2022}. \\

\subsubsection{Optimization Challenges} Further improvement can be achieved using complementary optimization methods: Firstly, the NQS can be pretrained with external data, see Sec. \ref{sec:LearningfromData}. Second, the energy resulting from the VMC optimization can be improved by applying a few \textit{Lanczos} steps after optimizing the network parameters using the techniques described above \cite{chen2022systematic}. Applying the Hamiltonian in the Lanczos algorithm to obtain the next Krylov vector corresponds to minimizing the infidelity to the corresponding state and thus necessitates a separate training, rendering typically only very few Krylov vectors accessible. Similar in spirit to Lanczos algorithms, power methods can be used to find the ground states of (gapped) Hamiltonians \cite{Giuliani2023learningground}.

A related approach, also based on applying imaginary time evolution, maximizes the fidelity to a target state at each iteration. In contrast to SR, this is done explicitly, i.e. a small imaginary time step $\Delta \tau$ has to be applied to the current state using e.g. the Euler \cite{Ledinauskas2023} or the Heun method \cite{donatella2023dynamics}. 
In Ref. \cite{kochkov2018variational}, the general idea of minimizing the difference between the current parameterization and an explicitly improved wave function is introduced as supervised wave function optimization (SWO). The improved wave function, which constitutes the target in this optimization, can e.g. be obtained through power methods or imaginary time evolution. 

Lastly, in Ref. \cite{Hristiana2023}  an optimization scheme based on stochastic representations of wave functions is proposed. In this representation, not the configurations, here in terms of particle positions $\{\vec{R}_i\}_{i=1,\dots,N_s}$ of a continuous system, but a set of $N_s$ samples $(\vec{R}_i, \psi_s(\vec{R}_i))$ is used for the optimization. The NQS is given by
\begin{align}
    \psi_s(\vec{R}_i) = e^{-\delta\tau \hat{H}} \hat{P}_{s/a} \psi_{\vec{\theta}}(\vec{R})\vert_{\vec{R}=\vec{R}_i},
\end{align}
with a variational function $\psi_{\vec{\theta}}$ parameterized by a FFNN and stochastic projection $\hat{P}_{s/a}$ onto the symmetric or antisymmetric subspace. In order to train the NQS, first a set of samples $(\vec{R}_i, \psi_s(\vec{R}_i))$ is generated and projected onto the target subspace. Then, simple regression is applied, with the goal of minimizing the sum of squared residuals between the projected samples and $ \hat{P}_{s/a}\psi_{\vec{\theta}}(\vec{R}_i)$. The updated trial function $\hat{P}_{s/a} \psi_{\vec{\theta}^\prime}$ is then used to generate new sample coordinates $\vec{R}_i^\prime$, and imaginary time evolution is performed on  $\hat{P}_{s/a} \psi_{\vec{\theta}^\prime}(\vec{R})\vert_{\vec{R}=\vec{R}_i^\prime}$. In contrast to VMC, this method does not require that the samples are distributed according to the wave functions' amplitudes. Furthermore, no evaluation of the energy or its gradients is required. \\

One reason why the optimization is so challenging is the intricate interplay between phase and amplitude parts during the optimization. In some cases, the optimization outcomes are improved by imposing a certain sign structure of the target state, e.g. the Marshall sign rule for the Heisenberg model (restricted to bipartite lattices) \cite{Szabo2020,roth_iterative_2020,hibat-allah_recurrent_2020}. Furthermore, the interplay between phase and amplitude during the optimization can be investigated by considering the partial optimization problem. In Refs. \cite{Bukov2021,lange2023neural}, either the exact phases or the exact amplitudes are set and kept constant during the training, such that only amplitude \textit{or} phase, respectively, have to be learned. In both works, considering the $J_1-J_2$ model and the bosonic and fermionic $t-J$ model, the authors find that none of the two optimization strategies can systematically improve the ground state representation, and hence conclude that the interplay between phase and amplitude seems to play a crucial role for the optimization.

\subsubsection{Results on the $J_1-J_2$ model}
\begin{figure}[t]
\centering
\includegraphics[width=1\textwidth]{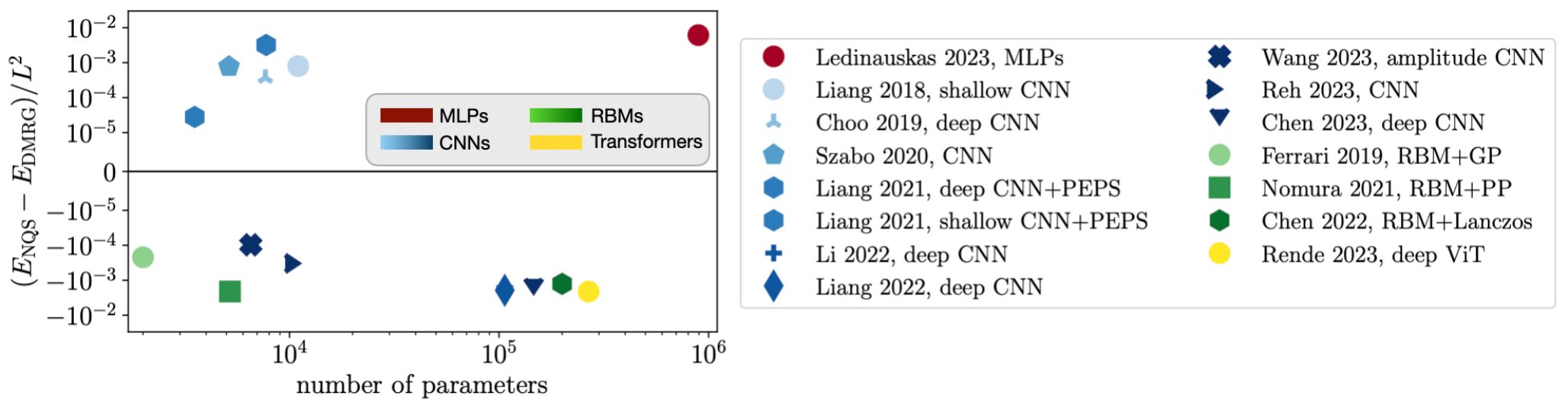}
\caption{Comparison of energies obtained with NQS, $E_\mathrm{NQS}$, and matrix-product states (MPS) using density matrix renormalization group (DMRG), $E_\mathrm{DMRG}$ with $8192$ $SU(2)$ states \cite{Gong2014}, for ground states of the prototypical $J_1-J_2$ model at $J_2/J_1=0.5$ on $10\times 10$ lattices: We show energies obtained from FFNNs/MLPs in red (Ledinauskas et al. (2023) \cite{Ledinauskas2023}), CNNs in blue (Liang et al. (2018) \cite{liang2018solving}, Choo et al. (2019) \cite{choo2019two}, Szabo et al. (2020) \cite{Szabo2020}, Liang et al. (2021) \cite{Liang2021}, Li et al. (2022) \cite{Li2022}, Liang et al. (2022) \cite{Liang_2023}, Wang et al. (2023) \cite{wang2023variational}, Reh et al. (2023) \cite{Reh2023}, Chen et al. (2023) \cite{chen2023efficient}), RBMs in green (Ferrari et al. (2019) \cite{Ferrari2019}, Nomura et al. (2021) \cite{Nomura2021_Dirac}, Chen et al. (2022) \cite{chen2022systematic}) and transformers in yellow (Rende et al. (2023) \cite{rende_simple_2023}). The color coding denotes the appearance of the respective works. This plot is adapted from a table in Ref. \cite{rende_simple_2023}.} 
\label{fig:overview}    
\end{figure}
A commonly used model for benchmarking new NQS architectures and comparing different optimization methods is the $J_1-J_2$ model,
\begin{align}
    \mathcal{H}_{J_1-J_2}=J_1\sum_{\langle i,j\rangle}\hat{\vec{S}}_i\cdot \hat{\vec{S}}_j+J_2\sum_{\langle \langle i,j\rangle\rangle}\hat{\vec{S}}_i\cdot \hat{\vec{S}}_j,
\end{align}
with spin-$1/2$ operators $\hat{\vec{S}}_i$ and $\langle i,j\rangle$ ($\langle\langle i,j\rangle\rangle$) denoting nearest (next-nearest) neighbors. In this model, the nearest neighbor $J_1$ and the next-nearest neighbor $J_2$ interactions compete. For $J_2\to 0$ and $J_1>0$, the system reduces to a Heisenberg antiferromagnet, whereas for $J_1\to0$ and $J_2>0$ the system favors antiferromagnetic stripes. In the intermediate regime around $J_2/J_1=0.5$, the nature of the ground state is not clear yet, ranging from numerical results for gapped or gapless quantum spin liquids, different types of valence bond solids or both of them \cite{Nomura2021_Dirac}. With the frustration controlled by the ratio $J_2/J_1$, the $J_1-J_2$ model has become a paradigmatic model for the evaluation of the performance of NQS architectures \cite{Ledinauskas2023,liang2018solving,choo2019two,Szabo2020,Liang2021,Li2022,Liang_2023,wang2023variational,Reh2023,chen2023efficient,Ferrari2019,Nomura2021_Dirac,chen2022systematic,rende_simple_2023,roth2023high}. 

The results obtained with variants of FFNNs, CNNs, RBMs and transformers for the $J_1-J_2$ model on $10\times10$ square lattices at $J_2/J_1=0.5$ are shown in Fig. \ref{fig:overview}. It can be seen that in recent years, results from CNNs, RBMs and Transformers have become competitive with or have even outperformed DMRG results obtained for a bond dimension of $8192$ with implemented $SU(2)$ symmetry (corresponding to a bond dimension $32000$ with only $U(1)$ symmetry). In particular, recent modifications of SR as in \cite{rende_simple_2023,chen2023efficient} have allowed to use more than $10^5$ parameters, systematically improving results obtained with smaller NQS with around $10^3-10^4$ parameters. However, also for $10^3-10^4$ parameters some works have obtained energy errors that are competitive with the NQS architectures using more than one order of magnitude more parameters \cite{Nomura2021_Dirac,wang2023variational,Reh2023}. It hence becomes evident that in principle many NQS architectures can achieve competitive results to conventional methods, but details on the implementation and the optimization procedure can have a significant impact on the performance. Consequently, architecture, its hyperparameters and optimization parameters have to be carefully chosen, but at the same time they can mostly only be determined by try and error.

\subsection{Excited States}\label{sec:excited}
\begin{figure}[t]
\centering
\includegraphics[width=0.75\textwidth]{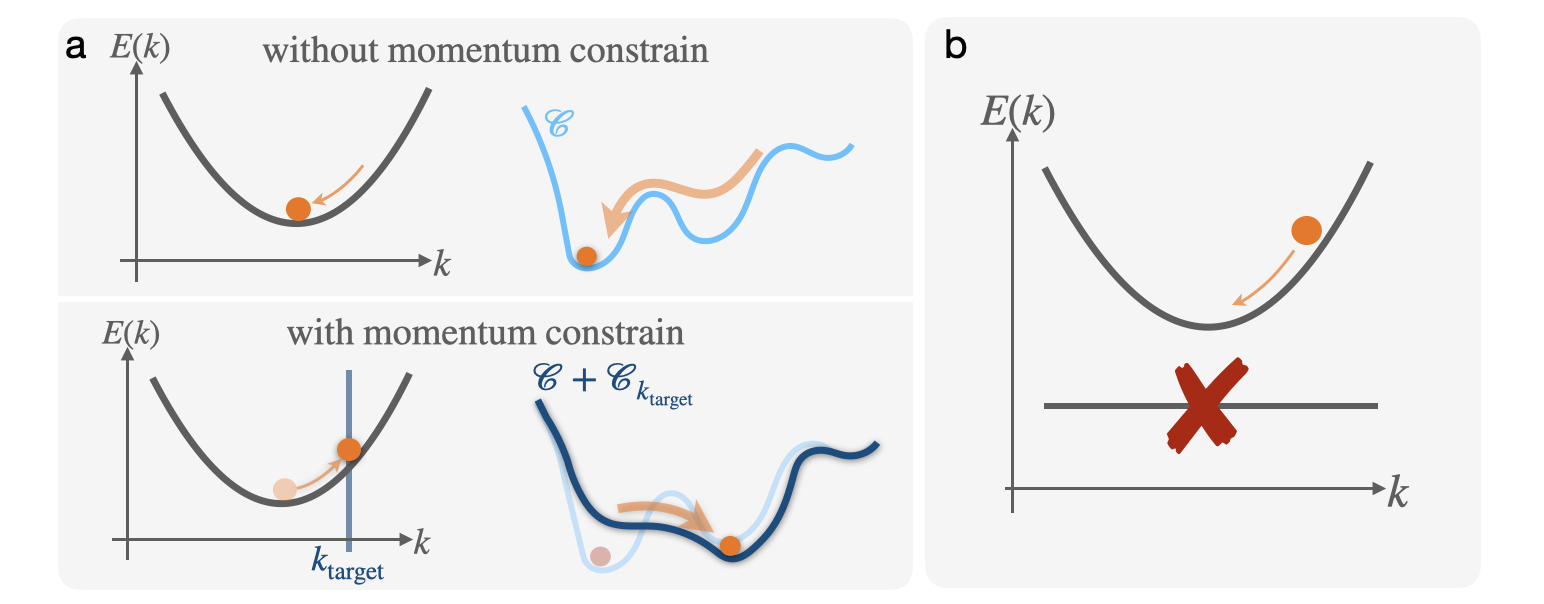}
\caption{Excited States: a) Optimization for different momentum sectors. Figure adapted from Ref. \cite{lange2023neural}. b) Low-energy states in the same symmetry by enforcing orthogonality to the lower excited states and the ground state as in Refs. \cite{Choo2018,Valenti2022}.} 
\label{fig:ExcitedStates}    
\end{figure}

The methods for calculating excited states using NQS fall into two categories, depending on the type of excited states that are targeted: $(i)$ Lowest energy states in a different symmetry sector than the ground state, e.g. different momentum or magnetization sectors. $(ii)$ Low-energy states in the same symmetry sector as the ground state.\\

The former usually rely on the same VMC scheme, but with a restriction to the targeted symmetry sector implemented in the wave function or the training loss \cite{Choo2018, lange2023neural,Nomura2020,Nomura_2021,Viteritti2022accuarcy,Lange_2023}. This enables e.g. the calculation of quasiparticle dispersions from NQS \cite{Choo2018, lange2023neural,Nomura2020,Nomura_2021,Valenti2022}:

In Ref. \cite{Nomura_2021}, excited states are targeted using quantum-number projections \cite{Mizusaki2004}, i.e. for a NQS parameterization $\psi_{\vec{\theta}}(\vec{\sigma})$ one defines the wave function using the total momentum $\vec{k}$ projection
\begin{align}
    \psi_{\vec{k}}(\vec{\sigma})=\sum_{\vec{R}}e^{-i\vec{k}\cdot \vec{R}}\psi_{\vec{\theta}}(\hat{T}_\mathrm{R}\vec{\sigma})
\end{align}
with the translation operator $\hat{T}_{\vec{R}}$, shifting the system by the vector $\vec{R}$. This can be combined with other quantum-number projections, including e.g. spin parity and spatial symmetries, to improve the accuracy. In another approach \cite{lange2023neural}, specific momentum $\vec{k}_\mathrm{target}$ sectors are targeted by adding a mean square error between $\vec{k}_\mathrm{target}$ and $\vec{k}_\mathrm{NQS}$, see Fig.~\ref{fig:ExcitedStates}a. Hereby, the momentum of the NQS is given by
\begin{align}
\vec{k}_\mathrm{NQS}^\mu = \frac{i}{a}  \mathrm{log}\bra{\psi_{\vec{\theta}}}\hat{T}_{\vec{e}_\mu}\ket{\psi_{\vec{\theta}}},
\end{align}
with the translation operator $\hat{T}_{\vec{e}_\mu}$ in direction of the unit vector $\vec{e}_\mu$ by a lattice constant $a$. In contrast to MPS calculations, this global operator can be evaluated at low computational cost. Another approach to target excited states in different symmetry sectors consists in the use of group convolutional neural networks, as discussed in Section \ref{sec:CNN}.\\

For the second category, Fig.~\ref{fig:ExcitedStates}b, usually more modifications need to be made. In Refs. \cite{Choo2018,Valenti2022} excited states are targeted by enforcing orthogonality to the ground state $\psi_0$ or lower lying excited states $\psi_i$. In Ref. \cite{Valenti2022}, the first excited state $\psi_{1,\vec{\theta}}$ is calculated by adding the normalized overlap with the ground state, 
\begin{align}
    \frac{\langle \psi_{0,\vec{\theta}^\prime} \vert \psi_{1,\vec{\theta}} \rangle}{\langle  \psi_{0,\vec{\theta}^\prime} \vert \psi_{0,\vec{\theta}^\prime}\rangle \langle  \psi_{1,\vec{\theta}} \vert \psi_{1,\vec{\theta}}\rangle},
\end{align}
to the cost function. In Ref. \cite{Choo2018}, the excited state is defined as \begin{align}
    \psi_1:= \Phi_{\vec{\theta}} -\lambda \psi_{0,\vec{\theta}^\prime}.
\end{align}
To enforce orthogonality, $\langle \psi_1\vert \psi_{0,\vec{\theta}^\prime}\rangle=0$, they use
$\lambda = \langle \Phi_{\vec{\theta}} / \psi_{0,\vec{\theta}^\prime} \rangle $. Both methods require the explicit representation of the ground state $\psi_{0,\vec{\theta}^\prime}$, rendering the calculation of higher excited states computationally demanding.

Another method, requiring no explicit orthogonalization of the different states, transforms the problem of finding the $K$ lowest excited states of a given system into that of finding the ground state of an expanded system given by all targeted excited states \cite{pfau2023natural}. The ansatz for the expanded ground state is written as
\begin{align}
    \Psi(\vec{x}) := \mathrm{det}\left(\begin{matrix}
        \psi_1(\vec{x^1}) & \dots& \psi_K(\vec{x^1})\\
        \vdots & &\vdots\\
        \psi_1(\vec{x^K}) & \dots& \psi_K(\vec{x^K})
    \end{matrix}\right),
\end{align}
i.e. an unnormalized Slater determinant of many-particle wave functions $\psi_i$ instead of single-particle orbitals. Here, $\vec{x}^i$ denotes a set of $N$ particles $x_1^i,...,x_N^i$. The Hamiltonian is correspondingly expanded to act on all $K$ particle sets, and VMC is performed to find the ground state of the expanded system. Subsequently, energies, expectation values, and overlaps in the excited states can be retrieved from $\Psi(\vec{x})$. 

Finally, in Ref. \cite{Yoshioka2021} the authors propose a method based on the assumption that one-particle excitations dominate the low-lying spectrum, allowing to construct the excited states by single-particle excitations on top of the ground state.

\subsection{Dynamics \label{sec:dynamics}}

\begin{figure}[t]
\centering
\includegraphics[width=0.85\textwidth]{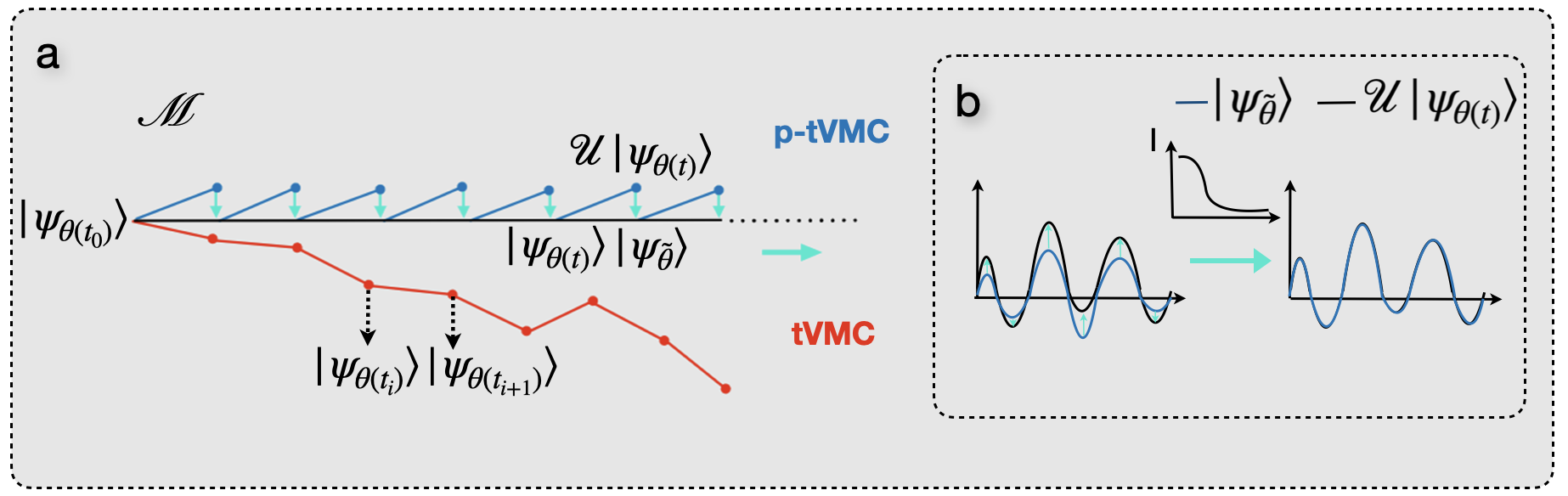}
\caption{a) Adapted visualisation from Ref. \cite{sinibaldi2023unbiasing} of the p-tVMC method. In the cases when the tVMC evolution of a state $\psi$ breaks down, p-tVMC can resolve this problem by projecting the exactly evolved state onto the variational manifold $\mathcal{M}$. The tVMC evolution of the state (red) starts to break down whereas the p-tVMC method projects the exactly evolved state $U |\psi_{\vec{\theta}}\rangle$ back onto the variational manifold $\mathcal{M}$ (blue) at each timestep. b) To project the exactly evolved state onto the variational wave function, the distance in the Hilbert space needs to be minimized, resulting in an optimization problem of the infidelity. } 
\label{fig:dynamics}    
\end{figure}

Neural network quantum states are furthermore capable of describing time-dependent systems. The quantum dynamics can be obtained using time-dependent network weights $\vec{\theta}(t)$ \cite{Carleo2017, becca_sorella_2017}. 
Numerically exact results for timescales comparable to or exceeding the capabilities of TN algorithms have been obtained for the paradigmatic two-dimensional transverse-field Ising model \cite{schmitt2020quantum}. However, even though NQS are able to capture strongly entangled states, the required number of parameters needed to represent a quantum state after a global quench can grow exponentially in time, according to Ref. \cite{Lin2022}. This can potentially render the use of NQS to represent the dynamics of a state inefficient. Dynamically increasing the network size and choosing a network architecture incorporating the symmetry of the state are promising approaches to overcome this problem. \\

Following the Dirac-Frenkel Variational principle, the time derivative of the neural network weights are to be optimized such that the variational residuals 
\begin{align}
 \text{R}(\vec{\dot{\theta}}(t)) = \text{dist}(\partial_{t}\psi_{\vec{\Tilde{\theta}}}, -i\hat{H}\psi_{\vec{\theta}}) 
\end{align} 
are minimized \cite{Carleo2017, gutierrez2022real}. This is achieved within the stochastic approach using time-dependent Variational Monte Carlo method (t-VMC). In most cases t-VMC is used in combination with stochastic reconfiguration, see Sec. \ref{sec:GS}. The iteration scheme to find the ground state energy can be interpreted as an effective imaginary time evolution, such that an iteration scheme to approximate the real-time evolution of the quantum spin system can be derived in an analogous way \cite{czischek2018quenches}. At each time step, $|\psi_{\vec{\theta}(t)}|^2$ is sampled and the variational residual is evaluated. Minimization of the variational residuals with respect to $\dot{\vec{\theta}}$ gives a first order differential equation for the weights $\vec{\theta}(t)$ \cite{schmitt2020quantum}. The final equation to be solved for the variational parameters $\vec{\theta}$ is then given by the time-dependent variational principle (TDVP) equation:
\begin{align}    
S(t) \dot{\vec{\theta}}(t) = -iF(t),
\label{eqofmot}
\end{align} 
with the covariance matrix $S$ and the vector of forces $F$ are the same as in section \ref{sec:GS} .   \\
To solve this equation for the variational parameters, the covariance matrix needs to be inverted. Since this matrix can be non-invertible, $S^{-1}$ denotes the Moore-Penrose pseudo-inverse. This inversion leads to various problems, see also Sec. \ref{sec:GS}. Moreover, the unstable Moore-Penrose pseudo-inverse of the $S$ matrix requires choosing a right cut-off tolerance for small singular values. In practice, one finds that the chosen cut-off for the pseudo-inverse of $S$ can alter t-VMC. Krylov subspace methods (i.e. the conjugate gradient method or MINRES algorithm) avoid this sensitivity problem, but are not always converging \cite{gutierrez2022real}. When calculating the dynamics of a system, this error accumulates with each time step. \cite{gutierrez2022real, donatella2023dynamics} \\

Regularization schemes for $S$ have been developed, see Sec. \ref{sec:GS}, but these still require a very large number of samples \cite{gutierrez2022real} and impact the accuracy \cite{donatella2023dynamics}. The choice of regularization method impacts the stability of the dynamics \cite{Hofmann2022}. 
 In Ref.~\cite{schmitt2020quantum}, this problem is approached by disregarding the contributions to the TDVP of which there is insufficient information available due to constraints of a finite number of samples $N_{MC}$. This still requires the diagonalization of $S$, which has a high computational cost ($O(N_{\vec{\theta}}^3)$) \cite{gutierrez2022real}. In Ref. \cite{zhang2024paths}, they adopt the minSR approach to reduce this computational cost to $O(N_{s}^3)$, as it allows to reshape $S$ into a $N_s \times N_s$ matrix.

Successful applications of t-VMC have been obtained for system sizes up to $N = 256$ spins using an RBM and for time scales up till 10s in Refs. \cite{Carleo2017, lee2021neural, Fabiani2019}. In Ref. \cite{Fabiani2021}, an RBM ansatz is used to capture the dynamics of the 2D Heisenberg model, predicting strong magnon-magnon interactions leading to supermagnonic propagation. In Ref. \cite{Schmitt2022_quantumphase} the dynamics of the 2D transverse-field Ising model around the quantum phase transition has been studied with t-VMC, leading to insights in the quantum Kibble-Zurek mechanism. Numerical simulations obtained with t-VMC have been used to validate experimental data interpretation, as is done in Ref. \cite{mendessantos2023wave} for a network description of wave function snapshots. However, accessing long times via t-VMC stays challenging. The stability of t-VMC strongly depends on the chosen variational ansatz \cite{donatella2023dynamics}, and is affected by systematic statistical bias or an exponential sample complexity when the wave-function contains zeros.  This is also the case for ground state calculations, but is less harmful due to the accumulation of error that affects dynamics \cite{sinibaldi2023unbiasing}.

A new method is proposed to circumvent these issues in Refs. \cite{donatella2023dynamics} and \cite{sinibaldi2023unbiasing}: Projected time-dependent Variational Monte Carlo (p-tVMC).  The scheme consists of casting a Runge-Kutta integration scheme into minimizing a variational distance at each time step. Starting again from the Dirac-Frenkel Variational Principle, an $s$-order Runge-Kutta approximant is to be used, instead of a first order expansion of the time propagator \cite{donatella2023dynamics}. 
In practice, a metric based upon infidelity can now be used, in contrast to the normally used Fubini-Study distance. This infidelity can be estimated through Monte Carlo sampling and should be considered the distance in the Hilbert space between $U|\psi_{\vec{\theta}}\rangle$ and $|\psi_{\Tilde{\vec{\theta}}}\rangle$ , which is to be optimized \cite{sinibaldi2023unbiasing}: \begin{align}
    \min_{\Tilde{\theta}} \text{I}(|\psi_{\vec{\Tilde{\theta}}} \rangle, U |\psi_{\vec{\theta}} \rangle)  .
\end{align}
No expansion of $|\psi_{\vec{\Tilde{\theta}}} \rangle $ with respect to the variational parameters is necessary for p-tVMC, opposed to tVMC where a second order expansion leads to Eq. \eqref{eqofmot}. However, the unitary time evolution $e^{-i\hat{H}\delta t}$ needs to be decomposed by the Trotter-Suzuki decomposition, leading to a scaling of the number of required samples with the Trotter order. This is costly and breaks translational invariance. Alternatively, an expansion of the time propagator into its Taylor decomposition is needed, typically up to second order in $\hat{H}$ in order to evaluate long time dynamics. This results in a quadratic increase of the connected matrix elements that need to be computed, and thereby the computational cost. Furthermore, for higher $s$-order integration schemes, the computational cost scales with the system size $N$ as $N^s$ \cite{donatella2023dynamics}. In Ref. \cite{sinibaldi2023unbiasing} the use of an RBM avoids this issue by computing the off-diagonal elements in the transverse field Ising model exactly. Because of this, the p-tVMC method only scales linearly with the number of parameters, which makes this a promising method to compute dynamics for large neural network architectures. As opposed to an update rule for the network parameters $\vec{\theta}$, standard gradient descent based techniques to minimize the infidelity can be used. Since p-tVMC is not affected by biases or vanishing SNR, it can simulate dynamics in cases where t-VMC fails or is inefficient \cite{sinibaldi2023unbiasing}. \\

 An alternative approach, using the implicit midpoint method, has been explored in Ref.~\cite{gutierrez2022real}. Here, the network parameters are optimized to minimize the error between the state at the next time step $t+\Delta t$ and the discrete flow of the implicit midpoint method applied to the Schr\"odinger equation.
This has shown some advantages in preserving the symplectic form of Hamiltonian dynamics while not complicating the network optimization with intermediate quantities \cite{gutierrez2022real}.  \\

In Ref.~\cite{Burau2021}, a dynamical strong disorder renormalization group approach is used to map the quantum dynamics of a disordered spin chain onto a quantum circuit generated by local unitaries. These local unitaries are applied to the NQS in a supervised scheme, similar to the SWO discussed in section \ref{sec:GS} and the infidelity minimization in p-tVMC. \\

Other methods to calculate the dynamics of a system consist of training with time evolved states that have been exactly calculated using ED, such that the evolution of new initial states can be predicted by a neural network without evolving the wave function explicitly with the Hamiltonian \cite{zhang2020predicting}. To speed up the simulation of the dynamics of many-body systems, hybrid methods are used such as using neural quantum states with calculations on quantum devices to determine expectation values with high computational cost \cite{lee2021neural}. Another approach are \textit{variational classical networks} \cite{Schmitt2018,Verdel2021,Karpov2021}, i.e. efficient and perturbatively controlled representations of (time-evolved) wave functions in terms of classical spins, where the latter are used to construct a NQS representation of the state under consideration.  \\

\subsubsection{Spectral Functions}

Most of the work discussed so far focus on global quenches. Time-dependent NQS can however also be used to simulate local quenches, such as the response of a system to a local perturbation, relevant for spectral functions. 
In Ref.~\cite{MendesSantos2023}, the dynamical spin structure factor of different two-dimensional quantum Ising models is calculated by applying t-VMC following a local perturbation (application of $\hat{S}^z$ operator) on top of the ground state represented by a convolutional neural network. Subsequent Fourier transformation yields the momentum- and frequency-resolved structure factor. 
A complementary approach is demonstrated in Ref.~\cite{Hendry2021}. Here, the dynamical structure factor of the one- and two-dimensional Heisenberg model is calculated explicitly using a Chebyshev expansion, where the corresponding wave functions are represented as RBMs. 
In Ref.~\cite{Hendry2019}, the Green's function
\begin{align}
    G_{ij}(z) = \left<\psi\right| \hat{A}_i^\dagger \frac{1}{z-\hat{H}}\hat{A}_j\left|\psi\right>
\end{align}
is directly calculated by an extension of the stochastic reconfiguration approach to obtain the {\it{correction vector} }
\begin{align}
    \left|\chi_j(z)\right> = \frac{1}{z-\hat{H}} \hat{A}_j \left|\psi\right>,
\end{align}
where the corresponding ground state $\ket{\psi}$ of the system has been obtained beforehand using SR.

\subsection{Finite Temperature States \label{sec:finiteT}}

In many experimentally relevant situations, we are dealing with quantum many-body systems at a finite temperature, and in order to compute thermodynamics properties of the system one needs to work with the thermal density matrix
\begin{equation}
\hat{\rho}=\frac{1}{Z} e^{-\beta \hat{H}}.
\end{equation}
Where $\hat{H}$ is the Hamiltonian of the system,  \(\beta = \frac{1}{k_B T}\) is the inverse temperature, and \(k_B\) is the Boltzmann constant. Hence, the task boils down to evaluating this density matrix efficiently. One approach that has been developed and is commonly applied in the context of MPS is the idea of purification, also known as the thermofield approach \cite{verstraete2004matrix,Zwolak2004,feiguin2005finite}. In the purification method, an additional auxiliary site is introduced for each physical site of the system, known as an ancilla. As a result, one deals with a pure instead of a mixed state, where said pure state lives in a higher dimensional Hilbert space. 
The algorithm then starts from an infinite temperature state, followed by imaginary time evolution to cool down the system to the desired temperature, where imaginary time \(\tau\) here denotes the inverse temperature \(\beta\).
The desired thermal density matrix is then obtained by tracing out the auxiliary degrees of freedom $a$, i.e.
\begin{align}
\hat{\rho}(\vec{\sigma},\vec{\sigma}^\prime) = \sum_a \langle \vec{\sigma},a \vert \psi \rangle \langle \psi \vert \vec{\sigma}^\prime,a\rangle.
\label{eq:purification}
\end{align}

\begin{figure}[t]
\centering
\includegraphics[width=0.99\textwidth]{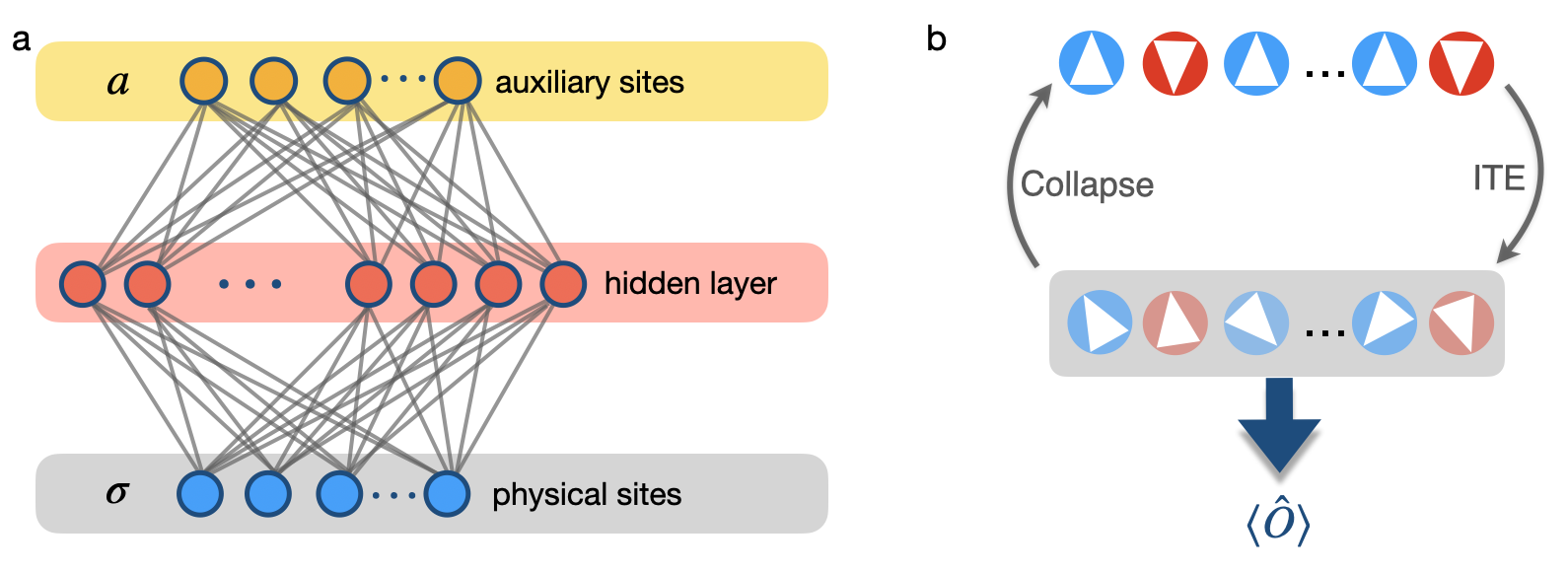}
\caption{Finite temperature state simulations: a) In the purification method, ancilla sites $a$ are introduced in addition to the physical sites $\sigma$, see Eq. \eqref{eq:purification}. The mixed state of the system is obtained by integrating out the auxiliary sites. b) METTS: Starting from a pure infinite temperature state (top), the state is evolved in imaginary time (ITE) to produce a METT state (bottom). Next, a projective measurement (collapse) to a random basis state is done in order to obtain a new pure state. The expectation value of desired observables is calculated through the ensemble of METT states.} 
\label{fig:finiteT}    
\end{figure}

In the context of neural network quantum states, one approach to purification is through a modified RBM, see Fig. \ref{fig:finiteT}a. A similar type of architecture was also used in Ref.~\cite{Torlai2018latent} to reconstruct mixed states. 
Refs. \cite{nomura2021purifying,wagner2023neural} employ the purification method to obtain finite temperature expectation values for a Heisenberg chain and a $6\times 6$ $J_1-J_2$ model. On top of the imaginary time evolution, Ref. \cite{nys2023real} deals with real-time evolution, leveraging an RNN architecture. \\

Another promising approach also developed in the context of tensor networks is the idea of minimally entangled typical thermal states (METTS) \cite{White2009,stoudenmire2010minimally}. METTS is designed to efficiently sample from the thermal ensemble instead of
dealing with the full complexity of the mixed state directly. The idea is to construct an ensemble of pure states, which provides a good approximation of the thermal equilibrium state. Concretely, the trace in the evaluation of finite temperature expectation values can be expanded in terms of an orthonormal basis as
\begin{align}
    \langle\hat{O}\rangle = \Tr \bigl(\hat{\rho}\hat{O}\bigr) = \frac{1}{Z} \sum_i \left<\boldsymbol{\sigma}_i\right| e^{-\beta\hat{H}/2} \hat{O} e^{-\beta\hat{H}/2} \left| \boldsymbol{\sigma}_i \right> = \frac{1}{Z} \sum_i P(i) \left< \psi_{\boldsymbol{\sigma}_i}(\beta)\right| \hat{O} \left| \psi_{\boldsymbol{\sigma}_i} (\beta)\right>
\end{align}
To this end, one starts from a pure product state $|\boldsymbol{\sigma}_0\rangle$. This product state is evolved in imaginary time to generate a state $\left| \psi_{\boldsymbol{\sigma}_i}(\beta)\right> = e^{-\beta\hat{H}/2} \left| \boldsymbol{\sigma}_i \right>$. This procedure gives us a so-called METTS state $\left|\psi_{\boldsymbol{\sigma}}(\beta)\right\rangle$. After this step, a projective measurement in the computational basis (collapse) is performed in order to produce a new pure state $|\boldsymbol{\sigma_1}\rangle$ to start over with imaginary time evolution. This procedure of sampling the states $\left|\boldsymbol{\sigma}_i\right>$ ensures that the resulting states represent the thermal ensemble accurately \cite{stoudenmire2010minimally}. At the end, we have a set of states $\{  \left|\psi_{\boldsymbol{\sigma}_0}(\beta) \right>, \left|\psi_{\boldsymbol{\sigma}_1}(\beta) \right>, \ldots, \left|\psi_{\boldsymbol{\sigma}_n}(\beta) \right> \}$ from which the thermal average of a given operator $\hat{O}$ can be estimated as:

\begin{equation}
\langle \hat{O} \rangle_{\beta} = \frac{1}{N_s} \sum_{i=1}^{n} \langle \psi_{\boldsymbol{\sigma}_i}(\beta) | \hat{O} | \psi_{\boldsymbol{\sigma}_i}(\beta) \rangle,
\end{equation}
where $N_s$ is the number of METTS state samples. In Refs. \cite{hendry2022neural,wagner2023neural}, the product states $\left|\boldsymbol{\sigma}_0\right>$ are prepared by adjusting the parameters of an RBM correspondingly. 

For the imaginary time evolution employed both in the purification and the METTS algorithm, the following equation must be solved:
\begin{equation}
\frac{\partial}{\partial \beta}|\psi(\beta)\rangle = -\frac{1}{2}\hat{H}|\psi(\beta)\rangle,
\end{equation}
where the new wave function after each imaginary time step $\delta \tau$ must stay within the variational manifold. This constraint can lead to a modification of $\beta$. 

Another approach to simulate finite temperature states is based on quantum typicality \cite{wagner2023neural} which utilizes the concept that a single pure state can accurately reproduce the expectation values of an observable in the Gibbs ensemble for large systems. This method approximates an infinite temperature state using a combination of a pair product (PP) wave function $\Phi_{P P}$ and a neural network component $\psi_{\vec{\theta}}$, i.e.
\begin{equation}
\Psi(\boldsymbol{\sigma})=\psi_{\mathbb{\vec{\theta}}}(\mathbf{\boldsymbol{\sigma}}) \Phi_{P P}(\mathbf{\boldsymbol{\sigma}}).
\label{eq:pp_nn}
\end{equation}
Pair product wave functions can model electron interactions within the system, including the prohibition of double occupancy through the use of the Gutzwiller projection. The typical state is then evolved in imaginary time to simulate finite temperatures.

In Ref. \cite{Irikura2020}, a CNN with two input channels $\vec{\sigma}$ and $\vec{\sigma}^\prime$ is used to represent a mixed state $\hat{\rho}(\vec{\sigma},\vec{\sigma}^\prime)$ of a one-dimensional bosonic system. Starting from an infinite temperature state, imaginary time evolution is performed, such that the output of the network is the corresponding matrix elements of density matrix at the desired temperature. As opposed to e.g. purification, this approach does not guarantee the hermiticity and positive definiteness of the density matrix.

In contrast to the works discussed so far, which all use imaginary time evolution, a recent paper \cite{lu2024variational} instead minimizes a modified free energy. Here, the von-Neumann entropy is replaced by the second Rényi entropy, which can be evaluated fairly efficiently. The optimization of neural network parameters is guided by the goal of minimizing this approximation to the free energy.

\subsection{Open Systems \label{sec:OpenSys}}

The state of an open quantum system is described by its density operator $\hat{\rho}$. This makes the simulation of open systems even more challenging than for closed systems, since for density matrices, the curse of dimensionality is even more pronounced as for wave functions, e.g. for a system of $N$ spin-$1/2$ particles the number of coefficients to parameterize $\hat{\rho}$ scales as $4^N$ coefficients  \cite{Reh2021}. The dynamics of open quantum systems is governed by the Lindblad master equation,
\begin{align}
    \dot{\hat{\rho}}=-i\left[ \hat{H}, \hat{\rho}\right]+\sum_i \gamma_i\left((\hat{L}^i)^\dagger\hat{\rho}\hat{L}^i - \frac{1}{2}\{ (\hat{L}^i)^\dagger\hat{L}^i , \hat{\rho}\} \right)
    \label{eq:Lindblad}
\end{align}
where $\left[\dots \right]$($\{\dots \}$) denote (anti-)commutators and $\hat{L}$ are so-called jump operators. The first term describes the unitary dynamics of the system given by $\hat{H}$, the second the non-unitary dynamics due to the dissipation to the environment with strength $\gamma$. Eq. \eqref{eq:Lindblad} can also be expressed as 
\begin{align}
    \frac{\mathrm{d}}{\mathrm{d}t}\hat{\rho}= \hat{\mathcal{L}}\hat{\rho}
    \label{eq:Liouvillian}
\end{align}
with the Liouvillian $\mathcal{\hat{L}}$.\\

\begin{figure}[t]
\centering
\includegraphics[width=0.6\textwidth]{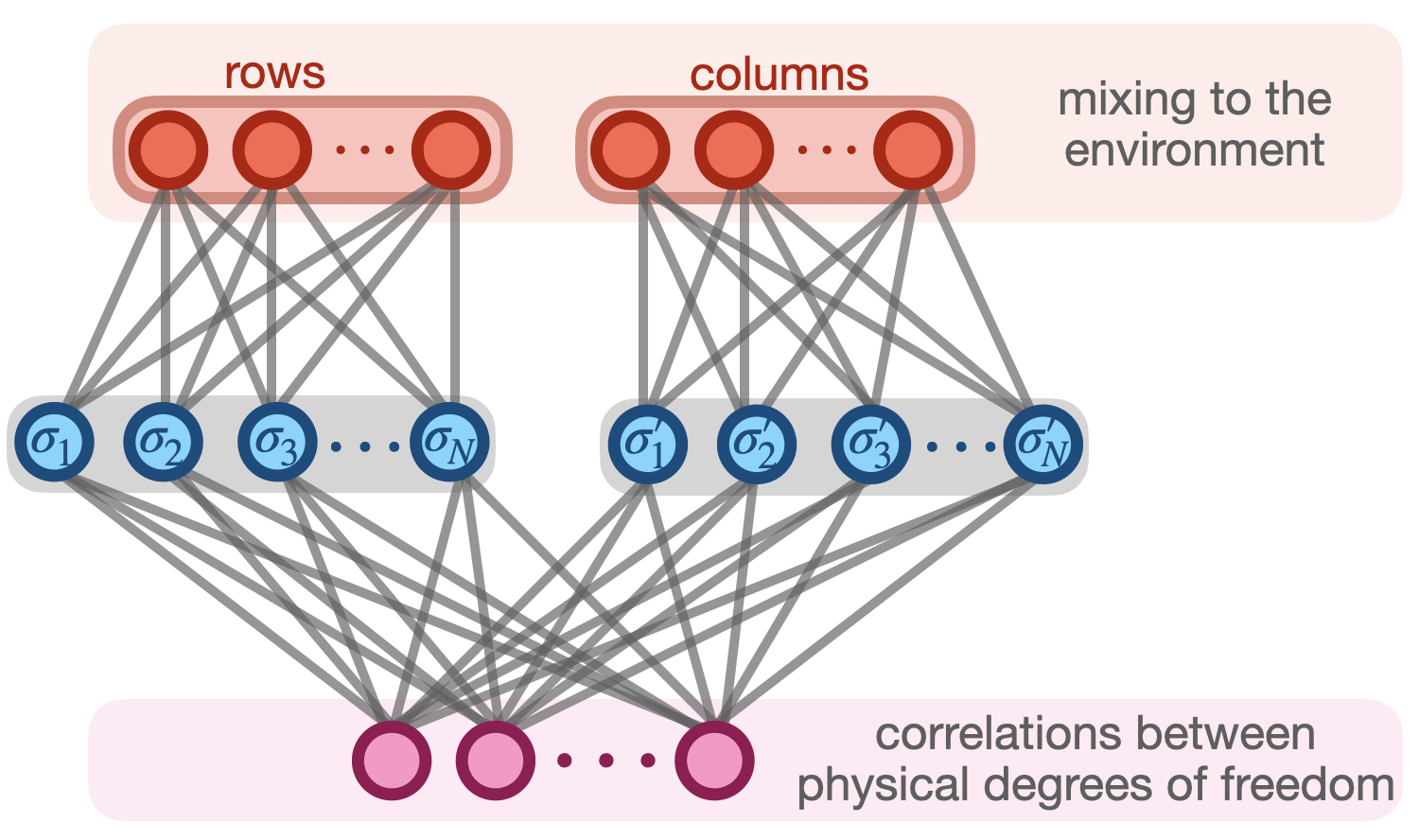}
\caption{Neural density operator based on a RBM as used in Ref. \cite{hartmann2019neural}: 
The state of the physical system is encoded by the visible (blue) layer, with correlations encoded in the bottom hidden layer (purple) and mixing due to the environment encoded in the top hidden layer (red). The left represents row, the right part column indices of the density matrix $\hat{\rho}(\vec{\sigma},\vec{\sigma}^\prime)$. } 
\label{fig:opensys}    
\end{figure}

In most cases, the second form is used to determine the solution $\hat{\rho}$ using NQS. In order to do so, a neural representation of the density operator is needed. This is typically realized by $(i)$ using positive operator valued measures (POVMs) \cite{carrasquilla_reconstructing_2019} or $(ii)$ introducing additional nodes that encode the mixing to the environment \cite{Torlai2018latent,hartmann2019neural}. We would like to point out that the works discussed here in the context of $(i)$ and $(ii)$ are inherently different from machine learning approaches to open quantum systems that e.g. aim learn a parameterization of $\hat{\mathcal{L}}$ and not $\hat{\rho}$ as e.g. in Refs. \cite{Mazza_2021,Carnazza_2022}. 

In the POVM approach \cite{carrasquilla_reconstructing_2019}, the density matrix is represented by a probability distribution over measurement outcomes $a$ of an informationally complete (IC) set of measurement operators $\hat{M}_a$, inspired from Born's rule
\begin{align}
    P_{\vec{\theta}}(a) = \mathrm{Tr}\left( \hat{M}_{a}\hat{\rho}_\mathrm{POVM}\right).
\end{align}
This leads to the definition
\begin{align}
    \hat{\rho}_\mathrm{POVM}= \sum_{a, a^\prime} P_{\vec{\theta}}(a)T^{-1}_{a, a^\prime}\hat{M}_{a^\prime}, 
    \label{eq:POVM}
\end{align}
with the overlap matrix $T_{a, a^\prime}=\mathrm{Tr}\hat{M}_{a}\hat{M}_{a^\prime}$ and different possible choices of $\hat{M}$. The advantage of the IC-POVM representation is that in Eq. \eqref{eq:POVM} only the positive amplitudes $P_{\vec{\theta}}(a)$ have to be modeled by a neural network. However, $\hat{\rho}_\mathrm{POVM}$ is, in general, not a positive-definite matrix. This problem does not occur in the purification ansatz \cite{vicentini2022positivedefinite}.

The second approach is e.g. taken in  Ref. \cite{hartmann2019neural}, where an RBM with an additional hidden layer is used, and both visible and hidden state representations are split into two representations for rows and columns of the density matrix $\hat{\rho}(\vec{\sigma}, \vec{\sigma}^\prime)$, see Fig. \ref{fig:opensys}. Other neural density operators exist, e.g. in terms of CNNs \cite{Herrera2021,mellak2024deep} or in form of an autoregressive network, see Ref. \cite{vicentini2022positivedefinite}. In the latter work, the density matrix is defined as
\begin{align}
     \hat{\rho}({\vec{\sigma}, \vec{\sigma}^\prime}) = \prod_{i=1}^N \sum_{a=1}^R \psi_{\vec{\sigma}_{\leq i},a} (\psi_{\vec{\sigma}^\prime_{\leq i},a})^*,
     \label{eq:GHDO}
 \end{align}
 with ancillas $a$ and neural network representations of $\psi_{\vec{\sigma}_{\leq i},a}$. This is an example of the purification approach discussed in Sec. \ref{sec:finiteT}. In contrast to the POVM approach, this purification via the ancilla nodes in Eq. \eqref{eq:GHDO} makes the density matrix positive semi-definite. Furthermore, each factor $\sum_{a=1}^R \psi_{\vec{\sigma}_{\leq i},a} (\psi_{\vec{\sigma}^\prime_{\leq i},a})^*$ in Eq. \eqref{eq:GHDO} can be normalized, making the neural network representation of $\hat{\rho}$ autoregressive. \\
 
 With these ansätze $(i)$ and $(ii)$, the solution of Eq. \eqref{eq:Liouvillian} can be obtained using different approaches:

\subsubsection*{Time dependent solution of the Lindblad equation:}
Eq. \eqref{eq:Liouvillian} can be solved directly by minimizing $\vert \vert \frac{\mathrm{d}}{\mathrm{d}t}\hat{\rho}- \hat{\mathcal{L}}_{\vec{\theta}}\vert\vert$ using SR, where $\vert \vert  \dots \vert \vert$ can e.g. be taken to be the Fubini-Study distance or the trace norm \cite{hartmann2019neural,Liu2022}. This is done e.g. in Ref. \cite{hartmann2019neural} using an RBM with additional nodes to simulate a 1D anisotropic Heisenberg model or in Ref. \cite{Liu2022} using a deep (quantum) FFNN for a dissipative 1D TFIM and 2D $J_1-J_2$ spin systems. For systems that lack translational invariance, more elaborate sampling and optimization procedures are necessary \cite{Kaestle2021,Mellak2023}. In Refs. \cite{Reh2021,Carrasquilla2021probabilistic,Luo2022} the POVM ansatz implemented with autoregressive networks is used for the time dependent solution of 1D and 2D dissipative Heisenberg models and prototypical states from quantum computing. In order to do so, the time evolution has to be represented in the stochastic representation \eqref{eq:POVM}, i.e. an operator $\hat{O}$ is calculated that time evolves $P_{\vec{\theta}_t}(a)$. Then, 
the parameters $\vec{\theta}_{t+1}$ are selected such that the distance between $\hat{O}P_{\vec{\theta}_t}(a)$ and $ P_{\vec{\theta}_{t+1}}(a)$ is minimal. In Refs. \cite{Carrasquilla2021probabilistic,Luo2022}, this distance is calculated explicitly, e.g. in Ref. \cite{Carrasquilla2021probabilistic} the
the Kullback-Leibler divergence $D_\text{KL}(\hat{O}P_{\vec{\theta}_t}(a)\vert P_{\vec{\theta}_{t+1}}(a))$, see Eq. \eqref{eq:KL}, is minimized. In Ref.~\cite{Luo2022}, the network parameters are optimized to minimize the error between the new, time evolved state and the target state given by the discrete flow according to a second-order forward-backward trapezoid method applied to the Lindblad equation.

In Ref. \cite{Reh2021}, the distance is measured by the Kullback-Leibler or the Hellinger distance, but in this work the distance metrics are expanded around small times, leading to the the time dependent variational principle update, see Eq. \eqref{eq:SR}, for $P_{\vec{\theta}_t}(a)$. This reduces the sampling cost and makes the optimization problem convex. 

\subsubsection*{Steady states:}
Other works use the fact that stationary states $\hat{\rho}_{ss}$ in open systems fulfill 
\begin{align}\frac{\mathrm{d}}{\mathrm{d}t}\hat{\rho}_{ss}=\hat{\mathcal{L}}\hat{\rho}_{ss}=0.
\label{eq:steadystate}
\end{align}
The neural density operator is trained to fulfill this condition by minimizing e.g. the expectation value of $\hat{\mathcal{L}}$ \cite{Weimer2015,Nagy2019} or the $L_2$-norm \cite{Vincentini2019,mellak2024deep}. Furthermore, $\hat{\mathcal{L}}^\dagger$ can be applied from the left to Eq. \eqref{eq:steadystate}, yielding an optimization problem of $\hat{\mathcal{L}}^\dagger \hat{\mathcal{L}}$ instead of $\hat{\mathcal{L}}$, with the advantage that the former operator is hermitean and hence has a real spectrum \cite{Yoshioka2019}. In Ref. \cite{Nagy2019}, the dynamics of a 2D dissipative XYZ spin model is simulated using an RBM with additional nodes. Refs. \cite{Yoshioka2019,Vincentini2019} consider 2D transverse-field Ising models and other similar spin systems.

\section{Learning from Data \label{sec:LearningfromData}}
\subsection{Quantum State Tomography \label{sec:QST}}

\begin{figure}[t]
\centering
\includegraphics[width=0.9\textwidth]{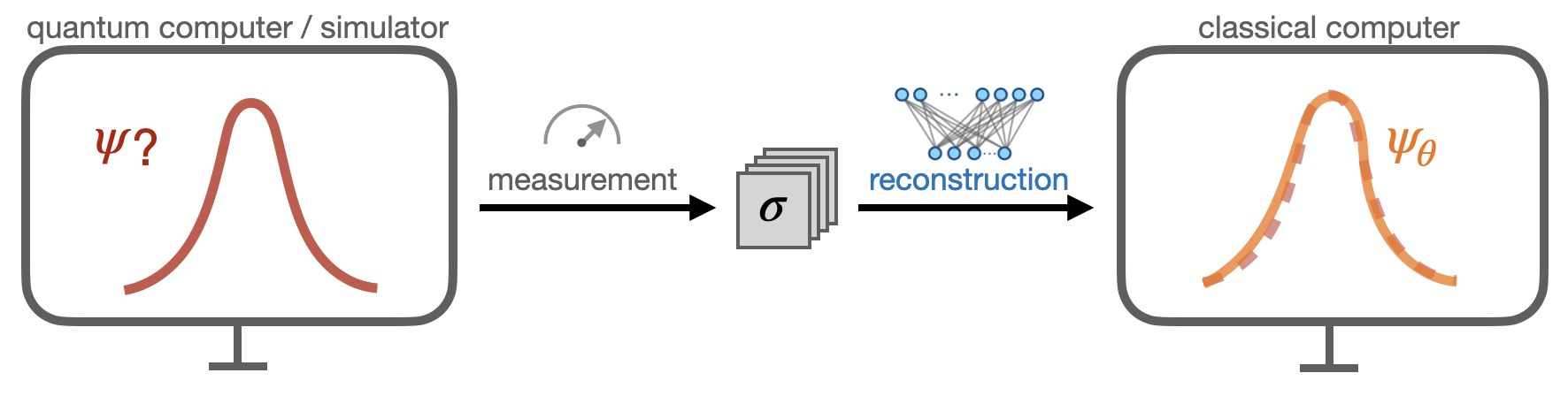}
\caption{Quantum state tomography: Reconstruction of an unknown quantum state $\psi$ on a quantum device, by taking measurements $\vec{\sigma}$ and reconstruction with a neural network. The network is trained such that the amplitudes of the reconstructed state $\psi_{\vec{\theta}}$ are as close as possible to the real measurement distribution generated from $\psi$, see Eq. \eqref{eq:KL}.} 
\label{fig:QST}    
\end{figure}

Quantum state tomography (QST), i.e. the reconstruction of a quantum state from measurement data, plays a crucial role for the characterization and verification of quantum devices \cite{Cramer2010}. For example, it can be applied to compare the experimentally prepared state against the target state to estimate the error of the quantum device under consideration, see Fig. \ref{fig:QST}. Furthermore, QST enables the evaluation of complex observables that would not be accessible directly from experiments \cite{torlai2020precise}.

Full QST relies on two assumptions: $(i)$ Since typically several measurements are needed to infer the quantum state, it is assumed that identical copies of the state can be prepared from which the measurements can be taken. $(ii)$ The set of measurements, described by positive operator valued measures, is informationally complete and hence the probability distribution over measurement outcomes uniquely determine the quantum state via Born's rule. Since these conditions are not fulfilled in most cases, approximate QST schemes are necessary.  \\

Conventional methods for QST, such as linear inversion and maximum likelihood estimation \cite{Haeffner2005, Hradil1997}, are based on inverting Born's rule and hence suffer from an exponential scaling with the system size, resulting from an exponential growth of both the sampling complexity and the number of parameters needed to represent the state. Under these aspects, machine learning techniques have enormous potential for QST: Firstly, machine learning models can learn the structure of a state under consideration, i.e. symmetries or correlations, allowing them to efficiently represent typical physical states with a reduced number of parameters \cite{Carrasquilla2021}. Furthermore, they have the ability to generalize from an incomplete dataset, tackling  the exponential scaling of the sample complexity \cite{Lohani_2020}. In Ref. \cite{Koutny2022}, the authors show that a simple FFNN can outperform conventional methods both in terms of reconstruction time and quality.

The potential of neural quantum states for QST has been explored for various pure and mixed quantum states. One of the first works reconstructs finite temperature states of the 1D and 2D Ising model using real-valued RBMs \cite{Torlai2016}. Furthermore,  highly entangled states with more than a hundred qubits are reconstructed in Ref. \cite{Torlai2018} using a complex RBM. In these works, the NQS is trained by minimizing the Kullback-Leibler divergence between the measurement distribution $q$ and the NQS amplitudes $p_{\vec{\vec{\theta}}}$,
\begin{align}
    D_\mathrm{KL}(q\vert p_{\vec{\vec{\theta}}}) = \sum_{\vec{\sigma}}
    q(\vec{\sigma})\, \mathrm{log}\left(
    \frac{q(\vec{\sigma})}{ p_{\vec{\vec{\theta}}}(\vec{\sigma})}\right)\,
    \label{eq:KL}
\end{align}
on the underlying dataset $\mathcal{D}$ with measurements $\vec{\sigma}\in \mathcal{D}$. In Ref. \cite{wei2023neuralshadow}, instead of $D_{KL}$ the classical shadow formalism (see below) is used to approximate the infidelity between target and reconstructed state.\\

This procedure assumes pure quantum states, which is typically not the case in experimental settings. In some cases, pure representations can nevertheless be used to approximately represent the states under consideration and effect of measurement errors in the training data can be mitigated by modifications of the NQS architecture, as shown in Ref. \cite{Torlai2019} for data from 1D Rydberg tweezer arrays using an RBM with an additional noise layer. However, in many cases, the reconstruction of the full density matrix is needed. Requirements on the density matrix, such as positivity, can be enforced using a purification scheme with additional latent units \cite{Torlai2018latent}. Other works considering mixed state representations involve generative models, neural density operators and POVM representatins \cite{carrasquilla_reconstructing_2019,torlai2020machine,zhao2023empirical}, see Sec. \ref{sec:OpenSys}. In Ref. \cite{melkani2020eigenstate}, an iterative scheme to promote any pure state reconstruction to a mixed state reconstruction is proposed. The reconstruction performance can be increased by filtering the experimental data \cite{palmieri2020experimental} or constraining the density matrix, e.g. to positivity or global symmetries, improve the performance, in particular in the presence of measurement imperfections  \cite{Neugebauer2020, morawetz_u1-symmetric_2021}. 

To further reduce the amount of measurement data needed for the reconstruction, an efficient evaluation of the typically incomplete set of measurements is needed. In Ref. \cite{Huang2020}, the authors present an efficient method for constructing approximate classical descriptions of quantum states from very few measurements, so-called classical shadows, with an information-theoretic bound for the precision of estimated expectation values. The procedure relies on randomly selecting unitaries $\hat{U}$ from an ensemble of particular unitaries that allow one to calculate 
\begin{align}
    \mathcal{M}(\hat{\rho}) = \mathbb{E}\left[ \hat{U}^\dagger \ket{\sigma}\bra{\sigma}\hat{U}\right]
\end{align}
from measurements in the computational basis $\ket{\sigma}$. Here, $\mathbb{E}\left[ \dots \right]$ denotes the average over both the choice of unitary and the outcome distribution. The density matrix $\hat{\rho}$ can be approximated by the so-called classical shadow $\mathcal{S}$ consisting of $N_s$ samples, with 
\begin{align}
    \mathcal{S} = \Bigg\{\hat{\rho}_i=\mathcal{M}^{-1}\left( \hat{U}_i^\dagger \ket{\sigma}_i\bra{\sigma}_i\hat{U}_i \right)  \quad \mathrm{with} \quad i\in 1,\dots,N_s\Bigg\}.
\end{align}
Further works consider the effect of local measurements \cite{Xin2019} or an efficient choice of measurement configurations \cite{Smith2021, Lange_2023}. Hereby, adaptive schemes, that incorporate the knowledge gained from previous measurements to propose the next measurement configuration, are of great interest \cite{quek2018adaptive,Lange_2023}. Moreover,  NQS can be pretrained with artificial data before the measurements to enhance the reconstruction \cite{Wu2022}.\\

Also in the setting of QST, the choice of NQS depends on the state under consideration, with potential advantages e.g. of autoregressive networks and their perfect sampling \cite{Schmale2022} or of networks which can represent a high degree of entanglement \cite{Cha_2022,ma2023attentionbased}. Typical architectures are RBMs \cite{Torlai2016,Torlai2018,tiunov2020experimental,Iouchtchenko_2023} (see Sec. \ref{sec:RBM}), RNNs \cite{morawetz_u1-symmetric_2021,Iouchtchenko_2023} (see Sec. \ref{sec:autoregressive}) and transformer networks \cite{Cha_2022,ma2023attentionbased,zhong2022quantum} (see Sec. \ref{sec:Transformers}) and CNNs \cite{Schmale2022} (see Sec. \ref{sec:CNN}). Furthermore, latent space representations like variational autoencoders are used \cite{Rocchetto2018,Luchnikov2019,Walker2020}. Since these architecture has not been introduced yet, we we shortly describe autoencoders and their application to quantum state reconstruction in the following.

\subsubsection{Latent Space Representations}

\begin{figure}[t]
\centering
\includegraphics[width=0.4\textwidth]{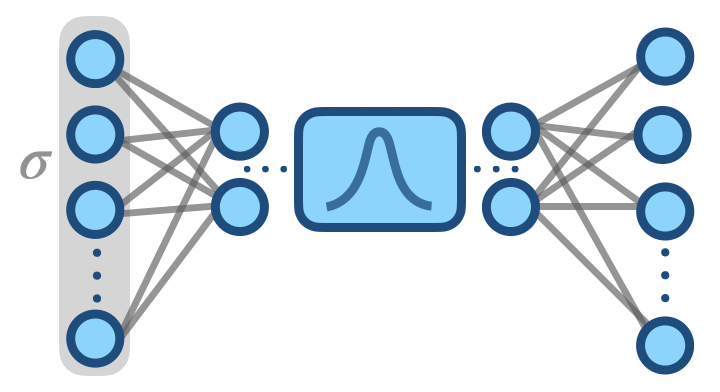}
\caption{Variational autoencoder consisting of an encoder, the latent space or bottleneck layer, and a decoder. The latent space representation is a multivariate Gaussian, i.e. the encoder generates the values for mean and variances of the Gaussian distribution. By sampling from this Gaussian, the input for the decoder layer is generated. } 
\label{fig:Autoencoder}    
\end{figure}

Latent space representations consist of an encoder, the latent space or bottleneck layer, and a decoder, see Fig.~\ref{fig:Autoencoder}. The encoder, typically several fully connected or convolutional layers, compresses the input into just a few nodes in the bottleneck layer. The decoder subsequently generates an output based on the information in the bottleneck layer. The network parameters are optimized such that the generated output is as close  to the input as possible.

In a variational autoencoder \cite{Kingma2022}, the encoder generates the values for mean and variances in the bottleneck layer, and the values used as input for the decoder are then sampled from a multivariate Gaussian. 

Variational autoencoders have been used in Refs. \cite{Rocchetto2018,Luchnikov2019,Walker2020} for (quantum) state reconstruction, where the input consists of the measured data, which the autoencoder learns to compress and de-compress using encoder and decoder. After training, the encoder can be dropped, and by sampling random numbers as input to the latent space, new, uncorrelated samples can be generated. 

In Ref. \cite{Rocchetto2018}, this approach is used to reconstruct positive wave functions, i.e. effectively, the probability distribution of the samples in the computational basis is learned. The efficiency of the compression is quantified by the ratio of the number of network parameters to the Hilbert space dimension. In this case, the size of the latent space is always chosen to correspond to the system size. 
Ref. \cite{Luchnikov2019} uses a conditional variational autoencoder to perform state reconstruction of ground states of the 1D transverse field Ising model based on informationally complete positive-operator valued measures. The magnetic field $h$ is the \textit{condition}, which is used as additional input to the decoder.  
Furthermore, the autoencoder representation of the quantum state is appealing since its latent space contains information on the state under consideration: In Ref. \cite{Walker2020}, the low-dimensional latent space representation of finite temperature samples of the 2D Ising model is used to extract physical features. The authors of Ref. \cite{Schmitt2022} determine the minimal size of the latent space needed to reproduce local observations to measure the local complexity of time-evolved states.

\subsection{Hybrid Training}

The ground state search described in section \ref{sec:GS} typically starts from an NQS with randomly initialized parameters $\vec{\theta}_0$. Assuming convergence of the variational Monte Carlo procedure, the details of this initialization should not matter. However, if the ground state search is challenging, e.g. due to local minima, the question of convergence itself, as well as how many iterations are needed to reach the ground state can crucially depend on the choice of $\vec{\theta}_0$. Using existing data, e.g. from other numerical simulations or an experimental realization, the initial parameters can be chosen such that the ground state search starts from a highly promising region of the parameter space. In this case, the data is used to perform state reconstruction as described in section \ref{sec:QST}. Subsequently, the parameters of the same neural quantum state are variationally minimized to find the ground state.

In an experiment, the ground state is typically not perfectly realized. The combination of experimental data from a state close to the ground state with a subsequent numerical ground state search can yield better results than either approach exhibits on its own, as demonstrated for large, interacting two-dimensional Rydberg atom arrays using quantum Monte Carlo data in \cite{czischek_data-enhanced_2022} and using experimental snapshots from large two-dimensional Rydberg atom arrays \cite{Ebadi2021} in \cite{moss_enhancing_2023}. This is done by minimizing the Kullback-Leibler (KL) divergence between the measurement distribution $q$ and the NQS amplitudes $p_{\vec{\vec{\theta}}}=\vert \psi_{\vec{\theta}}\vert^2$, see Eq. \eqref{eq:KL}. In both cases, a recurrent neural network was used to represent the quantum state. In these examples, the Hamiltonian under consideration is stoquastic, and thus a wave function with real coefficients can represent its ground state. This means in particular that measurements in the computational basis are sufficient for a faithful state reconstruction. 

In Ref. \cite{Bennewitz2022}, molecular Hamiltonians for $\text{LiH}$ and $\text{H}_2$, as well as the one-dimensional lattice Schwinger model are considered. 
The Hamiltonians under consideration are not stoquastic, and thus measurements in different bases are necessary to faitfhully reconstruct their ground states. The computational cost for a single iteration in the reconstruction training is $2^K$, where $K$ is the number of sites on which measurements outside of the computational basis (i.e. in the $x$- or $y$-basis) are performed. In order to constrain this computational cost, measurements with only a few rotations out of the $z$-basis are used for state reconstruction. 
The ground state is prepared using the variational quantum eigensolver on IBM quantum devices, based on superconducting qubits, as well as using numerical simulations. Again, the results show that this hybrid approach improves the numerical and experimental results to yield lower errors for the ground state energy. Moreover, accurate estimations of more complex observables, such as the entanglement entropy, are directly possible without requiring additional quantum resources.

The problem of the exponential cost for a reconstruction based on measurements from configurations away from the computational basis is overcome in Ref. \cite{lange2024transformer} by training on observables like spin-spin correlations instead of the snapshots in the rotated basis. In order to do so, instead of the KL divergence, a variant of the mean-square error loss is used for the rotated basis. This loss function does not incorporate information on the measurement statistics, but allows to compensate for systematic errors that are e.g. present in experimental data, for example by explicitly applying spatial symmetries when calculating the experimental observables. Furthermore, for the computational basis Ref. \cite{lange2024transformer} compares the KL divergence to the Wasserstein (or earth movers distance), with the advantage of the latter that it can incorporate information on the energy to the data-driven pretraining.

In Ref. \cite{montanaro2023accelerating} samples from a quantum state produced by the variational quantum eigensolver, a quantum algorithm which generates the ground state on a quantum device, are used. In contrast to the other works discussed in this context, the experimental samples directly replace the samples needed in the VMC, i.e. at the beginning of the training these samples are used to calculate the expectation values for the VMC optimization instead of samples generated from the trial function. The authors argue that this gives a better estimation of the expectation values at the beginning of the training, speeding up the convergence. After this first stage, the usual VMC algorithm is used. The method is applied for trial wave functions in form of an NQS and the Gutzwiller wave function to find the ground state of the 1D TFIM and Fermi-Hubbard model, respectively.

\section{Summary and Outlook}
In this review, we discuss neural quantum states (NQS), i.e. variational wave functions that are represented by neural networks. The extensive study of NQS since their proposal in 2017 \cite{Carleo2017}, has revealed that the strengths of neural networks -- namely their great expressive power, their ability to compress information very efficiently and their capability to generalize from a given dataset -- turn out to be extremely helpful for the simulation of quantum systems: 

On the one hand, their expressivity permits the representation of a broad range of quantum states, including a variety of (frustrated) spin systems as well as bosonic and fermionic quantum many-body states in one, two and even three dimensions that we review in this article. The limits of this expressive power are still topic of current research. Second, the efficiency of NQS allows to compress the exponential number of wave function coefficients w.r.t. the system size into a tractable number of network parameters, competitive with state of the art numerical methods. This makes NQS very versatile and applicable in many different contexts in the field of numerical simulation of quantum systems and quantum state tomography. However, finding the targeted state in the huge and complicated optimization landscape with many local minima remains one of the main challenges of the NQS approach, and the optimization depends sensitively on the choice of architecture, hyperparameters and the specific optimization strategies, as we discuss in this review. Advanced training strategies, among them stochastic reconfiguration (SR) which incorporates the geometric structure of the loss landscape, are crucial to overcome this problem. For example, recent results on spin systems obtained with a modified version of SR even start to reach numerical precision in terms of the obtained ground state energies \cite{chen2023efficient}. Furthermore, hybrid approaches, allowing to choose a good starting point of the optimization obtained from training on experimental or numerical data \cite{czischek_data-enhanced_2022,moss_enhancing_2023} or from the initialization of certain NQS from TNs \cite{Wu2023}, allow to combine the advantages of existing methods and NQS and are hence promising to overcome this challenge. Finally, the capacity of neural networks to generalize from the training data makes NQS an excellent platform for exploring innovative ideas that go beyond existing methods, such as the creation of toolboxes for simulating entire phase diagrams rather than individual states \cite{zhang_transformer_2023,Zhu2023}. Harnessed with these strengths and versatility, neural quantum states offer a new and promising perspective on the challenges posed by simulating quantum many-body systems.

\vspace{0.5cm}
\subsection*{Acknowledgements} We gratefully acknowledge discussions with 
Annika B\"ohler,
Fabian D\"oschl,
Fabian Grusdt,
Juan Carrasquilla,
Lukas Vetter,
Markus Schmitt,
Moritz Reh, 
Roeland Wiersema,
Roger Melko, 
Schuyler Moss,
Simon Linsel, 
Stefanie Czischek,
and Tizian Blatz.
We acknowledge funding by the Deutsche Forschungsgemeinschaft (DFG, German Research Foundation) under Germany’s Excellence Strategy—EXC-2111—390814868 and within Project-ID 314695032 – SFB 1277. HL acknowledges support by the International Max Planck Research School.


\section*{References}
\bibliographystyle{iopart-num}
\bibliography{sample}

\providecommand{\newblock}{}
\begin{thebibliography}{100}
\expandafter\ifx\csname url\endcsname\relax
  \def\url#1{{\tt #1}}\fi
\expandafter\ifx\csname urlprefix\endcsname\relax\def\urlprefix{URL }\fi
\providecommand{\eprint}[2][]{\url{#2}}

\bibitem{ORUS2014117}
Orús R 2014 {\em Annals of Physics\/} {\bf 349} 117--158 ISSN 0003-4916 \urlprefix\url{https://www.sciencedirect.com/science/article/pii/S0003491614001596}

\bibitem{Zwolak2004}
Zwolak M and Vidal G 2004 {\em Phys. Rev. Lett.\/} {\bf 93}(20) 207205 \urlprefix\url{https://link.aps.org/doi/10.1103/PhysRevLett.93.207205}

\bibitem{Evenbly2014}
Evenbly G and Vidal G 2014 {\em Phys. Rev. Lett.\/} {\bf 112}(24) 240502 \urlprefix\url{https://link.aps.org/doi/10.1103/PhysRevLett.112.240502}

\bibitem{Vidal2007}
Vidal G 2007 {\em Phys. Rev. Lett.\/} {\bf 99}(22) 220405 \urlprefix\url{https://link.aps.org/doi/10.1103/PhysRevLett.99.220405}

\bibitem{Shi2006}
Shi Y~Y, Duan L~M and Vidal G 2006 {\em Phys. Rev. A\/} {\bf 74}(2) 022320 \urlprefix\url{https://link.aps.org/doi/10.1103/PhysRevA.74.022320}

\bibitem{Cincio2008}
Cincio L, Dziarmaga J and Rams M~M 2008 {\em Phys. Rev. Lett.\/} {\bf 100}(24) 240603 \urlprefix\url{https://link.aps.org/doi/10.1103/PhysRevLett.100.240603}

\bibitem{verstraete2004renormalization}
Verstraete F and Cirac J~I 2004 Renormalization algorithms for quantum-many body systems in two and higher dimensions (\textit{Preprint} \eprint{{https://arxiv.org/abs/cond-mat/0407066}})

\bibitem{Klumper_1993}
Klümper A, Schadschneider A and Zittartz J 1993 {\em Europhysics Letters\/} {\bf 24} 293 \urlprefix\url{https://dx.doi.org/10.1209/0295-5075/24/4/010}

\bibitem{SCHOLLWOCK201196}
Schollwöck U 2011 {\em Annals of Physics\/} {\bf 326} 96--192 ISSN 0003-4916 january 2011 Special Issue \urlprefix\url{https://www.sciencedirect.com/science/article/pii/S0003491610001752}

\bibitem{White1992}
White S~R 1992 {\em Phys. Rev. Lett.\/} {\bf 69}(19) 2863--2866 \urlprefix\url{https://link.aps.org/doi/10.1103/PhysRevLett.69.2863}

\bibitem{Schollwock2005}
Schollw\"ock U 2005 {\em Rev. Mod. Phys.\/} {\bf 77}(1) 259--315 \urlprefix\url{https://link.aps.org/doi/10.1103/RevModPhys.77.259}

\bibitem{Hastings_2007}
Hastings M~B 2007 {\em Journal of Statistical Mechanics: Theory and Experiment\/} {\bf 2007} P08024 \urlprefix\url{https://dx.doi.org/10.1088/1742-5468/2007/08/P08024}

\bibitem{Eisert2010}
Eisert J, Cramer M and Plenio M~B 2010 {\em Rev. Mod. Phys.\/} {\bf 82}(1) 277--306 \urlprefix\url{https://link.aps.org/doi/10.1103/RevModPhys.82.277}

\bibitem{Orus2019}
Or{\'u}s R 2019 {\em Nature Reviews Physics\/} {\bf 1} 538--550 \urlprefix\url{https://doi.org/10.1038/s42254-019-0086-7}

\bibitem{becca_sorella_2017}
Becca F and Sorella S 2017 {\em Quantum Monte Carlo Approaches for Correlated Systems\/} (Cambridge University Press)

\bibitem{Ceperley1986}
Ceperley D and Alder B 1986 {\em Science\/} {\bf 231} 555--560 \urlprefix\url{https://www.science.org/doi/abs/10.1126/science.231.4738.555}

\bibitem{Foulkes2001}
Foulkes W~M~C, Mitas L, Needs R~J and Rajagopal G 2001 {\em Rev. Mod. Phys.\/} {\bf 73}(1) 33--83 \urlprefix\url{https://link.aps.org/doi/10.1103/RevModPhys.73.33}

\bibitem{PAN2024879}
Pan G and Meng Z~Y 2024 The sign problem in quantum monte carlo simulations {\em Encyclopedia of Condensed Matter Physics (Second Edition)\/} ed Chakraborty T (Oxford: Academic Press) pp 879--893 second edition ed ISBN 978-0-323-91408-6 \urlprefix\url{https://www.sciencedirect.com/science/article/pii/B9780323908009000950}

\bibitem{troyer2005computational}
Troyer M and Wiese U~J 2005 {\em Phys. Rev. Lett.\/} {\bf 94}(17) 170201 \urlprefix\url{https://link.aps.org/doi/10.1103/PhysRevLett.94.170201}

\bibitem{carrasquilla2020machine}
Carrasquilla J 2020 {\em Advances in Physics: X\/} {\bf 5} 1797528 (\textit{Preprint} \eprint{https://doi.org/10.1080/23746149.2020.1797528}) \urlprefix\url{https://doi.org/10.1080/23746149.2020.1797528}

\bibitem{Cybenco1989}
Cybenko G 1989 {\em Mathematics of Control, Signals and Systems\/} {\bf 2} 303--314 \urlprefix\url{https://doi.org/10.1007/BF02551274}

\bibitem{HORNIK1991251}
Hornik K 1991 {\em Neural Networks\/} {\bf 4} 251--257 ISSN 0893-6080 \urlprefix\url{https://www.sciencedirect.com/science/article/pii/089360809190009T}

\bibitem{Kim2003}
Kim T and Adalı T 2003 {\em Neural Computation\/} {\bf 15} 1641--1666 ISSN 0899-7667 \urlprefix\url{https://doi.org/10.1162/089976603321891846}

\bibitem{Roux2008}
Le~Roux N and Bengio Y 2008 {\em Neural Comput.\/} {\bf 20} 1631–1649 ISSN 0899-7667 \urlprefix\url{https://doi.org/10.1162/neco.2008.04-07-510}

\bibitem{Carleo2017}
Carleo G and Troyer M 2017 {\em Science\/} {\bf 355} 602--606 \urlprefix\url{https://www.science.org/doi/abs/10.1126/science.aag2302}

\bibitem{Sharir2022}
Sharir O, Shashua A and Carleo G 2022 {\em Phys. Rev. B\/} {\bf 106}(20) 205136 \urlprefix\url{https://link.aps.org/doi/10.1103/PhysRevB.106.205136}

\bibitem{Deng2017}
Deng D~L, Li X and Das~Sarma S 2017 {\em Phys. Rev. X\/} {\bf 7}(2) 021021 \urlprefix\url{https://link.aps.org/doi/10.1103/PhysRevX.7.021021}

\bibitem{Gao2017}
Gao X and Duan L~M 2017 {\em Nature Communications\/} {\bf 8} 662 \urlprefix\url{https://doi.org/10.1038/s41467-017-00705-2}

\bibitem{denis2023comment}
Denis Z, Sinibaldi A and Carleo G 2023 Comment on "can neural quantum states learn volume-law ground states?" (\textit{Preprint} \eprint{https://arxiv.org/pdf/2309.11534.pdf})

\bibitem{Levine2019}
Levine Y, Sharir O, Cohen N and Shashua A 2019 {\em Phys. Rev. Lett.\/} {\bf 122}(6) 065301 \urlprefix\url{https://link.aps.org/doi/10.1103/PhysRevLett.122.065301}

\bibitem{Lu2019}
Lu S, Gao X and Duan L~M 2019 {\em Phys. Rev. B\/} {\bf 99}(15) 155136 \urlprefix\url{https://link.aps.org/doi/10.1103/PhysRevB.99.155136}

\bibitem{luo2023gauge}
Luo D, Chen Z, Hu K, Zhao Z, Hur V~M and Clark B~K 2023 Gauge-invariant and anyonic-symmetric autoregressive neural network for quantum lattice models \urlprefix\url{https://link.aps.org/doi/10.1103/PhysRevResearch.5.013216}

\bibitem{Huang2021}
Huang Y and Moore J~E 2021 {\em Phys. Rev. Lett.\/} {\bf 127}(17) 170601 \urlprefix\url{https://link.aps.org/doi/10.1103/PhysRevLett.127.170601}

\bibitem{Sharir2020}
Sharir O, Levine Y, Wies N, Carleo G and Shashua A 2020 {\em Phys. Rev. Lett.\/} {\bf 124}(2) 020503 \urlprefix\url{https://link.aps.org/doi/10.1103/PhysRevLett.124.020503}

\bibitem{lange2023neural}
Lange H, Döschl F, Carrasquilla J and Bohrdt A 2023 Neural network approach to quasiparticle dispersions in doped antiferromagnets (\textit{Preprint} \eprint{https://arxiv.org/abs/2310.08578})

\bibitem{Torlai2020_}
Torlai G, Mazzola G, Carleo G and Mezzacapo A 2020 {\em Phys. Rev. Res.\/} {\bf 2}(2) 022060 \urlprefix\url{https://link.aps.org/doi/10.1103/PhysRevResearch.2.022060}

\bibitem{Iouchtchenko_2023}
Iouchtchenko D, Gonthier J~F, Perdomo-Ortiz A and Melko R~G 2023 {\em Machine Learning: Science and Technology\/} {\bf 4} 015016 \urlprefix\url{https://dx.doi.org/10.1088/2632-2153/acb4df}

\bibitem{Dawid2022}
Dawid A, Arnold J, Requena B, Gresch A, Płodzień M, Donatella K, Nicoli K~A, Stornati P, Koch R, Büttner M, Okuła R, Muñoz-Gil G, Vargas-Hernández R~A, Cervera-Lierta A, Carrasquilla J, Dunjko V, Gabrié M, Huembeli P, van Nieuwenburg E, Vicentini F, Wang L, Wetzel S~J, Carleo G, Greplová E, Krems R, Marquardt F, Tomza M, Lewenstein M and Dauphin A 2022 Modern applications of machine learning in quantum sciences \urlprefix\url{https://arxiv.org/abs/2204.04198}

\bibitem{Carleo2019}
Carleo G, Cirac I, Cranmer K, Daudet L, Schuld M, Tishby N, Vogt-Maranto L and Zdeborova L 2019 {\em Rev. Mod. Phys.\/} {\bf 91}(4) 045002 \urlprefix\url{https://link.aps.org/doi/10.1103/RevModPhys.91.045002}

\bibitem{Carrasquilla2021}
Carrasquilla J and Torlai G 2021 {\em PRX Quantum\/} {\bf 2}(4) 040201 \urlprefix\url{https://link.aps.org/doi/10.1103/PRXQuantum.2.040201}

\bibitem{Melko2024}
Melko R~G and Carrasquilla J 2024 {\em Nature Computational Science\/} \urlprefix\url{https://doi.org/10.1038/s43588-023-00578-0}

\bibitem{Jia2019}
Jia Z~A, Yi B, Zhai R, Wu Y~C, Guo G~C and Guo G~P 2019 {\em Advanced Quantum Technologies\/} {\bf 2} 1800077 \urlprefix\url{https://onlinelibrary.wiley.com/doi/abs/10.1002/qute.201800077}

\bibitem{Yang2019}
Yang Y, Cao H and Zhang Z 2019 {\em Science China Physics, Mechanics \& Astronomy\/} {\bf 63} 210312 \urlprefix\url{https://doi.org/10.1007/s11433-018-9407-5}

\bibitem{vivas2022neuralnetwork}
Vivas D~R, Madroñero J, Bucheli V, Gómez L~O and Reina J~H 2022 Neural-network quantum states: A systematic review (\textit{Preprint} \eprint{{https://arxiv.org/abs/2204.12966}})

\bibitem{Reh2023}
Reh M, Schmitt M and G\"arttner M 2023 {\em Phys. Rev. B\/} {\bf 107}(19) 195115 \urlprefix\url{https://link.aps.org/doi/10.1103/PhysRevB.107.195115}

\bibitem{medvidović2024neuralnetwork}
Medvidović M and Moreno J~R 2024 Neural-network quantum states for many-body physics (\textit{Preprint} \eprint{https://arxiv.org/abs/2402.11014})

\bibitem{Chen2018}
Chen J, Cheng S, Xie H, Wang L and Xiang T 2018 {\em Phys. Rev. B\/} {\bf 97}(8) 085104 \urlprefix\url{https://link.aps.org/doi/10.1103/PhysRevB.97.085104}

\bibitem{Glasser2018}
Glasser I, Pancotti N, August M, Rodriguez I~D and Cirac J~I 2018 {\em Phys. Rev. X\/} {\bf 8}(1) 011006 \urlprefix\url{https://link.aps.org/doi/10.1103/PhysRevX.8.011006}

\bibitem{Wu2023}
Wu D, Rossi R, Vicentini F and Carleo G 2023 {\em Phys. Rev. Res.\/} {\bf 5}(3) L032001 \urlprefix\url{https://link.aps.org/doi/10.1103/PhysRevResearch.5.L032001}

\bibitem{Passetti_2023}
Passetti G, Hofmann D, Neitemeier P, Grunwald L, Sentef M~A and Kennes D~M 2023 {\em Physical Review Letters\/} {\bf 131} \urlprefix\url{https://doi.org/10.1103%2Fphysrevlett.131.036502}

\bibitem{chen2023antn}
Chen Z, Newhouse L, Chen E, Luo D and Soljačić M 2023 Antn: Bridging autoregressive neural networks and tensor networks for quantum many-body simulation (\textit{Preprint} \eprint{https://arxiv.org/abs/2304.01996})

\bibitem{Nest}
Van Den~Nest M 2011 {\em Quantum Info. Comput.\/} {\bf 11} 784–812 ISSN 1533-7146 \urlprefix\url{https://dl.acm.org/doi/10.5555/2230936.2230941}

\bibitem{Choo2018}
Choo K, Carleo G, Regnault N and Neupert T 2018 {\em Phys. Rev. Lett.\/} {\bf 121}(16) 167204 \urlprefix\url{https://link.aps.org/doi/10.1103/PhysRevLett.121.167204}

\bibitem{Cai2018}
Cai Z and Liu J 2018 {\em Phys. Rev. B\/} {\bf 97}(3) 035116 \urlprefix\url{https://link.aps.org/doi/10.1103/PhysRevB.97.035116}

\bibitem{zhang2024paths}
Zhang W, Xing B, Xu X and Poletti D 2024 {\em arXiv preprint arXiv:2406.03381\/} (\textit{Preprint} \eprint{https://arxiv.org/abs/2406.03381})

\bibitem{ceven2022}
\ifmmode~\mbox{\c{C}}\else \c{C}\fi{}even K, Oktel M~O and Kele\ifmmode~\mbox{\c{s}}\else \c{s}\fi{} A 2022 {\em Phys. Rev. A\/} {\bf 106}(6) 063320 \urlprefix\url{https://link.aps.org/doi/10.1103/PhysRevA.106.063320}

\bibitem{Zhu2023}
Zhu Z, Mattheakis M, Pan W and Kaxiras E 2023 {\em Phys. Rev. Res.\/} {\bf 5}(4) 043084 \urlprefix\url{https://link.aps.org/doi/10.1103/PhysRevResearch.5.043084}

\bibitem{Saito2018}
Saito H and Kato M 2018 {\em Journal of the Physical Society of Japan\/} {\bf 87} 014001 (\textit{Preprint} \eprint{https://doi.org/10.7566/JPSJ.87.014001}) \urlprefix\url{https://doi.org/10.7566/JPSJ.87.014001}

\bibitem{Westerhout2020}
Westerhout T, Astrakhantsev N, Tikhonov K~S, Katsnelson M~I and Bagrov A~A 2020 {\em Nature Communications\/} {\bf 11} 1593 \urlprefix\url{https://doi.org/10.1038/s41467-020-15402-w}

\bibitem{Ferrari2019}
Ferrari F, Becca F and Carrasquilla J 2019 {\em Phys. Rev. B\/} {\bf 100}(12) 125131 \urlprefix\url{https://link.aps.org/doi/10.1103/PhysRevB.100.125131}

\bibitem{Nomura2021_Dirac}
Nomura Y and Imada M 2021 {\em Phys. Rev. X\/} {\bf 11}(3) 031034 \urlprefix\url{https://link.aps.org/doi/10.1103/PhysRevX.11.031034}

\bibitem{Nomura_2021}
Nomura Y 2021 {\em Journal of Physics: Condensed Matter\/} {\bf 33} 174003 \urlprefix\url{https://dx.doi.org/10.1088/1361-648X/abe268}

\bibitem{Vieijra2020}
Vieijra T, Casert C, Nys J, De~Neve W, Haegeman J, Ryckebusch J and Verstraete F 2020 {\em Phys. Rev. Lett.\/} {\bf 124}(9) 097201 \urlprefix\url{https://link.aps.org/doi/10.1103/PhysRevLett.124.097201}

\bibitem{Vieijra2021}
Vieijra T and Nys J 2021 {\em Phys. Rev. B\/} {\bf 104}(4) 045123 \urlprefix\url{https://link.aps.org/doi/10.1103/PhysRevB.104.045123}

\bibitem{Li2021}
Li C~X, Yang S and Xu J~B 2021 {\em Scientific Reports\/} {\bf 11} 16667 \urlprefix\url{https://doi.org/10.1038/s41598-021-95523-4}

\bibitem{Deng2017topological}
Deng D~L, Li X and Das~Sarma S 2017 {\em Phys. Rev. B\/} {\bf 96}(19) 195145 \urlprefix\url{https://link.aps.org/doi/10.1103/PhysRevB.96.195145}

\bibitem{Clark_2018}
Clark S~R 2018 {\em Journal of Physics A: Mathematical and Theoretical\/} {\bf 51} 135301 \urlprefix\url{https://dx.doi.org/10.1088/1751-8121/aaaaf2}

\bibitem{Kaubruegger2018}
Kaubruegger R, Pastori L and Budich J~C 2018 {\em Phys. Rev. B\/} {\bf 97}(19) 195136 \urlprefix\url{https://link.aps.org/doi/10.1103/PhysRevB.97.195136}

\bibitem{Valenti2022}
Valenti A, Greplova E, Lindner N~H and Huber S~D 2022 {\em Phys. Rev. Res.\/} {\bf 4}(1) L012010 \urlprefix\url{https://link.aps.org/doi/10.1103/PhysRevResearch.4.L012010}

\bibitem{czischek2018quenches}
Czischek S, G\"arttner M and Gasenzer T 2018 {\em Phys. Rev. B\/} {\bf 98}(2) 024311 \urlprefix\url{https://link.aps.org/doi/10.1103/PhysRevB.98.024311}

\bibitem{Hofmann2022}
Hofmann D, Fabiani G, Mentink J~H, Carleo G and Sentef M~A 2022 {\em SciPost Phys.\/} {\bf 12} 165 \urlprefix\url{https://scipost.org/10.21468/SciPostPhys.12.5.165}

\bibitem{Saito2017}
Saito H 2017 {\em Journal of the Physical Society of Japan\/} {\bf 86} 093001 (\textit{Preprint} \eprint{https://doi.org/10.7566/JPSJ.86.093001}) \urlprefix\url{https://doi.org/10.7566/JPSJ.86.093001}

\bibitem{McBrian_2019}
McBrian K, Carleo G and Khatami E 2019 {\em Journal of Physics: Conference Series\/} {\bf 1290} 012005 \urlprefix\url{https://dx.doi.org/10.1088/1742-6596/1290/1/012005}

\bibitem{Vargas2020}
Vargas-Calder\'{o}n V, Vinck-Posada H and Gonz\'{a}lez F~A 2020 {\em Journal of the Physical Society of Japan\/} {\bf 89} 094002 \urlprefix\url{https://doi.org/10.7566/JPSJ.89.094002}

\bibitem{schmitt2020quantum}
Schmitt M and Heyl M 2020 {\em Phys. Rev. Lett.\/} {\bf 125}(10) 100503 \urlprefix\url{https://link.aps.org/doi/10.1103/PhysRevLett.125.100503}

\bibitem{Fabiani2019}
Fabiani G and Mentink J~H 2019 {\em SciPost Phys.\/} {\bf 7} 004 \urlprefix\url{https://scipost.org/10.21468/SciPostPhys.7.1.004}

\bibitem{Fabiani2021}
Fabiani G, Bouman M~D and Mentink J~H 2021 {\em Phys. Rev. Lett.\/} {\bf 127}(9) 097202 \urlprefix\url{https://link.aps.org/doi/10.1103/PhysRevLett.127.097202}

\bibitem{Normura2017}
Nomura Y, Darmawan A~S, Yamaji Y and Imada M 2017 {\em Phys. Rev. B\/} {\bf 96}(20) 205152 \urlprefix\url{https://link.aps.org/doi/10.1103/PhysRevB.96.205152}

\bibitem{Xia2018}
Xia R and Kais S 2018 {\em Nature Communications\/} {\bf 9} 4195 \urlprefix\url{https://doi.org/10.1038/s41467-018-06598-z}

\bibitem{liang2018solving}
Liang X, Liu W~Y, Lin P~Z, Guo G~C, Zhang Y~S and He L 2018 {\em Phys. Rev. B\/} {\bf 98}(10) 104426 \urlprefix\url{https://link.aps.org/doi/10.1103/PhysRevB.98.104426}

\bibitem{choo2019two}
Choo K, Neupert T and Carleo G 2019 {\em Phys. Rev. B\/} {\bf 100}(12) 125124 \urlprefix\url{https://link.aps.org/doi/10.1103/PhysRevB.100.125124}

\bibitem{Szabo2020}
Szab\'o A and Castelnovo C 2020 {\em Phys. Rev. Res.\/} {\bf 2}(3) 033075 \urlprefix\url{https://link.aps.org/doi/10.1103/PhysRevResearch.2.033075}

\bibitem{Liang2021}
Liang X, Dong S~J and He L 2021 {\em Phys. Rev. B\/} {\bf 103}(3) 035138 \urlprefix\url{https://link.aps.org/doi/10.1103/PhysRevB.103.035138}

\bibitem{Li2022}
Li M, Chen J, Xiao Q, Wang F, Jiang Q, Zhao X, Lin R, An H, Liang X and He L 2022 {\em IEEE Transactions on Parallel and Distributed Systems\/} {\bf 33} 2846--2859 \urlprefix\url{https://ieeexplore.ieee.org/document/9693260}

\bibitem{Liang_2023}
Liang X, Li M, Xiao Q, Chen J, Yang C, An H and He L 2023 {\em Machine Learning: Science and Technology\/} {\bf 4} 015035 \urlprefix\url{https://dx.doi.org/10.1088/2632-2153/acc56a}

\bibitem{wang2023variational}
Wang J~Q, He R~Q and Lu Z~Y 2023 Variational optimization of the amplitude of neural-network quantum many-body ground states (\textit{Preprint} \eprint{https://arxiv.org/abs/2308.09664})

\bibitem{chen2023efficient}
Chen A and Heyl M 2023 Efficient optimization of deep neural quantum states toward machine precision (\textit{Preprint} \eprint{https://arxiv.org/abs/2302.01941})

\bibitem{MendesSantos2023}
Mendes-Santos T, Schmitt M and Heyl M 2023 {\em Phys. Rev. Lett.\/} {\bf 131}(4) 046501 \urlprefix\url{https://link.aps.org/doi/10.1103/PhysRevLett.131.046501}

\bibitem{Schmitt2022_quantumphase}
Schmitt M, Rams M~M, Dziarmaga J, Heyl M and Zurek W~H 2022 {\em Science Advances\/} {\bf 8} eabl6850 (\textit{Preprint} \eprint{https://www.science.org/doi/pdf/10.1126/sciadv.abl6850}) \urlprefix\url{https://www.science.org/doi/abs/10.1126/sciadv.abl6850}

\bibitem{Fu2022}
Fu C, Zhang X, Zhang H, Ling H, Xu S and Ji S 2022 Lattice convolutional networks for learning ground states of quantum many-body systems (\textit{Preprint} \eprint{https://arxiv.org/pdf/2206.07370.pdf})

\bibitem{roth2021group}
Roth C and MacDonald A~H 2021 Group convolutional neural networks improve quantum state accuracy (\textit{Preprint} \eprint{https://arxiv.org/abs/2104.05085})

\bibitem{roth2023high}
Roth C, Szab\'o A and MacDonald A~H 2023 {\em Phys. Rev. B\/} {\bf 108}(5) 054410 \urlprefix\url{https://link.aps.org/doi/10.1103/PhysRevB.108.054410}

\bibitem{duric2024spin12}
Duric T, Chung J~H, Yang B and Sengupta P 2024 Spin-1/2 kagome heisenberg antiferromagnet: Machine learning discovery of the spinon pair density wave ground state (\textit{Preprint} \eprint{{https://arxiv.org/abs/2401.02866}})

\bibitem{beck2024phase}
Beck J, Bodky J, Motruk J, Müller T, Thomale R and Ghosh P 2024 Phase diagram of the $j$-$j_d$ heisenberg model on the maple-leaf lattice: Neural networks and density matrix renormalization group (\textit{Preprint} \eprint{https://arxiv.org/abs/2401.04995})

\bibitem{kochkov2021learning}
Kochkov D, Pfaff T, Sanchez-Gonzalez A, Battaglia P and Clark B~K 2021 Learning ground states of quantum hamiltonians with graph networks (\textit{Preprint} \eprint{https://arxiv.org/abs/2110.06390})

\bibitem{yang2020scalable}
Yang L, Hu W and Li L 2020 Scalable variational monte carlo with graph neural ansatz (\textit{Preprint} \eprint{https://arxiv.org/abs/2011.12453})

\bibitem{viteritti_transformer_2023}
Viteritti L~L, Rende R and Becca F 2023 {\em Physical Review Letters\/} {\bf 130} 236401 ISSN 0031-9007, 1079-7114 arXiv:2211.05504 [cond-mat, physics:quant-ph] \urlprefix\url{http://arxiv.org/abs/2211.05504}

\bibitem{zhang_transformer_2023}
Zhang Y~H and Di~Ventra M 2023 {\em Physical Review B\/} {\bf 107} 075147 ISSN 2469-9950, 2469-9969 arXiv:2208.01758 [cond-mat, physics:physics, physics:quant-ph] \urlprefix\url{http://arxiv.org/abs/2208.01758}

\bibitem{rende_optimal_2023}
Rende R, Gerace F, Laio A and Goldt S 2023 Optimal inference of a generalised {Potts} model by single-layer transformers with factored attention arXiv:2304.07235 [cond-mat, stat] \urlprefix\url{http://arxiv.org/abs/2304.07235}

\bibitem{rende_simple_2023}
Rende R, Viteritti L~L, Bardone L, Becca F and Goldt S 2023 A simple linear algebra identity to optimize {Large}-{Scale} {Neural} {Network} {Quantum} {States} arXiv:2310.05715 [cond-mat] (\textit{Preprint} \eprint{http://arxiv.org/abs/2310.05715})

\bibitem{sprague2023variational}
Sprague K and Czischek S 2023 Variational monte carlo with large patched transformers (\textit{Preprint} \eprint{{https://arxiv.org/abs/2306.03921}})

\bibitem{fitzek2024rydberggpt}
Fitzek D, Teoh Y~H, Fung H~P, Dagnew G~A, Merali E, Moss M~S, MacLellan B and Melko R~G 2024 {\em arXiv preprint arXiv:2405.21052\/}

\bibitem{lange2024transformer}
Lange H, Bornet G, Emperauger G, Chen C, Lahaye T, Kienle S, Browaeys A and Bohrdt A 2024 Transformer neural networks and quantum simulators: a hybrid approach for simulating strongly correlated systems (\textit{Preprint} \eprint{https://arxiv.org/abs/2406.00091})

\bibitem{cha2021attention}
Cha P, Ginsparg P, Wu F, Carrasquilla J, McMahon P~L and Kim E~A 2021 {\em Machine Learning: Science and Technology\/} {\bf 3} 01LT01 \urlprefix\url{https://iopscience.iop.org/article/10.1088/2632-2153/ac362b}

\bibitem{Luo2022}
Luo D, Chen Z, Carrasquilla J and Clark B~K 2022 {\em Phys. Rev. Lett.\/} {\bf 128}(9) 090501 \urlprefix\url{https://link.aps.org/doi/10.1103/PhysRevLett.128.090501}

\bibitem{Cha_2022}
Cha P, Ginsparg P, Wu F, Carrasquilla J, McMahon P~L and Kim E~A 2021 {\em Machine Learning: Science and Technology\/} {\bf 3} 01LT01 \urlprefix\url{https://dx.doi.org/10.1088/2632-2153/ac362b}

\bibitem{vonglehn2023selfattention}
von Glehn I, Spencer J~S and Pfau D 2023 A self-attention ansatz for ab-initio quantum chemistry (\textit{Preprint} \eprint{https://arxiv.org/abs/2211.13672})

\bibitem{shang2023solving}
Shang H, Guo C, Wu Y, Li Z and Yang J 2023 Solving schr\"odinger equation with a language model (\textit{Preprint} \eprint{https://arxiv.org/abs/2307.09343})

\bibitem{wu2023nnqstransformer}
Wu Y, Guo C, Fan Y, Zhou P and Shang H 2023 Nnqs-transformer: an efficient and scalable neural network quantum states approach for ab initio quantum chemistry (\textit{Preprint} \eprint{{https://arxiv.org/abs/2306.16705}})

\bibitem{Schmale2022}
Schmale T, Reh M and G{\"a}rttner M 2022 {\em npj Quantum Information\/} {\bf 8} 115 \urlprefix\url{https://doi.org/10.1038/s41534-022-00621-4}

\bibitem{hibat-allah_recurrent_2020}
Hibat-Allah M, Ganahl M, Hayward L~E, Melko R~G and Carrasquilla J 2020 {\em Phys. Rev. Res.\/} {\bf 2} 23358 publisher: American Physical Society \urlprefix\url{https://link.aps.org/doi/10.1103/PhysRevResearch.2.023358}

\bibitem{roth_iterative_2020}
Roth C 2020 Iterative {Retraining} of {Quantum} {Spin} {Models} {Using} {Recurrent} {Neural} {Networks} arXiv:2003.06228 [cond-mat, physics:physics] \urlprefix\url{http://arxiv.org/abs/2003.06228}

\bibitem{morawetz_u1-symmetric_2021}
Morawetz S, Vlugt I~J~S~D, Carrasquilla J and Melko R~G 2021 {\em Phys. Rev. A\/} {\bf 104} 12401 publisher: American Physical Society \urlprefix\url{https://link.aps.org/doi/10.1103/PhysRevA.104.012401}

\bibitem{hibat-allah_variational_2021}
Hibat-Allah M, Inack E~M, Wiersema R, G J~M~R and {Carrasquilla} 2021 {\em Nature Machine Intelligence\/} {\bf 3} 2522--5839 \urlprefix\url{https://doi.org/10.1038/s42256-021-00401-3}

\bibitem{Reh2021}
Reh M, Schmitt M and G\"arttner M 2021 {\em Phys. Rev. Lett.\/} {\bf 127}(23) 230501 \urlprefix\url{https://link.aps.org/doi/10.1103/PhysRevLett.127.230501}

\bibitem{HibatAllah2023}
Hibat-Allah M, Melko R~G and Carrasquilla J 2023 {\em Phys. Rev. B\/} {\bf 108}(7) 075152 \urlprefix\url{https://link.aps.org/doi/10.1103/PhysRevB.108.075152}

\bibitem{Doeschl2023}
Döschl F, Palm F~A, Lange H, Grusdt F and Bohrdt A 2024 Neural network quantum states for the interacting hofstadter model with higher local occupations and long-range interactions (\textit{Preprint} \eprint{https://arxiv.org/abs/2405.04472})

\bibitem{moss_enhancing_2023}
Moss M~S, Ebadi S, Wang T~T, Semeghini G, Bohrdt A, Lukin M~D and Melko R~G 2023 Enhancing variational {Monte} {Carlo} using a programmable quantum simulator arXiv:2308.02647 [cond-mat, physics:quant-ph] \urlprefix\url{http://arxiv.org/abs/2308.02647}

\bibitem{malyshev2023autoregressive}
Malyshev A, Arrazola J~M and Lvovsky A~I 2023 Autoregressive neural quantum states with quantum number symmetries (\textit{Preprint} \eprint{https://arxiv.org/abs/2310.04166})

\bibitem{mendessantos2023wave}
Mendes-Santos T, Schmitt M, Angelone A, Rodriguez A, Scholl P, Williams H~J, Barredo D, Lahaye T, Browaeys A, Heyl M and Dalmonte M 2024 {\em Phys. Rev. X\/} {\bf 14}(2) 021029 \urlprefix\url{https://link.aps.org/doi/10.1103/PhysRevX.14.021029}

\bibitem{MEHTA20191}
Mehta P, Bukov M, Wang C~H, Day A~G, Richardson C, Fisher C~K and Schwab D~J 2019 {\em Physics Reports\/} {\bf 810} 1--124 ISSN 0370-1573 a high-bias, low-variance introduction to Machine Learning for physicists \urlprefix\url{https://www.sciencedirect.com/science/article/pii/S0370157319300766}

\bibitem{Teng2018}
Teng P 2018 {\em Phys. Rev. E\/} {\bf 98}(3) 033305 \urlprefix\url{https://link.aps.org/doi/10.1103/PhysRevE.98.033305}

\bibitem{Hopfield}
Hopfield J~J 1982 {\em Proc Natl Acad Sci U S A\/} {\bf 79} 2554--2558 ISSN 0027-8424 (Print); 1091-6490 (Electronic); 0027-8424 (Linking)

\bibitem{Melko2019}
Melko R~G, Carleo G, Carrasquilla J and Cirac J~I 2019 {\em Nature Physics\/} {\bf 15} 887--892 \urlprefix\url{https://doi.org/10.1038/s41567-019-0545-1}

\bibitem{Torlai2018}
Torlai G, Mazzola G, Carrasquilla J, Troyer M, Melko R and Carleo G 2018 {\em Nature Phys\/} {\bf 14} 447–450 ISSN 0370-1573 \urlprefix\url{https://doi.org/10.1038/s41567-018-0048-5}

\bibitem{aoki2016restricted}
Aoki K~I and Kobayashi T 2016 {\em Modern Physics Letters B\/} {\bf 30} 1650401 (\textit{Preprint} \eprint{https://doi.org/10.1142/S0217984916504017}) \urlprefix\url{https://doi.org/10.1142/S0217984916504017}

\bibitem{Liu_2019}
Liu D, Ran S~J, Wittek P, Peng C, Garc{\'{\i}}a R~B, Su G and Lewenstein M 2019 {\em New Journal of Physics\/} {\bf 21} 073059 \urlprefix\url{https://doi.org/10.1088%2F1367-2630%2Fab31ef}

\bibitem{gan2017holography}
Gan W~C and Shu F~W 2017 {\em International Journal of Modern Physics D\/} {\bf 26} 1743020 \urlprefix\url{https://doi.org/10.1142/S0218271817430209}

\bibitem{Borin2020}
Borin A and Abanin D~A 2020 {\em Phys. Rev. B\/} {\bf 101}(19) 195141 \urlprefix\url{https://link.aps.org/doi/10.1103/PhysRevB.101.195141}

\bibitem{Pastori2019}
Pastori L, Kaubruegger R and Budich J~C 2019 {\em Phys. Rev. B\/} {\bf 99}(16) 165123 \urlprefix\url{https://link.aps.org/doi/10.1103/PhysRevB.99.165123}

\bibitem{Sfondrini2010}
Sfondrini A, Cerrillo J, Schuch N and Cirac J~I 2010 {\em Phys. Rev. B\/} {\bf 81}(21) 214426 \urlprefix\url{https://link.aps.org/doi/10.1103/PhysRevB.81.214426}

\bibitem{Nomura2020}
Nomura Y 2020 {\em Journal of the Physical Society of Japan\/} {\bf 89} 054706 (\textit{Preprint} \eprint{https://doi.org/10.7566/JPSJ.89.054706}) \urlprefix\url{https://doi.org/10.7566/JPSJ.89.054706}

\bibitem{Puente2020}
Alcalde~Puente D and Eremin I~M 2020 {\em Phys. Rev. B\/} {\bf 102}(19) 195148 \urlprefix\url{https://link.aps.org/doi/10.1103/PhysRevB.102.195148}

\bibitem{Karthikconvolutional}
Karthik V and Medhi A 2024 Convolutional restricted boltzmann machine (crbm) correlated variational wave function for the hubbard model on a square lattice: Mott metal-insulator transition (\textit{Preprint} \eprint{https://arxiv.org/abs/2402.02794})

\bibitem{bronstein2021geometric}
Bronstein M~M, Bruna J, Cohen T and Veličković P 2021 Geometric deep learning: Grids, groups, graphs, geodesics, and gauges (\textit{Preprint} \eprint{https://arxiv.org/abs/2104.13478})

\bibitem{kipf2017semisupervised}
Kipf T~N and Welling M 2017 Semi-supervised classification with graph convolutional networks (\textit{Preprint} \eprint{https://arxiv.org/abs/1609.02907})

\bibitem{li2017gated}
Li Y, Tarlow D, Brockschmidt M and Zemel R 2017 Gated graph sequence neural networks (\textit{Preprint} \eprint{https://arxiv.org/abs/1511.05493})

\bibitem{pescia2023}
Pescia G, Nys J, Kim J, Lovato A and Carleo G 2023 Message-passing neural quantum states for the homogeneous electron gas (\textit{Preprint} \eprint{https://arxiv.org/abs/2305.07240})

\bibitem{kim2023neuralnetwork}
Kim J, Pescia G, Fore B, Nys J, Carleo G, Gandolfi S, Hjorth-Jensen M and Lovato A 2023 Neural-network quantum states for ultra-cold fermi gases (\textit{Preprint} \eprint{https://arxiv.org/abs/2305.08831})

\bibitem{luo2023pairingbased}
Luo D, Dai D~D and Fu L 2023 Pairing-based graph neural network for simulating quantum materials (\textit{Preprint} \eprint{https://arxiv.org/pdf/2311.02143.pdf})

\bibitem{bortone2023impact}
Bortone M, Rath Y and Booth G~H 2023 Impact of conditional modelling for universal autoregressive quantum states (\textit{Preprint} \eprint{https://arxiv.org/abs/2306.05917})

\bibitem{steffen_learning_2006}
{Steffen}, Maximilian Z~H~G~S~A and {Udluft} 2006 Learning {Long} {Term} {Dependencies} with {Recurrent} {Neural} {Networks} {\em Artificial {Neural} {Networks} – {ICANN} 2006\/} ed {Andreas}, Włodzisław D, D O~E~K~S and {Stafylopatis} (Springer Berlin Heidelberg) pp 71--80 ISBN 978-3-540-38627-8

\bibitem{chung2014empirical}
Chung J, Gulcehre C, Cho K and Bengio Y 2014 Empirical evaluation of gated recurrent neural networks on sequence modeling (\textit{Preprint} \eprint{{https://arxiv.org/abs/1412.3555}})

\bibitem{graves2007multidimensional}
Graves A, Fernandez S and Schmidhuber J 2007 Multi-dimensional recurrent neural networks (\textit{Preprint} \eprint{https://arxiv.org/abs/0705.2011})

\bibitem{hibat-allah_supplementing_2022}
Hibat-Allah M, Melko R~G and Carrasquilla J 2022 Supplementing {Recurrent} {Neural} {Network} {Wave} {Functions} with {Symmetry} and {Annealing} to {Improve} {Accuracy} arXiv:2207.14314 [cond-mat, physics:physics] \urlprefix\url{http://arxiv.org/abs/2207.14314}

\bibitem{carrasquilla_reconstructing_2019}
Carrasquilla J, Torlai G, Melko R~G and Aolita L 2019 {\em Nature Machine Intelligence\/} {\bf 1} 155--161 ISSN 2522-5839 number: 3 Publisher: Nature Publishing Group \urlprefix\url{https://www.nature.com/articles/s42256-019-0028-1}

\bibitem{verstraete2004matrix}
Verstraete F, Garc\'{\i}a-Ripoll J~J and Cirac J~I 2004 {\em Phys. Rev. Lett.\/} {\bf 93}(20) 207204 \urlprefix\url{https://link.aps.org/doi/10.1103/PhysRevLett.93.207204}

\bibitem{vaswani2017attention}
Vaswani A, Shazeer N, Parmar N, Uszkoreit J, Jones L, Gomez A~N, Kaiser {\L} and Polosukhin I 2017 {\em Advances in neural information processing systems\/} {\bf 30} \urlprefix\url{https://papers.nips.cc/paper_files/paper/2017}

\bibitem{rende2024queries}
Rende R and Viteritti L~L 2024 Are queries and keys always relevant? a case study on transformer wave functions (\textit{Preprint} \eprint{https://arxiv.org/abs/2405.18874})

\bibitem{viteritti2024transformer}
Viteritti L~L, Rende R, Parola A, Goldt S and Becca F 2024 Transformer wave function for the shastry-sutherland model: emergence of a spin-liquid phase (\textit{Preprint} \eprint{https://arxiv.org/abs/2311.16889})

\bibitem{pfau2020ferminet}
Pfau D, Spencer J, de~G~Matthews A and Foulkes W 2020 {\em Phys. Rev. Research\/} {\bf 2}(3) 033429 \urlprefix\url{https://link.aps.org/doi/10.1103/PhysRevResearch.2.033429}

\bibitem{Hermann2020}
Hermann J, Schätzle Z and Noé F 2020 {\em Nature Chemistry\/} {\bf 12}(10) 1755--4349 \urlprefix\url{https://doi.org/10.1038/s41557-020-0544-y}

\bibitem{Schaetzle2021}
Schätzle Z, Hermann J and Noé F 2021 {\em The Journal of Chemical Physics\/} {\bf 154} 124108 ISSN 0021-9606 \urlprefix\url{https://doi.org/10.1063/5.0032836}

\bibitem{Han_2019}
Han J, Zhang L and E W 2019 {\em Journal of Computational Physics\/} {\bf 399} 108929 \urlprefix\url{https://doi.org/10.1016%2Fj.jcp.2019.108929}

\bibitem{pang2022on2}
Pang T, Yan S and Lin M 2022 $o(n^2)$ universal antisymmetry in fermionic neural networks (\textit{Preprint} \eprint{https://arxiv.org/abs/2205.13205})

\bibitem{Stokes2020}
Stokes J, Moreno J~R, Pnevmatikakis E~A and Carleo G 2020 {\em Phys. Rev. B\/} {\bf 102}(20) 205122 \urlprefix\url{https://link.aps.org/doi/10.1103/PhysRevB.102.205122}

\bibitem{Nomura2017}
Nomura Y, Darmawan A~S, Yamaji Y and Imada M 2017 {\em Phys. Rev. B\/} {\bf 96}(20) 205152 \urlprefix\url{https://link.aps.org/doi/10.1103/PhysRevB.96.205152}

\bibitem{liu2023unifying}
Liu Z and Clark B~K 2023 A unifying view of fermionic neural network quantum states: From neural network backflow to hidden fermion determinant states (\textit{Preprint} \eprint{https://arxiv.org/abs/2311.09450})

\bibitem{Moreno2022}
Moreno J~R, Carleo G, Georges A and Stokes J 2022 {\em Proceedings of the National Academy of Sciences\/} {\bf 119} e2122059119 (\textit{Preprint} \eprint{https://www.pnas.org/doi/pdf/10.1073/pnas.2122059119}) \urlprefix\url{https://www.pnas.org/doi/abs/10.1073/pnas.2122059119}

\bibitem{gauvinndiaye2023mott}
Gauvin-Ndiaye C, Tindall J, Moreno J~R and Georges A 2023 Mott transition and volume law entanglement with neural quantum states (\textit{Preprint} \eprint{https://arxiv.org/abs/2311.05749})

\bibitem{Feynman1956}
Feynman R~P and Cohen M 1956 {\em Phys. Rev.\/} {\bf 102}(5) 1189--1204 \urlprefix\url{https://link.aps.org/doi/10.1103/PhysRev.102.1189}

\bibitem{Luo2019}
Luo D and Clark B~K 2019 {\em Phys. Rev. Lett.\/} {\bf 122}(22) 226401 \urlprefix\url{https://link.aps.org/doi/10.1103/PhysRevLett.122.226401}

\bibitem{romero2024spectroscopy}
Romero I, Nys J and Carleo G 2024 Spectroscopy of two-dimensional interacting lattice electrons using symmetry-aware neural backflow transformations (\textit{Preprint} \eprint{https://arxiv.org/pdf/2406.09077})

\bibitem{li2024emergentwignerphasesmoire}
Li X, Qian Y, Ren W, Xu Y and Chen J 2024 Emergent wigner phases in moir\'e superlattice from deep learning (\textit{Preprint} \eprint{https://arxiv.org/abs/2406.11134}) \urlprefix\url{https://arxiv.org/abs/2406.11134}

\bibitem{luo2024simulatingmoirequantummatter}
Luo D, Dai D~D and Fu L 2024 Simulating moir\'e quantum matter with neural network (\textit{Preprint} \eprint{https://arxiv.org/abs/2406.17645}) \urlprefix\url{https://arxiv.org/abs/2406.17645}

\bibitem{Humeniuk2022}
Humeniuk S, Wan Y and Wang L 2022 Autoregressive neural slater-jastrow ansatz for variational monte carlo simulation (\textit{Preprint} \eprint{https://arxiv.org/abs/2210.05871})

\bibitem{Barret2022}
Barrett T~D, Malyshev A and Lvovsky A~I 2022 {\em Nature Machine Intelligence\/} {\bf 4} 351--358 \urlprefix\url{https://doi.org/10.1038/s42256-022-00461-z}

\bibitem{Inui2021}
Inui K, Kato Y and Motome Y 2021 {\em Phys. Rev. Res.\/} {\bf 3}(4) 043126 \urlprefix\url{https://link.aps.org/doi/10.1103/PhysRevResearch.3.043126}

\bibitem{Yoshioka2021}
Yoshioka N, Mizukami W and Nori F 2021 {\em Communications Physics\/} {\bf 4}(1) 2399--3650 \urlprefix\url{https://doi.org/10.1038/s42005-021-00609-0}

\bibitem{Viteritti2022accuarcy}
Viteritti L~L, Ferrari F and Becca F 2022 {\em SciPost Phys.\/} {\bf 12} 166 \urlprefix\url{https://scipost.org/10.21468/SciPostPhys.12.5.166}

\bibitem{Luo2021}
Luo D, Carleo G, Clark B~K and Stokes J 2021 {\em Phys. Rev. Lett.\/} {\bf 127}(27) 276402 \urlprefix\url{https://link.aps.org/doi/10.1103/PhysRevLett.127.276402}

\bibitem{Sehayek2019}
Sehayek D, Golubeva A, Albergo M~S, Kulchytskyy B, Torlai G and Melko R~G 2019 {\em Phys. Rev. B\/} {\bf 100}(19) 195125 \urlprefix\url{https://link.aps.org/doi/10.1103/PhysRevB.100.195125}

\bibitem{Golubeva2022}
Golubeva A and Melko R~G 2022 {\em Phys. Rev. B\/} {\bf 105}(12) 125124 \urlprefix\url{https://link.aps.org/doi/10.1103/PhysRevB.105.125124}

\bibitem{dash2024efficiency}
Dash S, Vicentini F, Ferrero M and Georges A 2024 Efficiency of neural quantum states in light of the quantum geometric tensor (\textit{Preprint} \eprint{https://arxiv.org/abs/2402.01565})

\bibitem{KLASSERT2022104707}
Klassert R, Baumbach A, Petrovici M~A and Gärttner M 2022 {\em iScience\/} {\bf 25} 104707 ISSN 2589-0042 \urlprefix\url{https://www.sciencedirect.com/science/article/pii/S2589004222009798}

\bibitem{Czischek_spiking2022}
Czischek S, Baumbach A, Billaudelle S, Cramer B, Kades L, Pawlowski J~M, Oberthaler M~K, Schemmel J, Petrovici M~A, Gasenzer T and Gärttner M 2022 {\em SciPost Phys.\/} {\bf 12} 039 \urlprefix\url{https://scipost.org/10.21468/SciPostPhys.12.1.039}

\bibitem{Czischek2019sampling}
Czischek S, Pawlowski J~M, Gasenzer T and G\"arttner M 2019 {\em Phys. Rev. B\/} {\bf 100}(19) 195120 \urlprefix\url{https://link.aps.org/doi/10.1103/PhysRevB.100.195120}

\bibitem{netket}
Vicentini F, Hofmann D, Szabó A, Wu D, Roth C, Giuliani C, Pescia G, Nys J, Vargas-Calderón V, Astrakhantsev N and Carleo G 2022 {\em SciPost Phys. Codebases\/}  7 \urlprefix\url{https://scipost.org/10.21468/SciPostPhysCodeb.7}

\bibitem{netket_fidelity}
Sinibaldi A and Vicentini F 2023 Netket fidelity package \urlprefix\url{https://github.com/netket/netket_fidelity}

\bibitem{jVMC}
Schmitt M and Reh M 2022 {\em SciPost Phys. Codebases\/}  2 \urlprefix\url{https://scipost.org/10.21468/SciPostPhysCodeb.2}

\bibitem{jax2018github}
Bradbury J, Frostig R, Hawkins P, Johnson M~J, Leary C, Maclaurin D, Necula G, Paszke A, Vander{P}las J, Wanderman-{M}ilne S and Zhang Q 2018 {JAX}: composable transformations of {P}ython+{N}um{P}y programs \urlprefix\url{http://github.com/google/jax}

\bibitem{Qucumber}
Beach M~J~S, Vlugt I~D, Golubeva A, Huembeli P, Kulchytskyy B, Luo X, Melko R~G, Merali E and Torlai G 2019 {\em SciPost Phys.\/} {\bf 7} 009 \urlprefix\url{https://scipost.org/10.21468/SciPostPhys.7.1.009}

\bibitem{McMillan1965}
McMillan W~L 1965 {\em Phys. Rev.\/} {\bf 138}(2A) A442--A451 \urlprefix\url{https://link.aps.org/doi/10.1103/PhysRev.138.A442}

\bibitem{huang2017accelerated}
Huang L and Wang L 2017 {\em Phys. Rev. B\/} {\bf 95}(3) 035105 \urlprefix\url{https://link.aps.org/doi/10.1103/PhysRevB.95.035105}

\bibitem{Assaraf1999}
Assaraf R and Caffarel M 1999 {\em Phys. Rev. Lett.\/} {\bf 83}(23) 4682--4685 \urlprefix\url{https://link.aps.org/doi/10.1103/PhysRevLett.83.4682}

\bibitem{kingma2017adam}
Kingma D~P and Ba J 2017 Adam: A method for stochastic optimization (\textit{Preprint} \eprint{https://arxiv.org/abs/1412.6980})

\bibitem{loshchilov2019adamW}
Loshchilov I and Hutter F 2019 Decoupled weight decay regularization (\textit{Preprint} \eprint{https://arxiv.org/abs/1711.05101})

\bibitem{Sorella1998}
Sorella S 1998 {\em Phys. Rev. Lett.\/} {\bf 80}(20) 4558--4561 \urlprefix\url{https://link.aps.org/doi/10.1103/PhysRevLett.80.4558}

\bibitem{Sorella2001}
Sorella S 2001 {\em Phys. Rev. B\/} {\bf 64}(2) 024512 \urlprefix\url{https://link.aps.org/doi/10.1103/PhysRevB.64.024512}

\bibitem{Amari1998}
Amari S and Douglas S 1998 Why natural gradient? {\em Proceedings of the 1998 IEEE International Conference on Acoustics, Speech and Signal Processing, ICASSP '98\/} vol~2 pp 1213--1216 vol.2

\bibitem{Amari2019}
Amari S, Karakida R and Oizumi M 2019 Fisher information and natural gradient learning in random deep networks {\em Proceedings of the 22nd International Conference on Artificial Intelligence and Statistics\/} ({\em Proceedings of Machine Learning Research\/} vol~89) ed Chaudhuri K and Sugiyama M (PMLR) pp 694--702 \urlprefix\url{https://proceedings.mlr.press/v89/amari19a.html}

\bibitem{Hackl2020}
Hackl L, Guaita T, Shi T, Haegeman J, Demler E and Cirac J~I 2020 {\em SciPost Phys.\/} {\bf 9} 048 \urlprefix\url{https://scipost.org/10.21468/SciPostPhys.9.4.048}

\bibitem{wagner2023neural}
Wagner D, Klümper A and Sirker J 2024  (\textit{Preprint} \eprint{https://arxiv.org/abs/2311.13799})

\bibitem{Park2020}
Park C~Y and Kastoryano M~J 2020 {\em Phys. Rev. Res.\/} {\bf 2}(2) 023232 \urlprefix\url{https://link.aps.org/doi/10.1103/PhysRevResearch.2.023232}

\bibitem{donatella2023dynamics}
Donatella K, Denis Z, Le~Boit\'e A and Ciuti C 2023 {\em Phys. Rev. A\/} {\bf 108}(2) 022210 \urlprefix\url{https://link.aps.org/doi/10.1103/PhysRevA.108.022210}

\bibitem{Zhao2022}
Zhao X, Li M, Xiao Q, Chen J, Wang F, Shen L, Zhao M, Wu W, An H, He L and Liang X 2022 Ai for quantum mechanics: High performance quantum many-body simulations via deep learning {\em SC22: International Conference for High Performance Computing, Networking, Storage and Analysis\/} pp 1--15

\bibitem{minres}
Choi S~C~T and Saunders M~A 2014 {\em ACM Trans. Math. Softw.\/} {\bf 40} ISSN 0098-3500 \urlprefix\url{https://doi.org/10.1145/2527267}

\bibitem{Zhang2023}
Zhang W, Xu X, Wu Z, Balachandran V and Poletti D 2023 {\em Phys. Rev. B\/} {\bf 107}(16) 165149 \urlprefix\url{https://link.aps.org/doi/10.1103/PhysRevB.107.165149}

\bibitem{Bukov2021}
Bukov M, Schmitt M and Dupont M 2021 {\em SciPost Phys.\/} {\bf 10} 147 \urlprefix\url{https://scipost.org/10.21468/SciPostPhys.10.6.147}

\bibitem{Inack2022}
Inack E~M, Morawetz S and Melko R~G 2022 {\em Condensed Matter\/} {\bf 7} ISSN 2410-3896 \urlprefix\url{https://www.mdpi.com/2410-3896/7/2/38}

\bibitem{khandoker_supplementing_2023}
Khandoker S~A, Abedin J~M and Hibat-Allah M 2023 {\em Machine Learning: Science and Technology\/} {\bf 4} 15026 publisher: IOP Publishing \urlprefix\url{https://dx.doi.org/10.1088/2632-2153/acb895}

\bibitem{Zen2020}
Zen R, My L, Tan R, H\'ebert F, Gattobigio M, Miniatura C, Poletti D and Bressan S 2020 {\em Phys. Rev. E\/} {\bf 101}(5) 053301 \urlprefix\url{https://link.aps.org/doi/10.1103/PhysRevE.101.053301}

\bibitem{Efthymiou}
Efthymiou S, Beach M~J~S and Melko R~G 2019 {\em Phys. Rev. B\/} {\bf 99}(7) 075113 \urlprefix\url{https://link.aps.org/doi/10.1103/PhysRevB.99.075113}

\bibitem{chen2022systematic}
Chen H, Hendry D, Weinberg P and Feiguin A~E 2022 Systematic improvement of neural network quantum states using a lanczos recursion (\textit{Preprint} \eprint{https://arxiv.org/abs/2206.14307})

\bibitem{Giuliani2023learningground}
Giuliani C, Vicentini F, Rossi R and Carleo G 2023 {\em {Quantum}\/} {\bf 7} 1096 ISSN 2521-327X \urlprefix\url{https://doi.org/10.22331/q-2023-08-29-1096}

\bibitem{Ledinauskas2023}
Ledinauskas E and Anisimovas E 2023 {\em SciPost Phys.\/} {\bf 15} 229 \urlprefix\url{https://scipost.org/10.21468/SciPostPhys.15.6.229}

\bibitem{kochkov2018variational}
Kochkov D and Clark B~K 2018 Variational optimization in the ai era: Computational graph states and supervised wave-function optimization (\textit{Preprint} \eprint{https://arxiv.org/abs/1811.12423})

\bibitem{Hristiana2023}
Atanasova H, Bernheimer L and Cohen G 2023 {\em Nature Communications\/} {\bf 14} 3601 \urlprefix\url{https://doi.org/10.1038/s41467-023-39244-4}

\bibitem{Gong2014}
Gong S~S, Zhu W, Sheng D~N, Motrunich O~I and Fisher M~P~A 2014 {\em Phys. Rev. Lett.\/} {\bf 113}(2) 027201 \urlprefix\url{https://link.aps.org/doi/10.1103/PhysRevLett.113.027201}

\bibitem{Lange_2023}
Lange H, Kebri{\v{c}} M, Buser M, Schollwöck U, Grusdt F and Bohrdt A 2023 {\em Quantum\/} {\bf 7} 1129 \urlprefix\url{https://doi.org/10.22331%2Fq-2023-10-09-1129}

\bibitem{Mizusaki2004}
Mizusaki T and Imada M 2004 {\em Phys. Rev. B\/} {\bf 69}(12) 125110 \urlprefix\url{https://link.aps.org/doi/10.1103/PhysRevB.69.125110}

\bibitem{pfau2023natural}
Pfau D, Axelrod S, Sutterud H, von Glehn I and Spencer J~S 2023 Natural quantum monte carlo computation of excited states (\textit{Preprint} \eprint{https://arxiv.org/abs/2308.16848})

\bibitem{sinibaldi2023unbiasing}
Sinibaldi A, Giuliani C, Carleo G and Vicentini F 2023 {\em arXiv preprint arXiv:2305.14294\/}

\bibitem{Lin2022}
Lin S~H and Pollmann F 2022 {\em physica status solidi (b)\/} {\bf 259} 2100172 (\textit{Preprint} \eprint{https://onlinelibrary.wiley.com/doi/pdf/10.1002/pssb.202100172}) \urlprefix\url{https://onlinelibrary.wiley.com/doi/abs/10.1002/pssb.202100172}

\bibitem{gutierrez2022real}
Guti{\'e}rrez I~L and Mendl C~B 2022 {\em Quantum\/} {\bf 6} 627

\bibitem{lee2021neural}
Lee C~K, Patil P, Zhang S and Hsieh C~Y 2021 {\em Phys. Rev. Res.\/} {\bf 3}(2) 023095 \urlprefix\url{https://link.aps.org/doi/10.1103/PhysRevResearch.3.023095}

\bibitem{Burau2021}
Burau H and Heyl M 2021 {\em Phys. Rev. Lett.\/} {\bf 127}(5) 050601 \urlprefix\url{https://link.aps.org/doi/10.1103/PhysRevLett.127.050601}

\bibitem{zhang2020predicting}
Zhang Z~W, Yang S, Wu Y~H, Liu C~X, Han Y~M, Lee C~H, Sun Z, Li G~J and Zhang X 2020 {\em Chinese Physics Letters\/} {\bf 37} 018401

\bibitem{Schmitt2018}
Schmitt M and Heyl M 2018 {\em SciPost Phys.\/} {\bf 4} 013 \urlprefix\url{https://scipost.org/10.21468/SciPostPhys.4.2.013}

\bibitem{Verdel2021}
Verdel R, Schmitt M, Huang Y~P, Karpov P and Heyl M 2021 {\em Phys. Rev. B\/} {\bf 103}(16) 165103 \urlprefix\url{https://link.aps.org/doi/10.1103/PhysRevB.103.165103}

\bibitem{Karpov2021}
Karpov P, Verdel R, Huang Y~P, Schmitt M and Heyl M 2021 {\em Phys. Rev. Lett.\/} {\bf 126}(13) 130401 \urlprefix\url{https://link.aps.org/doi/10.1103/PhysRevLett.126.130401}

\bibitem{Hendry2021}
Hendry D, Chen H, Weinberg P and Feiguin A~E 2021 {\em Phys. Rev. B\/} {\bf 104}(20) 205130 \urlprefix\url{https://link.aps.org/doi/10.1103/PhysRevB.104.205130}

\bibitem{Hendry2019}
Hendry D and Feiguin A~E 2019 {\em Phys. Rev. B\/} {\bf 100}(24) 245123 \urlprefix\url{https://link.aps.org/doi/10.1103/PhysRevB.100.245123}

\bibitem{feiguin2005finite}
Feiguin A~E and White S~R 2005 {\em Phys. Rev. B\/} {\bf 72}(22) 220401 \urlprefix\url{https://link.aps.org/doi/10.1103/PhysRevB.72.220401}

\bibitem{Torlai2018latent}
Torlai G and Melko R~G 2018 {\em Phys. Rev. Lett.\/} {\bf 120}(24) 240503 \urlprefix\url{https://link.aps.org/doi/10.1103/PhysRevLett.120.240503}

\bibitem{nomura2021purifying}
Nomura Y, Yoshioka N and Nori F 2021 {\em Phys. Rev. Lett.\/} {\bf 127}(6) 060601 \urlprefix\url{https://link.aps.org/doi/10.1103/PhysRevLett.127.060601}

\bibitem{nys2023real}
Nys J, Denis Z and Carleo G 2024 {\em Phys. Rev. B\/} {\bf 109}(23) 235120 \urlprefix\url{https://link.aps.org/doi/10.1103/PhysRevB.109.235120}

\bibitem{White2009}
White S~R 2009 {\em Phys. Rev. Lett.\/} {\bf 102}(19) 190601 \urlprefix\url{https://link.aps.org/doi/10.1103/PhysRevLett.102.190601}

\bibitem{stoudenmire2010minimally}
Stoudenmire E~M and White S~R 2010 {\em New Journal of Physics\/} {\bf 12} 055026 \urlprefix\url{https://dx.doi.org/10.1088/1367-2630/12/5/055026}

\bibitem{hendry2022neural}
Hendry D, Chen H and Feiguin A 2022 {\em Phys. Rev. B\/} {\bf 106}(16) 165111 \urlprefix\url{https://link.aps.org/doi/10.1103/PhysRevB.106.165111}

\bibitem{Irikura2020}
Irikura N and Saito H 2020 {\em Phys. Rev. Res.\/} {\bf 2}(1) 013284 \urlprefix\url{https://link.aps.org/doi/10.1103/PhysRevResearch.2.013284}

\bibitem{lu2024variational}
Lu S, Giudice G and Cirac J~I 2024 Variational neural and tensor network approximations of thermal states (\textit{Preprint} \eprint{https://arxiv.org/abs/2401.14243})

\bibitem{hartmann2019neural}
Hartmann M~J and Carleo G 2019 {\em Phys. Rev. Lett.\/} {\bf 122}(25) 250502 \urlprefix\url{https://link.aps.org/doi/10.1103/PhysRevLett.122.250502}

\bibitem{Mazza_2021}
Mazza P~P, Zietlow D, Carollo F, Andergassen S, Martius G and Lesanovsky I 2021 {\em Physical Review Research\/} {\bf 3} ISSN 2643-1564 \urlprefix\url{http://dx.doi.org/10.1103/PhysRevResearch.3.023084}

\bibitem{Carnazza_2022}
Carnazza F, Carollo F, Zietlow D, Andergassen S, Martius G and Lesanovsky I 2022 {\em New Journal of Physics\/} {\bf 24} 073033 \urlprefix\url{https://dx.doi.org/10.1088/1367-2630/ac7df6}

\bibitem{vicentini2022positivedefinite}
Vicentini F, Rossi R and Carleo G 2022 Positive-definite parametrization of mixed quantum states with deep neural networks (\textit{Preprint} \eprint{https://arxiv.org/abs/2206.13488})

\bibitem{Herrera2021}
Herrera~Rodr{\'\i}guez L~E and Kananenka A~A 2021 {\em The Journal of Physical Chemistry Letters\/} {\bf 12} 2476--2483 \urlprefix\url{https://doi.org/10.1021/acs.jpclett.1c00079}

\bibitem{mellak2024deep}
Mellak J, Arrigoni E and von~der Linden W 2024 Deep neural networks as variational solutions for correlated open quantum systems (\textit{Preprint} \eprint{https://arxiv.org/abs/2401.14179})

\bibitem{Liu2022}
Liu Z, Duan L~M and Deng D~L 2022 {\em Phys. Rev. Res.\/} {\bf 4}(1) 013097 \urlprefix\url{https://link.aps.org/doi/10.1103/PhysRevResearch.4.013097}

\bibitem{Kaestle2021}
Kaestle O and Carmele A 2021 {\em Phys. Rev. B\/} {\bf 103}(19) 195420 \urlprefix\url{https://link.aps.org/doi/10.1103/PhysRevB.103.195420}

\bibitem{Mellak2023}
Mellak J, Arrigoni E, Pock T and von~der Linden W 2023 {\em Phys. Rev. B\/} {\bf 107}(20) 205102 \urlprefix\url{https://link.aps.org/doi/10.1103/PhysRevB.107.205102}

\bibitem{Carrasquilla2021probabilistic}
Carrasquilla J, Luo D, P\'erez F, Milsted A, Clark B~K, Volkovs M and Aolita L 2021 {\em Phys. Rev. A\/} {\bf 104}(3) 032610 \urlprefix\url{https://link.aps.org/doi/10.1103/PhysRevA.104.032610}

\bibitem{Weimer2015}
Weimer H 2015 {\em Phys. Rev. Lett.\/} {\bf 114}(4) 040402 \urlprefix\url{https://link.aps.org/doi/10.1103/PhysRevLett.114.040402}

\bibitem{Nagy2019}
Nagy A and Savona V 2019 {\em Phys. Rev. Lett.\/} {\bf 122}(25) 250501 \urlprefix\url{https://link.aps.org/doi/10.1103/PhysRevLett.122.250501}

\bibitem{Vincentini2019}
Vicentini F, Biella A, Regnault N and Ciuti C 2019 {\em Phys. Rev. Lett.\/} {\bf 122}(25) 250503 \urlprefix\url{https://link.aps.org/doi/10.1103/PhysRevLett.122.250503}

\bibitem{Yoshioka2019}
Yoshioka N and Hamazaki R 2019 {\em Phys. Rev. B\/} {\bf 99}(21) 214306 \urlprefix\url{https://link.aps.org/doi/10.1103/PhysRevB.99.214306}

\bibitem{Cramer2010}
Cramer M, Plenio M~B, Flammia S~T, Somma R, Gross D, Bartlett S~D, Landon-Cardinal O, Poulin D and Liu Y~K 2010 {\em Nature Communications\/} {\bf 1} 149 \urlprefix\url{https://doi.org/10.1038/ncomms1147}

\bibitem{torlai2020precise}
Torlai G, Mazzola G, Carleo G and Mezzacapo A 2020 {\em Phys. Rev. Res.\/} {\bf 2}(2) 022060 \urlprefix\url{https://link.aps.org/doi/10.1103/PhysRevResearch.2.022060}

\bibitem{Haeffner2005}
H{\"a}ffner H, H{\"a}nsel W, Roos C~F, Benhelm J, Chek-al kar D, Chwalla M, K{\"o}rber T, Rapol U~D, Riebe M, Schmidt P~O, Becher C, G{\"u}hne O, D{\"u}r W and Blatt R 2005 {\em Nature\/} {\bf 438} 643--646 \urlprefix\url{https://doi.org/10.1038/nature04279}

\bibitem{Hradil1997}
Hradil Z 1997 {\em Phys. Rev. A\/} {\bf 55}(3) R1561--R1564 \urlprefix\url{https://link.aps.org/doi/10.1103/PhysRevA.55.R1561}

\bibitem{Lohani_2020}
Lohani S, Kirby B~T, Brodsky M, Danaci O and Glasser R~T 2020 {\em Machine Learning: Science and Technology\/} {\bf 1} 035007 \urlprefix\url{https://dx.doi.org/10.1088/2632-2153/ab9a21}

\bibitem{Koutny2022}
Koutn\'y D, Motka L, Hradil Z~c~v, \ifmmode \check{R}\else \v{R}\fi{}eh\'a\ifmmode~\check{c}\else \v{c}\fi{}ek J and S\'anchez-Soto L~L 2022 {\em Phys. Rev. A\/} {\bf 106}(1) 012409 \urlprefix\url{https://link.aps.org/doi/10.1103/PhysRevA.106.012409}

\bibitem{Torlai2016}
Torlai G and Melko R~G 2016 {\em Phys. Rev. B\/} {\bf 94}(16) 165134 \urlprefix\url{https://link.aps.org/doi/10.1103/PhysRevB.94.165134}

\bibitem{wei2023neuralshadow}
Wei V, Coish W~A, Ronagh P and Muschik C~A 2023 Neural-shadow quantum state tomography (\textit{Preprint} \eprint{https://arxiv.org/abs/2305.01078})

\bibitem{Torlai2019}
Torlai G, Timar B, van Nieuwenburg E~P~L, Levine H, Omran A, Keesling A, Bernien H, Greiner M, Vuleti\ifmmode~\acute{c}\else \'{c}\fi{} V, Lukin M~D, Melko R~G and Endres M 2019 {\em Phys. Rev. Lett.\/} {\bf 123}(23) 230504 \urlprefix\url{https://link.aps.org/doi/10.1103/PhysRevLett.123.230504}

\bibitem{torlai2020machine}
Torlai G and Melko R~G 2020 {\em Annual Review of Condensed Matter Physics\/} {\bf 11} 325--344 \urlprefix\url{https://www.annualreviews.org/content/journals/10.1146/annurev-conmatphys-031119-050651}

\bibitem{zhao2023empirical}
Zhao H, Carleo G and Vicentini F 2023 Empirical sample complexity of neural network mixed state reconstruction (\textit{Preprint} \eprint{https://arxiv.org/abs/2307.01840})

\bibitem{melkani2020eigenstate}
Melkani A, Gneiting C and Nori F 2020 {\em Phys. Rev. A\/} {\bf 102}(2) 022412 \urlprefix\url{https://link.aps.org/doi/10.1103/PhysRevA.102.022412}

\bibitem{palmieri2020experimental}
Palmieri A~M, Kovlakov E, Bianchi F, Yudin D, Straupe S, Biamonte J~D and Kulik S 2020 {\em npj Quantum Information\/} {\bf 6} 20 \urlprefix\url{https://www.nature.com/articles/s41534-020-0248-6}

\bibitem{Neugebauer2020}
Neugebauer M, Fischer L, J\"ager A, Czischek S, Jochim S, Weidem\"uller M and G\"arttner M 2020 {\em Phys. Rev. A\/} {\bf 102}(4) 042604 \urlprefix\url{https://link.aps.org/doi/10.1103/PhysRevA.102.042604}

\bibitem{Huang2020}
Huang H~Y, Kueng R and Preskill J 2020 {\em Nature Physics\/} {\bf 16} 1050--1057 \urlprefix\url{https://doi.org/10.1038/s41567-020-0932-7}

\bibitem{Xin2019}
Xin T, Lu S, Cao N, Anikeeva G, Lu D, Li J, Long G and Zeng B 2019 {\em npj Quantum Information\/} {\bf 5} 109 \urlprefix\url{https://doi.org/10.1038/s41534-019-0222-3}

\bibitem{Smith2021}
Smith A~W~R, Gray J and Kim M~S 2021 {\em PRX Quantum\/} {\bf 2}(2) 020348 \urlprefix\url{https://link.aps.org/doi/10.1103/PRXQuantum.2.020348}

\bibitem{quek2018adaptive}
Quek Y, Fort S and Ng H~K 2021 Adaptive quantum state tomography with neural networks \urlprefix\url{https://doi.org/10.1038/s41534-021-00436-9}

\bibitem{Wu2022}
Zhu Y, Wu Y~D, Bai G, Wang D~S, Wang Y and Chiribella G 2022 {\em Nature Communications\/} {\bf 13} 6222 \urlprefix\url{https://doi.org/10.1038/s41467-022-33928-z}

\bibitem{ma2023attentionbased}
Ma H, Sun Z, Dong D, Chen C and Rabitz H 2023 Attention-based transformer networks for quantum state tomography (\textit{Preprint} \eprint{https://arxiv.org/abs/2305.05433})

\bibitem{tiunov2020experimental}
Tiunov E~S, Tiunova V, Ulanov A~E, Lvovsky A and Fedorov A~K 2020 {\em Optica\/} {\bf 7} 448--454 \urlprefix\url{https://opg.optica.org/optica/fulltext.cfm?uri=optica-7-5-448&id=431506}

\bibitem{zhong2022quantum}
Zhong L, Guo C and Wang X 2022 Quantum state tomography inspired by language modeling (\textit{Preprint} \eprint{https://arxiv.org/abs/2212.04940})

\bibitem{Rocchetto2018}
Rocchetto A, Grant E, Strelchuk S, Carleo G and Severini S 2018 {\em npj Quantum Information\/} {\bf 4} 28 \urlprefix\url{https://doi.org/10.1038/s41534-018-0077-z}

\bibitem{Luchnikov2019}
Luchnikov I~A, Ryzhov A, Stas P~J, Filippov S~N and Ouerdane H 2019 {\em Entropy\/} {\bf 21} ISSN 1099-4300 \urlprefix\url{https://www.mdpi.com/1099-4300/21/11/1091}

\bibitem{Walker2020}
Walker N, Tam K~M and Jarrell M 2020 {\em Scientific Reports\/} {\bf 10} 13047 \urlprefix\url{https://doi.org/10.1038/s41598-020-69848-5}

\bibitem{Kingma2022}
Kingma D~P and Welling M 2022 Auto-encoding variational bayes (\textit{Preprint} \eprint{https://arxiv.org/abs/1312.6114})

\bibitem{Schmitt2022}
Schmitt M and Lenar\ifmmode \check{c}\else \v{c}\fi{}i\ifmmode~\check{c}\else \v{c}\fi{} Z 2022 {\em Phys. Rev. B\/} {\bf 106}(4) L041110 \urlprefix\url{https://link.aps.org/doi/10.1103/PhysRevB.106.L041110}

\bibitem{czischek_data-enhanced_2022}
Czischek S, Moss M~S, Radzihovsky M, Merali E and Melko R~G 2022 {\em Phys. Rev. B\/} {\bf 105} 205108 publisher: American Physical Society \urlprefix\url{https://link.aps.org/doi/10.1103/PhysRevB.105.205108}

\bibitem{Ebadi2021}
Ebadi S, Wang T~T, Levine H, Keesling A, Semeghini G, Omran A, Bluvstein D, Samajdar R, Pichler H, Ho W~W, Choi S, Sachdev S, Greiner M, Vuleti{\'{c}} V and Lukin M~D 2021 {\em Nature\/} {\bf 595} 227--232 \urlprefix\url{https://doi.org/10.1038%2Fs41586-021-03582-4}

\bibitem{Bennewitz2022}
Bennewitz E~R, Hopfmueller F, Kulchytskyy B, Carrasquilla J and Ronagh P 2022 {\em Nature Machine Intelligence\/} {\bf 4} 618--624 \urlprefix\url{https://doi.org/10.1038/s42256-022-00509-0}

\bibitem{montanaro2023accelerating}
Montanaro A and Stanisic S 2023 Accelerating variational quantum monte carlo using the variational quantum eigensolver (\textit{Preprint} \eprint{https://arxiv.org/abs/2307.07719})

\end{thebibliography}

\end{document}